\documentclass[amsmath,
amssymb,
nofootinbib,
notitlepage,
prd,
superscriptaddress,
twocolumn,
longbibliography,
]{revtex4-1}
\usepackage[T1]{fontenc}
\usepackage[usenames,dvipsnames]{xcolor}
\usepackage{graphicx,hyperref,orcidlink,float}
\hypersetup{colorlinks=true,citecolor=RoyalBlue,linkcolor=RoyalBlue,urlcolor=RoyalBlue}
\usepackage{lipsum}
\usepackage{comment}
\usepackage{appendix}
\usepackage{graphicx}%
\usepackage{enumitem}
\usepackage{adjustbox}

\usepackage{soul}
\usepackage{tikz}
\usepackage{pgfplots}
\usepackage{dcolumn}%
\usepackage{bm}%
\usepackage{amsmath}
\usepackage{anyfontsize}
\usepackage{multirow}
\usepackage{makecell}
\usepackage{listings}
\usepackage[capitalize]{cleveref}
\usepackage{slashed}
\newcommand{\msun}{M_{\odot}}

\newcommand{\tr}[1]{{#1}_{\mathrm{tr}}}
\newcommand{\BF}[1]{\boldsymbol{#1}}

\newcommand{\bftr}[1]{\boldsymbol{#1}_{\mathrm{tr}}}
\newcommand{\ml}[1]{{#1}_{\mathrm{ML}}}
\newcommand{\bfml}[1]{\boldsymbol{#1}_{\mathrm{ML}}}

\newcommand\norm[1]{\left\lVert#1\right\rVert}

\newcommand{\betappe}[1]{\beta_{\mathrm{#1 PN}}}

\begin{document}
\title{
Systematic biases due to waveform mismodeling in parametrized post-Einsteinian tests of general relativity: The impact of neglecting spin precession and higher modes
}
\author{Rohit S. Chandramouli}%
\email{rsc4@illinois.edu}%
  \affiliation{Illinois Center for Advanced Studies of the Universe \& Department of Physics,\\
   University of Illinois Urbana-Champaign, Urbana, Illinois 61801, USA}%
\author{Kaitlyn Prokup}%
  \affiliation{Carthage College, Kenosha, Wisconsin 53140, USA}%
  \author{Emanuele Berti}%
  \affiliation{William H. Miller III Department of Physics and Astronomy, Johns Hopkins
University, Baltimore, Maryland 21218, USA}%
  \author{Nicol\'as Yunes}%
  \affiliation{Illinois Center for Advanced Studies of the Universe \& Department of Physics,\\
   University of Illinois Urbana-Champaign, Urbana, Illinois 61801, USA}%

\date{\today}%
\begin{abstract}

    We study the robustness of parametrized post-Einsteinian (ppE) tests of General Relativity (GR) with gravitational waves, due to waveform inaccuracy.
    In particular, we determine the properties of the signal -- signal-to-noise ratio (SNR) and source parameters -- such that we are led to falsely identify a ppE deviation in the post-Newtonian (PN) inspiral phase at -1PN, 1PN, or 2PN order, due to neglecting spin precession or higher models in the recovery model.
    To characterize the statistical significance of the biases, we compute the Bayes factor between the ppE and GR models, and the fitting factor of the ppE model.
    For highly-precessing, edge-on signals, we find that mismodeling the signal leads to a significant systematic bias in the recovery of the ppE parameters, even at an SNR of 30.
    However, these biased inferences are characterized by a significant loss of SNR and a weak preference for the ppE model.
    At a higher SNR, the biased inferences display a strong preference for the ppE model and a significant loss of SNR.
    For edge-on signals containing asymmetric masses, at an SNR of 30, we find that excluding higher modes does not impact the ppE tests as much as excluding spin precession.
    Our analysis, therefore, identifies the spin-precessing and mass-asymmetric systems for which parametrized tests of GR are robust. 
    With a toy model and using the linear signal approximation, we illustrate these regimes of bias and characterize them by obtaining bounds on the ratio of systematic to statistical error and the effective cycles incurred due to mismodeling.
    As a by-product of our analysis, we connect various measures and techniques commonly used to estimate systematic errors -- linear-signal approximation, Laplace approximation, fitting factor, effective cycles, and Bayes factor -- that apply to all studies of systematic uncertainties in gravitational wave parameter estimation.

\end{abstract}

\maketitle

\section{Introduction} \label{sec:introduction}

Parameter estimation of compact binary source parameters from gravitational-wave (GW) observations~\cite{gwtc-1,gwtc-2,LIGOScientific:2021djp,LIGOScientific:2021usb} has allowed for advances in several  problems at the forefront of gravitational physics, astrophysics, nuclear physics and cosmology. 
As examples, GW measurements have probed deviations from general relativity~\cite{LIGOScientific:2021sio,Sanger:2024axs,LIGOScientific:2016lio,Perkins:2021mhb, Schumacher:2023cxh,Carullo:2021oxn, Gupta:2020lxa, 
Okounkova:2021xjv,Shoom:2021mdj, 
Wang:2021yll, 
Lagos:2024boe,Alexander:2017jmt,Callister:2023tws,Chung:2021rcu,Berti:2018cxi}, nuclear physics~\cite{LIGOScientific:2018cki,Chatziioannou:2018vzf,Raithel:2019ejc,Annala:2017llu,Landry:2020vaw,Tews:2019cap,Bauswein:2019skm,Gamba:2019kwu,Chatziioannou:2020pqz,Yunes:2022ldq,Ripley:2023lsq}, astrophysical formation channels and the rate of mergers~\cite{Antonelli:2023gpu,prop_GW190521_ligo_virgo,KAGRA:2021duu,Zevin:2020gbd}, dark matter~\cite{Zhang:2021mks,Alexander:2018qzg,Guo:2023gfc,Tsutsui:2023jbk,Basak:2021ten,LIGOScientific:2021ffg}, and cosmology~\cite{LIGOScientific:2021aug,Chatterjee:2021xrm}.
In order to make accurate and robust inferences on the underlying physics and astrophysics of GW sources, it is essential to mitigate systematic errors that can bias the inference. 
Understanding these different systematic biases and mitigating them is an essential problem in GW data analysis, especially with upgrades to ground-based detectors~\cite{KAGRA:2013rdx}, upcoming next-generation detectors, such as Cosmic Explorer~\cite{Reitze:2019iox} and the Einstein Telescope~\cite{Punturo:2010zz}, and space-based detectors such as the Laser Interferometer Space Antenna (LISA)~\cite{lisa-white-paper}.

\begin{figure*}
    \centering
    \includegraphics[width=0.7\textwidth]{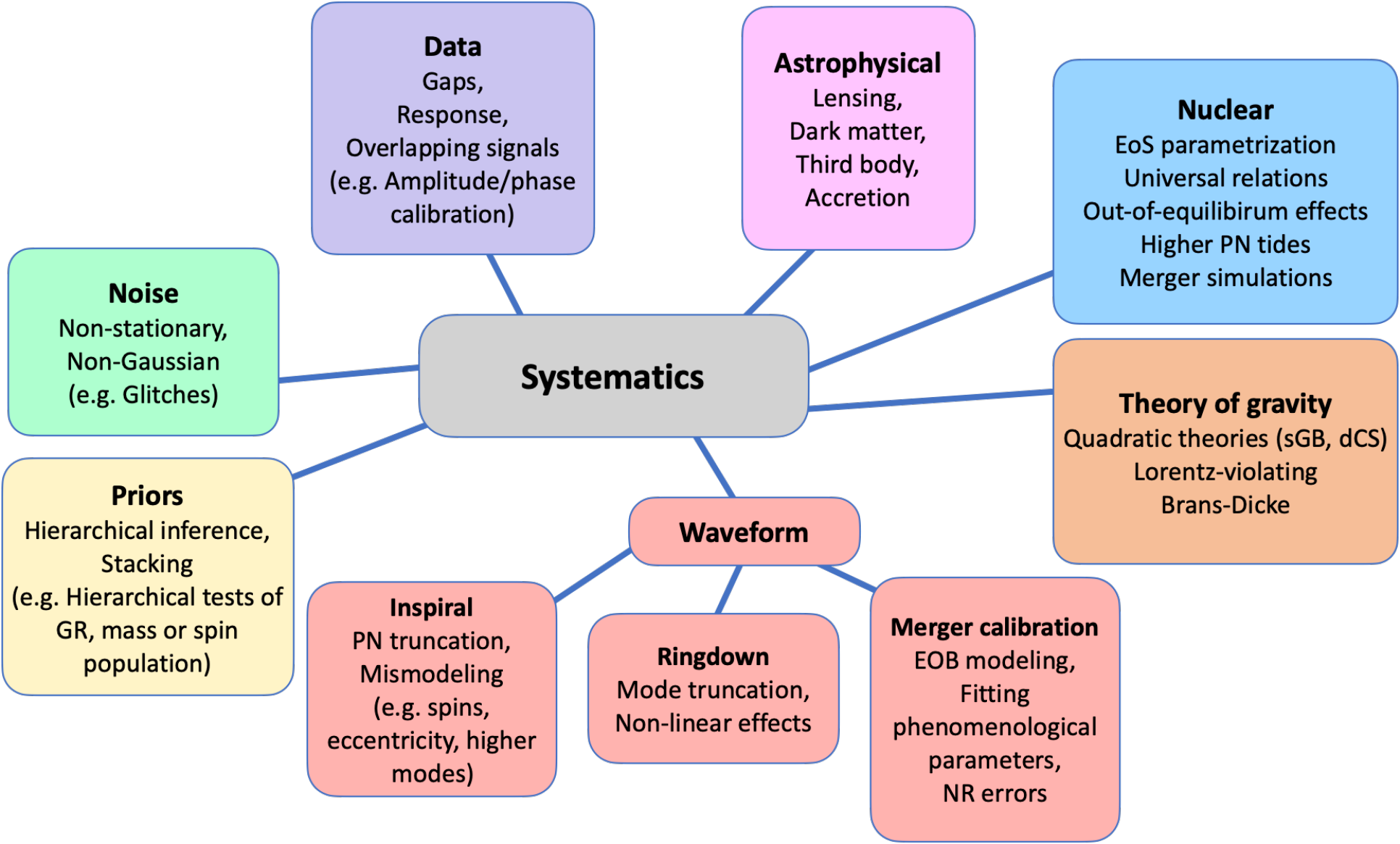}
    \caption{Overview of systematic biases caused by various aspects of GW inference. 
    Different categories of systematics are shown with a few examples (by no means exhaustive).
    In principle, biases can be incurred due to all of the different categories of systematics.
    For example, unmodeled effects due to spins or eccentricity in the waveform will add on to the systematics caused by unmodeled astrophysics, nuclear physics and/or modified gravity (refer to the text for further details).
    In our work, we focus on systematic biases arising from inaccurate inspiral waveforms.
    }
    \label{fig:systematics-overview}
\end{figure*}

As summarized in~\cref{fig:systematics-overview} and Appendix~\ref{sec:lit_review}, the different sources of systematic error in GW inference can be broadly classified into biases induced by the inaccuracies in the data, noise, priors, waveforms, astrophysics, nuclear physics, and the underlying theory of gravity. As detectors become more sensitive, the observed signals will typically become louder, thus resulting in smaller statistical errors.
Consequently, as signals become louder, the different systematic errors can no longer hide behind statistical errors. 
Taming these different sources of systematic error is, thus, a major challenge for performing ``precision'' inferences of physics and astrophysics using GWs.
An important inference we would like to make using GWs is the underlying theory of gravity in play during compact binary mergers.
Mismodeling the noise, response, waveform, astrophysics (among other aspects shown in~\cref{fig:systematics-overview}) can lead to a ``false positive'' inference of a GR deviation, i.e.~the inference that there is a GR deviation when, in reality, there is none.
Reference~\citep{Gupta:2024gun} has recently reviewed the different systematics that lead to such false positive GR deviations.

In this paper, we focus on characterizing systematic biases in parametric tests of GR due to waveform systematics.
More concretely, we study the robustness of tests of GR with models that deviate from GR parametrically but lack certain physics in the GR sector of the model.  
In particular, we study how waveform inaccuracies in the GR sector caused by neglecting spin precession and higher modes can induce systematic biases in the parametrized post-Einsteinian (ppE) tests of GR~\cite{Yunes_2009}. 
The ppE waveform model is constructed by introducing deviations in both the phase and amplitude of the GR waveform.
In general, the deviations are an asymptotic PN series that first enters at a specific PN order.
We will restrict to the simplest formulation, where the deviation is a single PN term.
The ppE deviation is then of the form $\beta u^b$, where $\beta$ is the \emph{ppE parameter}, $b$ is the \emph{ppE index}, and $u$ is the orbital velocity.
For a given ppE index, the ppE parameter maps to specific theories of gravity.
In this work, we focus on phase ppE corrections, as they are typically better constrained than amplitude ppE corrections~\cite{Perkins:2022fh}.
We provide an overview of the ppE model in Sec.~\ref{subsec:rec-models}, and elaborate on additional details in Appendix~\ref{sec:review_ppE}.

An \emph{incorrect inference of a GR deviation}, or equivalently a \emph{false-positive inference of a GR deviation}, can occur when one uses a ppE model to analyze a GR signal, and the ppE parameter recovery is biased away from GR due to a neglected effect in the GR sector of the ppE model. 
It is important to characterize such false positive GR deviations based on their statistical significance, and to better understand how the biases depend on the type of neglected GR effect. 
Understanding such systematic biases in tests of GR will help distinguish true GR deviations from false ones, thus making probes of fundamental physics more robust.
We can illustrate the importance of this work using a historical (if pedagogical) example: pinning down systematics within the Solar System was crucial to reveal a fundamental bias in using Newtonian gravity to describe the pericenter precession of Mercury in the early 1900s.
These systematics included corrections to the pericenter precession that came from higher-order multipole moments of the tidal fields, induced by the planets (such as Jupiter), as well as accurately modeling the multipole moments induced by the bulge of the Sun, among other corrections (see~\cite{poisson_will_2014,Will:2014kxa} for a review).
With GWs, if such fundamental theoretical biases~\cite{Yunes_2009} are to reveal themselves from the data, we need to exhaustively understand all possible systematic biases (detection, priors, waveform modeling etc.) that may incorrectly point to a false GR deviation 
(see for example~\cite{Gupta:2024gun}).

We center our analysis around the following two sets of questions and provide two sets of corresponding results in our paper:

\noindent {1.} \emph{How do we assess the statistical significance of systematic biases in ppE tests of GR? When can we claim that the biases point to a {\underline{false}} GR deviation?}
    
\begin{figure}[t]
    \centering
    \includegraphics[width=0.5\textwidth, clip=true]{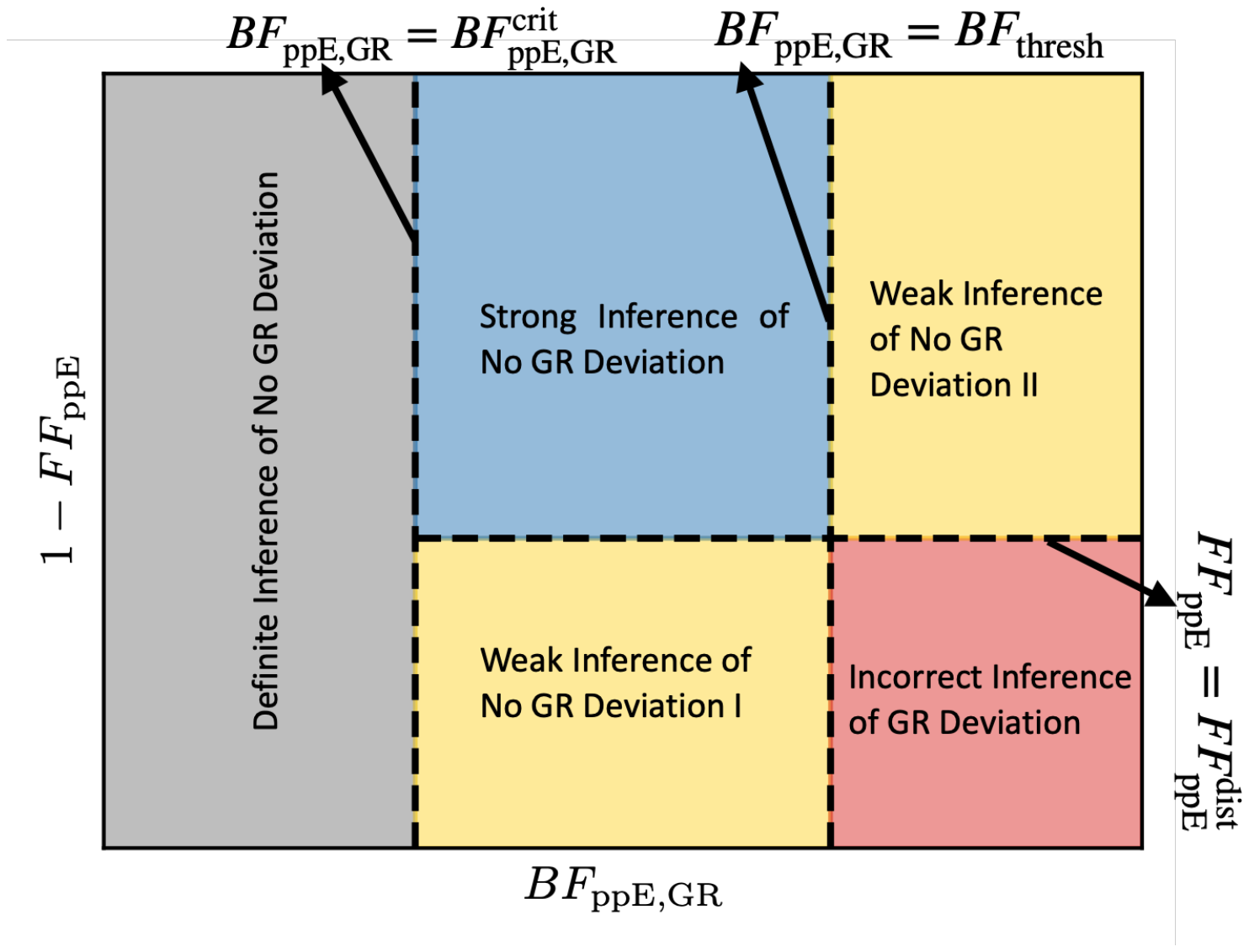}
    \caption{Schematic representation of the distinct regimes of systematic bias determined by $BF_{\mathrm{ppE,GR}}$ and $1-FF_{\mathrm{ppE}}$.
    The gray shaded region is where there is insignificant systematic bias in the ppE parameter, so one would correctly identify that there is no GR deviation in the signal. 
    The regime of significant bias is further divided into four subregions based on how strong the ppE model is preferred over the GR model by the data, and based on how much SNR is lost when using the ppE model: Strong Inference of No GR Deviation (blue), Weak Inference of No GR Deviation I and II (yellow), and Incorrect Inference of GR Deviation (red). 
    See Sec.~\ref{subsec:claiming-false-GR-deviation} for more details.
    }
    \label{fig:stealth-overt}
\end{figure}

The systematic bias in the inference of a GR deviation is significant when the systematic error in the ppE parameter exceeds the statistical error.
When this occurs, we say there is a \emph{Definite Inference of No GR Deviation}.
When the systematic bias in the ppE parameter is significant, there is a \emph{Possible Inference of a False GR Deviation}. 
We assess the statistical significance of a Possible Inference of a False GR Deviation by computing the fitting factor of the ppE model, $FF_{\rm ppE}$, and the Bayes factor between the ppE and the GR models, $BF_{\rm ppE/GR}$, given a signal.
The fitting factor is a proxy for the loss in SNR due to model inaccuracies.
Given a signal SNR, when the fitting factor of the ppE model is above (below) a distinguishability threshold $FF_{\rm ppE}^{\rm dist}$, the signal is accurately (inaccurately) captured by the ppE model with negligible (significant) loss of SNR, and the residual SNR test is then passed (failed).
The Bayes factor $BF_{\rm ppE,GR}$ tells us whether the ppE model is preferred over the GR model.
When $BF_{\rm ppE,GR}$ is above (below) a given threshold $BF_{\rm ppE,GR}^{\rm thresh}$, the ppE model is said to be strongly preferred (disfavored) over the GR model, and the model selection test is then passed (failed).
Based on how $FF_{\rm ppE}$ and $BF_{\rm ppE,GR}$ compare to $FF_{\rm ppE}^{\rm dist}$ and $BF_{\rm ppE,GR}^{\rm thresh}$ respectively, one can define four subregimes within the space of a Possible Inference of a False GR Deviation, as shown in Fig.~\ref{fig:stealth-overt}.

These different subregimes can be dealt with differently within the inference pipeline. 
In the Strong Inference of No GR Deviation (blue shaded region in~\cref{fig:stealth-overt}), there is significant loss of SNR and only a weak preference for the ppE model over the GR model.
When the loss in SNR is insignificant (significant), but the ppE model is weakly (strongly) preferred over the GR model, the systematic bias leads to a Weak Inference of No GR Deviation I (II) (yellow shaded regions of~\cref{fig:stealth-overt}). In all three of these subregimes, either the residual SNR test or the model selection test is not passed, suggesting that any inference of a GR deviation is false. 
However, there can be instances when there is insignificant loss of SNR and a strong preference for the ppE model over the GR model, even though the signal contains no GR deviation; this defines the Incorrect Inference of GR Deviation subregime (the red shaded region of~\cref{fig:stealth-overt}). The existence of this regime is unavoidable if the GR sector of the ppE model is inaccurate and degenerate with a ppE deviation, as may be the case when considering environmental effects, the neglect of spin precession or eccentricity.  
In the context of detecting true GR deviations (which is the flip side of our focus in this paper), an Incorrect Inference of a GR deviation is analogous to a \emph{stealth bias}, as described by~\cite{Vallisneri:2012qq,Cornish:2011ys,Vallisneri:2013rc}.
We elaborate on these regimes in Sec.~\ref{subsec:claiming-false-GR-deviation}, which addresses the first question.

As a by-product of our analysis, we show, from first principles, the explicit connection between different statistical measures and techniques used to quantify systematic error --the linear signal approximation (LSA), the Laplace approximation, the fitting factor, the effective cycles, and the Bayes factor (see Sec.~\ref{subsec:distinguishability}). These connections are applicable to and may be useful for studies that go beyond tests of general relativity.  
These explicit connections are established by studying a toy model, in which we can explicitly compute all of the above measures and use them to define the different regimes of systematic bias in ppE tests described above (see Sec.~\ref{subsec:toy}). 
As a rule of thumb for waveform modeling efforts, we find that the regime of Definite Inference of No GR deviation is conservatively defined when the effective cycles satisfy 
\begin{align}
\mathcal{N}_e \lesssim 0.02 \sqrt{\frac{N_M}{15}} \left(\frac{30}{\rho}\right)\,,
\end{align}
where $N_M$ are the number of waveform parameters and $\rho$ is the SNR. When this occurs, the systematic errors in GW inference are smaller than the statistical errors at 1-sigma confidence. 
If the effective cycles of dephasing incurred due to mismodeling are larger than this threshold, one enters the regime of Possible Inference of False GR deviation. 
When the effective cycles satisfy 
\begin{align}
\mathcal{N}_e \gtrsim \mathcal{N}_e^{\rm thresh} = 0.03
\left(\frac{30}{\rho}\right) 
\left[\log \left(\frac{BF_{\rm thresh}}{\sqrt{2 \pi} {\cal{O}}_p}\right)  \right]^{1/2} \!\!\!\! \left(\frac{{\cal{I}}_{\rm thresh} }{4}\right), \label{eqn:Ne_thresh_char}
\end{align}
then the ppE model is strongly preferred over the GR model. The above equation depends on the Occam penalty ${\cal{O}}_p \equiv \sigma_\beta/V_\beta$ (effectively the ratio of the posterior volume to the prior volume, where $V_\beta$ is the prior volume of the ppE parameter and $\sigma_\beta$ is the accuracy to which the ppE parameter can be inferred), the threshold Bayes factor chosen $BF_{\rm thresh}$, and ${\cal{I}}_{\rm thresh}$ (another function that depends on the ppE index and the type of mismodeling in the GR sector).
In~\cref{eqn:Ne_thresh_char}, for a typical value of $BF_{\rm thresh}\sim 10$ and ${\cal{O}}_p \sim 0.1$, we have that $\mathcal{N}_{e} \gtrsim 0.05$ and the model selection test is passed.
When the effective cycles satisfy 
\begin{align}
\mathcal{N}_e \gtrsim \mathcal{N}_e^{\rm dist} = 0.2 \sqrt{\frac{N_M}{15}} \left(\frac{30}{\rho}\right) \left(\frac{{\cal{I}}_{\rm dist} }{10}\right)\,,
\end{align}
with ${\cal{I}}_{\rm dist}$ (a function that depends on the ppE index and the type of mismodeling in the GR sector), then the fitting factor is below threshold ($FF_{\rm ppE} < 1-N_M/(2\rho^2)$), there is significant loss of SNR, and the SNR residual test would not be passed.

\vspace{0.2cm}
\noindent {2.} \emph{How do the systematic biases in ppE parameters depend on specific parameters of the neglected GR effect, the SNR of the signal, and the PN order of the ppE deviation (holding other parameters fixed)?}

To address this second question, we perform detailed Bayesian parameter estimation studies of biases induced by either neglecting spin precession or higher modes in the recovery waveform model.
We do so by varying the injected masses, spins, SNR, the inclination angle, and the PN order of the ppE parameter (one parameter at a time), and holding all other injection parameters fixed. 
Even at an SNR of 30, we find precession-induced systematic biases in ppE tests performed at $-1$PN, 1PN and 2PN, when the spins are largely in the orbital plane and when the binary is viewed nearly edge-on.
At the same SNR, for the same mass ratio, neglecting higher modes does not affect ppE tests as much as neglecting spin precession.
However, we do find that the biases get larger with increasing mass ratio, especially for the lower PN ppE tests.    
Our study thus identifies systems for which it is crucial to perform ppE tests with spin precession included. 
On the flip side, our study also identifies systems for which ppE tests are robust even with precession effects neglected.
These results can be helpful for waveform modeling efforts that are trying to incorporate simultaneously spin precession, eccentricity, and higher modes into the waveforms.
Our results are notably distinct from previous work as we systematically study the dependence of the biases on particular injection parameters, thereby addressing the second question in detail.

The remainder of this paper presents the details that led us to the conclusions summarized above, and it is organized as follows.
We review the geometrical approach to systematic error and the statistical criteria for assessing systematic biases in Sec.~\ref{subsec:definitions} and Sec.~\ref{subsec:distinguishability}, respectively. 
We also detail the scheme for characterizing the systematic bias in the inference of a ppE deviation in Sec.~\ref{subsec:claiming-false-GR-deviation}. 
In Sec.~\ref{subsec:toy}, we introduce a toy model to illustrate systematic biases (within the LSA) in ppE tests.
In Sec.~\ref{sec:bayesian_framework} we describe the Bayesian framework and the injection-recovery set up used in our work.
In Sec.~\ref{sec:bias_prec} and Sec.~\ref{sec:bias_HM} we provide the results of biases due to neglecting spin precession and higher modes, respectively.
In Sec.~\ref{sec:conclusion} we summarize and discuss future work.
For the reader's convenience, we provide an overview of the different causes of systematic bias in GW inference in Appendix~\ref{sec:lit_review}.
In other Appendices we review
our chosen waveform models (\cref{sec:waveform_review}),
ppE tests of GR with GWs (\cref{sec:review_ppE}),
and the use of the linear signal approximation and of the linear waveform difference approximation to compute systematic errors (\cref{sec:LSA}).
We also provide technical details of a toy model that explains most features observed in systematic biases (\cref{app:bias_char_toy}),
the Parallel Tempering Markov Chain Monte Carlo (PTMCMC)/nested sampling methods used in parameter estimation (\cref{sec:additional-figures}), as well as
the sampler settings and various convergence checks~(\cref{sec:convergence}).
Throughout the paper we adopt geometrical units $(G=c=1)$. 

\section{General problem  of systematic biases due to inaccurate waveforms} \label{sec:general_problem}

We first review the basics of GW data analysis to introduce relevant methods, definitions, notation and interpretations of systematic bias following~\cite{Flanagan:1997kp,Cutler:1994ys,Cutler:2007mi,Lindblom:2008cm,Kejriwal:2023djc}.
We then discuss the methods of computing systematic bias. In particular, we review the connection between systematic error and other criteria for distinguishability between models (within the linear signal and Laplace approximation). 
We also connect systematic errors to effective cycles, the Bayes factor and the fitting factor, which are all helpful statistical tools in data analysis.
We examine a toy example that captures several features of the general problem of systematic errors due to inaccurate waveforms. 
The toy example sheds light on the main results that will follow in Sec.~\ref{sec:bias_prec} and Sec.~\ref{sec:bias_HM}.

\subsection{Review of the geometrical approach \\ to systematic bias} \label{subsec:definitions}
Consider the hypothesis $H_S$ that the data $d$ in a detector is described by a GW signal $s$, i.e.,~$d = s + n$, where $n$ is the detector noise. 
Let the GW model that can be used to represent the signal $s$ be given by $h_S (\BF{\theta})= A_S (\BF{\theta}) e^{i \Psi_S (\BF{\theta})}$, where $A_S$ and $\Psi_S$ are the waveform amplitude and phase functions respectively, and $\BF{\theta}$ is a vector of $N_S$ waveform parameters.
The observed signal $s$ is a particular realization of the model $h_S$, so $s = h_S (\bftr{\theta})$, where $\bftr{\theta}$ is a vector of the ``true'' parameters of the signal.
We can interpret this in a geometrical setting as follows.
The model $h_S$ spans a \emph{signal space}, a finite dimensional Riemannian manifold $\mathcal{S}$, which is a submanifold of an infinite dimensional Euclidean vector space $V$ of all possible signals~\cite{Cutler:1994ys}.
The observed signal $s$ is then a point on $\mathcal{S}$.

\begin{figure}[t]
    \centering
    \includegraphics[width=0.49\textwidth]{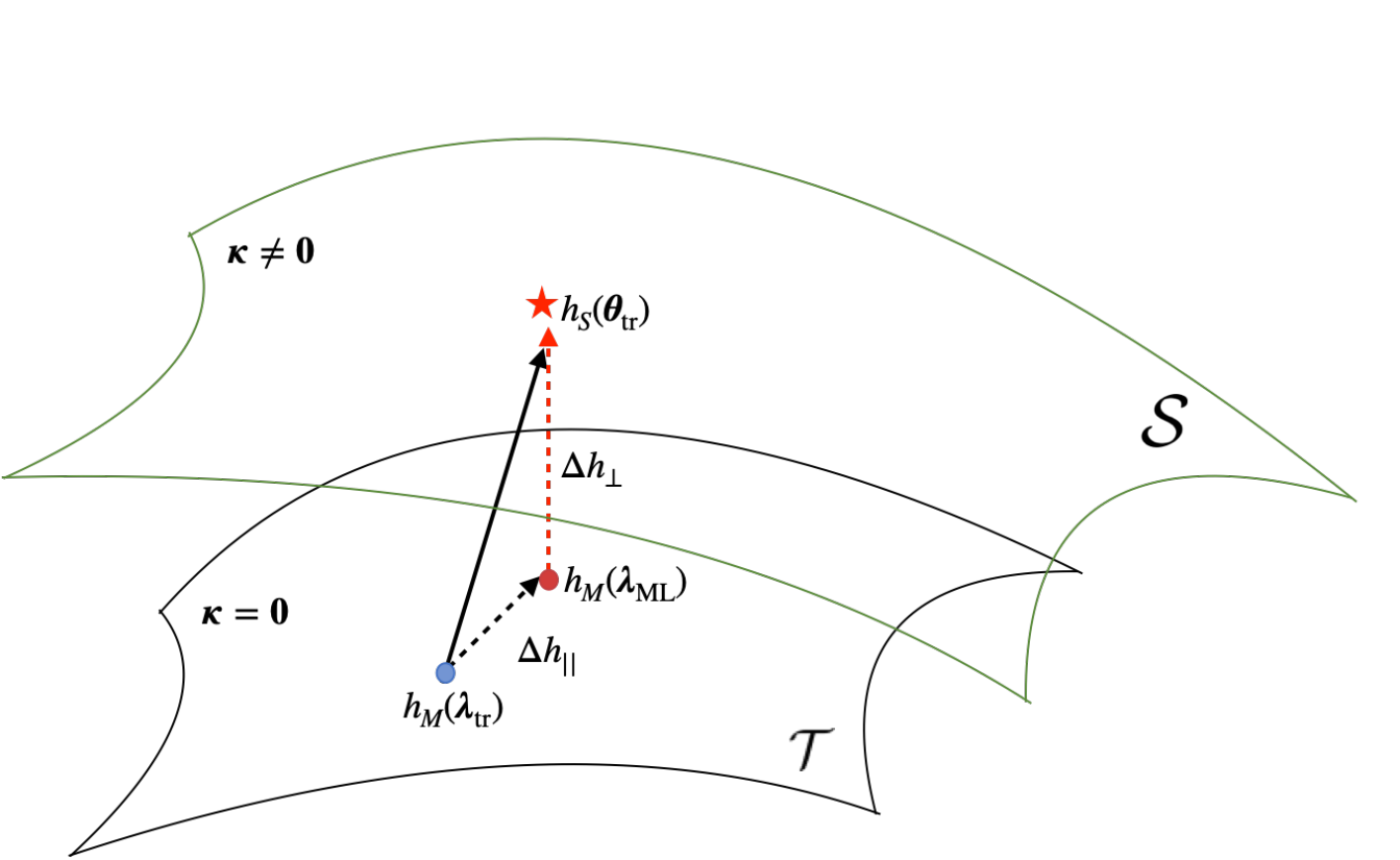}
    \caption{
    Sketch depicting the geometric view of systematic error. 
    The 2d-hypersurfaces are the template submanifolds $\mathcal{S}$ and $\mathcal{T}$. 
    For simplicity, we have shown a particular 2d-hypersurface of $\mathcal{S}$ with $\BF{\kappa} \neq \BF{0}$ and a 2d-hypersurface of $\mathcal{T}$ with $\BF{\kappa} = \BF{0}$, although in reality $\mathcal{S}$ and $\mathcal{T}$ are of higher dimension.
    The injected signal, denoted by the red star, resides on the hypersurface $\mathcal{S}$, while
    the waveform models, denoted by circles, reside on the hypersurface $\mathcal{T}$.
    The blue dot corresponds to the projection of the injected signal onto $\mathcal{T}$, while the red dot corresponds to the maximum likelihood waveform.
    The difference vector $\Delta h_{\perp}$ between the signal and the maximum likelihood waveform is denoted by the red dashed line.
    Meanwhile, the difference vector $\Delta h_{||}$ between the maximum likelihood waveform and the projection of the signal onto $\mathcal{T}$ is denoted by the black dashed line. 
    Observe that $\Delta h_{||}$ lies along the tangent subspace of $\mathcal{T}$ at $\bfml{\lambda}$. Meanwhile, $\Delta h_{\perp}$ is orthogonal to it, and it defines the shortest distance between the waveform model and the signal. 
    }
    \label{fig:geom_sys_error}
\end{figure}

The goal of GW data analysis is to estimate $\bftr{\theta}$ from the data as accurately and precisely as possible using a waveform model, but, in practice, the model is never exactly equal to the signal.
Let the recovery waveform model be given by $ h_M(\BF{\lambda})=A_M (\BF{\lambda}) e^{i \Psi_M(\BF{\lambda}) }$, where $\BF{\lambda}$ is a vector of $N_M$ waveform parameters.
The hypothesis that the signal is described by the model $h_M$ is then $H_M$.
Geometrically, the model $h_M(\BF{\lambda})$ spans a Riemannian manifold $\mathcal{T}$, called the \emph{template submanifold}, which is also a submanifold of $V$.
In general, $\mathcal{S}$ and $\mathcal{T}$ are distinct submanifolds of $V$ (as illustrated in~\cref{fig:geom_sys_error}) because $h_M$ is only an approximation to $h_S$, i.e., the amplitude function $A_M$ and the phase function $\Psi_M$ are approximations to $A_S$ and $\Psi_S$, respectively.
In fact, in general, $h_M$ will always be an approximation to $h_S$, because we can never know what $h_S$ is with infinite precision and accuracy. 
Expressed another way, the hypothesis $H_M$ is nested within the hypothesis $H_S$, or $H_S \supseteq H_M$, and the problem of systematic bias then reduces to understanding the errors caused by approximating $h_S$ with $h_M$.
Throughout this paper, we work with data without specific noise realizations, and thus we set $n=0$, so that we may understand errors due to waveform inaccuracies and isolate them from errors induced by a particular realization of $n$. 
Having said this, we do include the spectral noise density $S_n$ in our analysis, to represent the behavior of Gaussian and stationary noise in the frequency domain. 
The spectral noise density depends on the properties of the detector, as we specify in Secs.~\ref{subsec:toy} and~\ref{sec:bayesian_framework} below.

We will parametrize $h_M$ and $h_S$ in such a way that they share $N_M$ common parameters $\BF{\lambda}$, leaving $h_S$ with $N_S-N_M$ additional parameters $\BF{\kappa}$.  
Succinctly, this means that $\BF{\theta} =  \BF{\lambda}  \cup \BF{\kappa} $ and $h_S (\BF{\theta}) \equiv h_S(\BF{\lambda},\BF{\kappa})$.
In this paper, we will assume the model $h_M$ is equal to $h_S$ when the latter is evaluated on $\BF{\kappa} = \BF{0}$ (see for eg.~\cite{Kejriwal:2023djc}). 
Moreover, we will be interested in two specific cases: the case when $h_M$ has less parameters than $h_S$, $N_S > N_M$, and the case when $h_M$ has the same number of parameters as $h_S$, $N_M = N_S$.
For example, when we study the use of the spin-aligned, dominant mode waveform \texttt{IMRPhenomD}~\cite{Khan:2015jqa} to recover parameters from a spin-precessing, dominant mode waveform \texttt{IMRPhenomPv2}~\cite{Hannam:2013oca,Schmidt:2012rh,Khan:2019kot} injection, we will be focusing on the $N_S > N_M$ case. 
In this example, setting the orbital spin components to zero will reduce the true waveform to the nonprecessing waveform, which is in general an approximation.
On the other hand, when we study the use of \texttt{IMRPhenomD} to perform inference on the injection of a spin-aligned waveform with higher modes, \texttt{IMRPhenomHM}~\cite{London:2017bcn}, we will be focusing on the $N_S = N_M$ case.
In this example, the approximation of using \texttt{IMRPhenomD} neglects the physics of subdominant higher modes, but there is no change in the dimensionality of the parameter space.
This example is similar to that studied in Ref.~\cite{Owen:2023mid}, where the PN series was truncated, while in our study the mode expansion of the waveform will be truncated.

Evaluating $h_M$ at the injected parameters $\bftr{\lambda}$ corresponds to projecting $h_S$ onto the template submanifold, as illustrated in~\cref{fig:geom_sys_error}.
To obtain further geometrical insight, consider the case when the effects induced by $\BF{\kappa}$ are infinitesimally small and the signal waveform model reduces to the approximate waveform model when $\BF{\kappa}$ vanishes.
In this case, the projection is performed by $h_M(\bftr{\lambda}) \rightarrow h_S(\bftr{\theta})$ and Taylor expanding $h_S (\bftr{\theta})$ for small $\kappa^i$ to linear order, i.e., 
\begin{align}
h_S(\bftr{\theta}) \sim h_S(\bftr{\lambda},0)+ \delta_{\kappa} h_S(\bftr{\theta})\,,
\end{align}    
where we recognize that $h_S(\bftr{\lambda},0) = h_M(\bftr{\lambda})$. 
We denote the difference between the signal and the projection as $\delta_{\kappa} h_S(\bftr{\theta}) = h_S(\bftr{\theta}) - h_M(\bftr{\lambda})$, where
\begin{align}
    \delta_{\kappa} h_S(\bftr{\theta}) = \kappa^i \partial_{\kappa^i} h_S (\BF{\theta})\Big |_{\BF{\theta} = \bftr{\theta}}, \quad \partial_{\kappa^i}= \dfrac{\partial }{\partial \kappa^i} .
\end{align}
In other words, the hypersurface of $\mathcal{S}$ defined by $\BF{\kappa} = \bftr{\kappa}$ is related to the hypersurface of $\mathcal{T}$ defined by $\BF{\kappa} = \BF{0}$ through a \emph{lapse}~\cite{Poisson:2009pwt} $\delta_{\kappa} h_S(\bftr{\theta})$, which is also the \emph{directional derivative} of the model along $\BF{\kappa}$. 

Generally, when the priors on the waveform parameters are flat, the matched filtering problem~\cite{Cutler:1994ys} of determining the ``best-fit'' parameters translates to finding the parameters $\bfml{\lambda}$ at which the distance between the data $d$ and the waveform model $h_M$ is minimum. 
The distance between waveforms $h_1$ and $h_2$ is defined in terms of the inner product $(h_1 | h_2)$ given by
\begin{align}
       (h_1|h_2) &\equiv 4 \Re \int \limits_{f_{\min}}^{f_{\max}} \dfrac{df}{S_n(f)} h_1^{*}(f) h_2(f),
\end{align}
where the norm of the waveforms is their SNR, namely $\rho_1^2 = \norm{h_1}^2 = (h_1|h_1)$ and $\rho_2^2 = \norm{h_2}^2 = (h_2|h_2)$.

The distance between the waveform model and the data is then $\norm{\Delta h (\BF{\lambda},\bftr{\theta})}$, with 
\begin{equation}
    \Delta h(\BF{\lambda},\bftr{\theta}) \equiv h_S (\bftr{\theta}) - h_M (\BF{\lambda})\,.
    \label{eq:Deltah-def}
\end{equation} 
For a Gaussian and stationary noise model, the log-likelihood for parameters $\BF{\lambda}$ given the data (for a given detector)  and hypothesis $H_M$ is given by
\begin{align}
    \begin{split}
    \log \mathcal{L} (d | \BF{\lambda}, H_M) &\equiv -\dfrac{1}{2} \norm{h_S (\bftr{\theta}) - h_M (\BF{\lambda})}^2  \\
    & = -\frac{1}{2} (h_S-h_M|h_S-h_M) \\
     &= -\dfrac{1}{2}\rho^2_{S}(\bftr{\theta})-\dfrac{1}{2}\rho^2_{M}(\BF{\lambda}) + \rho^2_{MS} (\BF{\lambda}, \bftr{\theta}),
    \end{split} \label{eqn:log-likelihood}
\end{align}
where the SNRs of the signal and of the model are
\begin{subequations}
\begin{align}
      \rho^2_{S}(\bftr{\theta}) &= (h_S|h_S) = 4 \int \limits_{f_{\min}}^{f_{\max}} \dfrac{df}{S_n(f)} A_S^2(\bftr{\theta} ; f), \\
   \rho^2_{M}(\BF{\lambda}) &= (h_M|h_M) = 4 \int \limits_{f_{\min}}^{f_{\max}} \dfrac{df}{S_n(f)} A_M^2(\BF{\lambda} ; f),
\end{align}    
\end{subequations}
while the last term of Eq.~\eqref{eqn:log-likelihood} is
\begin{align}
\begin{split}
    \rho^2_{MS} (\BF{\lambda},\bftr{\theta}) &= 4  \int \limits_{f_{\min}}^{f_{\max}} \dfrac{df}{S_n(f)} \Big ( A_S(\bftr{\theta} ; f)A_M(\BF{\lambda} ; f) \\
    & \times \cos \left[ \Psi_M(\BF{\lambda} ; f)-\Psi_S(\bftr{\theta} ; f) \right] \Big ).    
\end{split}
\end{align}
The inner products above also define the matched-filter SNR $\rho_{\mathrm{match}} (\BF{\lambda},\bftr{\theta})$ and the match (or faithfulness) $\mathcal{F}(\BF{\lambda},\bftr{\theta})$ between the waveform model $h_M$ and the signal:
\begin{subequations}
\begin{align}
    \rho_{\mathrm{match}} (\BF{\lambda},\bftr{\theta})  & \equiv \max_{\Delta \phi,\Delta t} \dfrac{\rho^2_{MS}(\BF{\lambda},\bftr{\theta})}{\rho_M(\BF{\lambda})} \\
    \begin{split}
     \mathcal{F} (\BF{\lambda},\bftr{\theta}) &\equiv \dfrac{\rho_{\mathrm{match}}(\BF{\lambda},\bftr{\theta})}{\rho_S(\bftr{\theta})}\\
     &= \max_{\Delta \phi,\Delta t}\dfrac{(h_S(\bftr{\theta})| h_M(\BF{\lambda}))}{\rho_S(\bftr{\theta}) \rho_M (\BF{\lambda}) },        
    \end{split}
\end{align} \label{eqn:match}%
\end{subequations}
where the maximization is done\footnote{In practice, the maximization is done by recasting the integral as an inverse FFT and performing \emph{implicit} maximization. The values of $\Delta t$ and $\Delta \phi$ are then found using \emph{explicit} maximization~\cite{blake_2018}.} by introducing a time and phase shift to the phase difference between the data and the waveform. More concretely, 
if we let $\Delta \Psi (\bftr{\theta} ; f) \equiv \Psi_S (\bftr{\theta} ; f)-\Psi_M (\bftr{\lambda};f)$, then the shift is done by 
\begin{align}
\label{eq:DeltaPhiinDeltaPsi}
\Delta \Psi \rightarrow  \Delta \Phi (\bftr{\theta} ; f) \equiv \Delta \Psi (\bftr{\theta} ; f) + 2\pi f \Delta t - \Delta \phi.    
\end{align}
Maximizing the match this way essentially aligns the entire waveform model with the data. 
The alignment can also be carried out per individual waveform mode, as done for eccentric waveforms in~\cite{blake_2018}.

Minimizing the distance between the data and the waveform model is identical to maximizing the likelihood. 
At the maximum likelihood, the gradient with respect to the parameters $\BF{\lambda}$ will vanish, or $\partial_{\lambda^i} \mathcal{L} = 0$, where $\partial_{\lambda^i} \equiv \partial/ \partial \lambda^i$, and $\lambda^i$ are the components of the waveform parameter vector $\BF{\lambda}$.
This translates to 
\begin{align}
    \left( \partial_{\lambda^i} h_{M} (\bfml{\lambda}) | h_S(\bftr{\theta}) - h_{M}(\bfml{\lambda}) \right) =0, \label{eqn:max_likelihood_condition_main}
\end{align}
where $\bfml{\lambda}$ are the parameters that maximize the likelihood, and, by definition,  $\log \mathcal{L} (d | \bfml{\lambda}, H_M) \leq 0$. 
Ideally\footnote{Some parameters are never measured well even when using the same model as the injection. 
This can be due to extremely strong correlations, typically resulting from the choice of parametrization (i.e., the choice of coordinates in the geometrical viewpoint).}, the equality holds when $h_S = h_M$, which results in $\bfml{\lambda} = \bftr{\lambda}$. 
When using an approximate model, it follows that $\log \mathcal{L} (d | \bfml{\lambda}, H_M) < 0$ and the resulting \emph{likelihood-based systematic error} is defined as $ \Delta_L 
 \BF{\lambda} \equiv \bfml{\lambda} - \bftr{\lambda}$.
Geometrically, shifting the parameters $\bftr{\lambda} \rightarrow \bfml{\lambda} = \bftr{\lambda} + \Delta_L \BF{\lambda} $ results in a null projection of the true signal along the tangent space at $\bfml{\lambda}$, reflected by~\cref{eqn:max_likelihood_condition_main}.
At $\bfml{\lambda}$, the maximum likelihood is given by
\begin{subequations}
\begin{align}
    \log \mathcal{L} (d|\bfml{\lambda}, H_M) &= -\dfrac{1}{2} \norm{\Delta h_{\perp} }^2,
    \label{eqn:log-likelihood-max1}
    \\  
    \Delta h_{\perp} &\equiv \Delta h (\bfml{\lambda},\bftr{\theta}).
    \label{eqn:log-likelihood-max}
\end{align} 
\end{subequations}

As pointed out in Ref.~\cite{Vallisneri:2012qq}, $\Delta h_{\perp}$ is the residual signal that cannot be captured by the shift $\Delta_L \BF{\lambda}$ and is normal to the tangent space at the maximum likelihood point (i.e., the geometric interpretation of~\cref{eqn:max_likelihood_condition_main}).
We can quantify this residual further by introducing the fitting factor\footnote{Note that the maximum likelihood waveform might need to be further aligned with the signal by shifting the time and phase shift. 
The phase shift may not be well measured in practice, making this alignment important.} as $FF = \mathcal{F}(\bfml{\lambda},\bftr{\theta})$, i.e., the faithfulness or match evaluated at the maximum likelihood parameters.
The loss in SNR due to the imperfect recovery is then 
$\rho^2_{\mathrm{res}} = (1-FF) \rho^2_S(\bftr{\theta})$, with $1-FF$ representing the fraction of SNR lost. 
When amplitude corrections are negligible between the two models, we have that $\rho_S (\bftr{\theta}) \approx \rho_M (\bfml{\lambda}) \approx \rho_M (\bftr{\lambda}) \equiv \rho$. 
Using~\cref{eqn:log-likelihood,eqn:match,eqn:log-likelihood-max}, we have that 
\begin{align}
    1-FF &\approx \dfrac{\norm{\Delta h_{\perp}}^2}{2 \rho^2} + \mathcal{O} \left( 1- \dfrac{\rho_M(\bfml{\lambda})}{\rho_S(\bftr{\theta})} \right), \label{eqn:FF_hperp}
\end{align}
showing that the fitting factor is SNR scale-invariant. 
Note that $FF = 1$ does not imply that there is no systematic error~\cite{Vallisneri:2012qq}, but only that $\Delta_L \BF{\lambda}$ perfectly absorbs the difference between the injected signal and the waveform model.
In Sec.~\ref{subsec:toy} we provide an explicit example to illustrate this subtlety.

Having established the definition of the systematic error and the connection to quantities such as the fitting factor, how do we actually compute the systematic error? 
There is no general closed-form solution to the root finding problem defined by~\cref{eqn:max_likelihood_condition_main}, and solving it numerically on a grid can be expensive and unstable due to the ``curse of dimensionality''~\cite{numerical_recipes}.
However, when the signal is loud and $h_M$ is a ``good'' approximation of $h_S$, the systematic errors will be small, and~\cref{eqn:max_likelihood_condition_main} can be solved perturbatively using the linear signal approximation (LSA)~\cite{Flanagan:1997kp,Cutler:2007mi}.
In the LSA, the waveform is expanded about the maximum likelihood point, and the systematic error on the waveform parameters $\BF{\lambda}$ is given by (see Appendix~\ref{sec:LSA} for our rederivation)~\cite{Flanagan:1997kp,Cutler:2007mi}
\begin{align}
    \Delta_L \lambda^i &= C^{ij} (\bftr{\lambda}) \left(\partial_{\lambda^j} h_M(\bftr{\lambda}) |  d- h_M(\bftr{\lambda}) \right), \label{eqn:systematic_error_explicit_main_LO_unexpanded}
\end{align}
where $C^{ij}(\bftr{\lambda}) = (\Gamma^{-1}(\bftr{\lambda}))^{ij}$ is the inverse of the Fisher matrix $\Gamma^{ij}(\bftr{\lambda})$ corresponding to the waveform model, defined by $\Gamma^{ij} \equiv (\partial_{\lambda^i} h_M | \partial_{\lambda^j} h_M)$, and we recall that $d = h_S (\bftr{\theta})$. 
Additionally, we have used the Einstein summation rule for repeated dummy indices.\footnote{In Ref.~\cite{Cutler:2007mi}, the systematic error was evaluated at the maximum likelihood (or ``best-fit'') parameters because the goal of that work was to \emph{postdict} the systematic error given $\bfml{\lambda}$ and $h_M$. 
On the other hand, in our study (and in most studies that have used the LSA) we are interested in \emph{predicting} the systematic error given $h_M$ and $\bftr{\lambda}$.
For the sake of completeness, in Appendix~\ref{sec:LSA} we rederive~\cref{eqn:systematic_error_explicit_main_LO_unexpanded}.}
Linearizing~\cref{eqn:systematic_error_explicit_main_LO_unexpanded} to leading order in $\Delta \Psi$, and neglecting the amplitude difference between $A_S$ and $A_M$ for simplicity, we obtain
\begin{align}
\begin{split}
\Delta_L \lambda^i &= C^{ij} (\bftr{\lambda}) \ 4 \Re \int \limits_{f_{\min}}^{f_{\max}} \Big [ \dfrac{df}{S_n(f)} \partial_{\lambda^j} h_M 
\\
& \times
\Big(-i A_M(\bftr{\lambda} ; f) \Delta \Psi (\bftr{\theta} ; f) \Big ) e^{-i \Psi_M(\bftr{\lambda} ; f)} \Big ], \label{eqn:systematic_error_explicit_main_LO_expanded} 
\end{split}
\end{align}
which is one of the main results of Ref.~\cite{Cutler:2007mi} when the amplitudes $A_S$ and $A_M$ are comparable.
For completeness, we rederive~\cref{eqn:systematic_error_explicit_main_LO_expanded} with a nonzero $\Delta A = A_S - A_M$ in Appendix~\ref{sec:LSA}.

The difference $\Delta h_{||} = h_M (\bfml{\lambda}) - h_M({\bftr{\lambda}})$, when Taylor-expanded about $\bfml{\lambda} = \bftr{\lambda}$ to linear order, then reduces to
\begin{align}
    \Delta h_{||} = \left(\Delta_L \lambda^i\right) \left(\partial_{\lambda^i} h_M\right)\,, \label{eqn:h_parallel}
\end{align}
where $\Delta_L \lambda^i$ is defined in~\cref{eqn:systematic_error_explicit_main_LO_unexpanded}. 
Geometrically, $\Delta h_{||}$ is the parallel component of $h_S(\bftr{\theta}) - h_M (\bftr{\lambda})$ along the tangent subspace of $\mathcal{T}$ at $\bftr{\lambda}$ (i.e.,~for an infinitesimal $\Delta_L \lambda^i$, it is the directional derivative of $h_M$ at $\bftr{\lambda}$, or the \emph{shift}~\cite{Poisson:2009pwt}).
The infinitesimal case can be seen as the triangle law (Pythagorean theorem) 
$\norm{\Delta h}^2 = \norm{\Delta h_{||}}^2 + \norm{\Delta h_{\perp}}^2$ (see~\cref{fig:geom_sys_error} for a visualization).

As pointed out by Ref.~\cite{Cutler:2007mi}, the systematic error is invariant under a rescaling of the SNR. This invariance can be seen by showing that the systematic error is unchanged under $S_n(f) \rightarrow k S_n(f)$ or $D_L \rightarrow \sqrt{k} D_L$, where $k$ is some constant and $D_L$ is the luminosity distance to the source.
This invariance holds even without the LSA because~\cref{eqn:systematic_error_explicit_main_LO_unexpanded} is invariant under a scaling of the SNR, implying that solutions to~\cref{eqn:systematic_error_explicit_main_LO_unexpanded} must also be SNR-scale invariant.
However, the systematic error will be detector-dependent through the frequency window $[f_{\min},f_{\max}]$, the ``shape'' of the noise curve (such as its slope, its minimum, etc.), and the number of detectors.\footnote{For simplicity and clarity, we have presented equations restricted to a single detector, but this analysis can be trivially generalized to a detector network by summing over all relevant inner products: see e.g. Appendix~A of Ref.~\cite{Kapil:2024zdn}.}
For a given source, changing these detector properties will affect both the SNR and the systematic error.

The goal of parameter estimation is not just to compute $\bfml{\lambda}$, but also to determine parameter covariances and obtain the \emph{posterior} probability $\mathcal{P}(\BF{\lambda} | d,H_M)$ of model parameters $\BF{\lambda}$, given the data and hypothesis $H_M$, across the \emph{prior} range of the parameter space.
The posterior is computed using Bayes' theorem:
\begin{subequations}
\begin{align}
    \mathcal{P} (\BF{\lambda} | d, H_M) &= \dfrac{\mathcal{L} (d|\BF{\lambda},H_M) \Pi (\BF{\lambda}|H_M)}{Z_M},\\
    Z_M &=  \int \mathcal{P} (\BF{\lambda} | d) d\lambda_1 \cdots d\lambda_{N_M},   
\end{align}    
\end{subequations}
where $\Pi (\BF{\lambda}|H_M)$ is the prior on the waveform parameters under the hypothesis $H_M$, and $Z_M$ is the \emph{model evidence}.
In the LSA approach, it is possible to incorporate priors and perform relatively inexpensive calculations of the posterior.
Typically this is the case for flat or Gaussian priors.
However, when using the LSA approach, the posteriors can be unfaithful if the true covariances between parameters are not captured by the Fisher-covariance matrix or if the priors are highly nontrivial.
When there are strong parameter correlations and multimodal features, the Fisher matrix can become singular or at the very least ill-conditioned~\cite{Berti:2004bd,Vallisneri:2007ev}, and the LSA approach can fail.

A more robust way to compute the posteriors is to stochastically sample it using Nested Sampling or Markov Chain Monte Carlo (MCMC) methods~\cite{gregory_2005,Skilling:2006gxv} that are now routinely employed in GW inference.
Sampling from the posterior distribution directly lets one compute likelihood samples, from which $\bfml{\lambda}$ can be estimated. 
Furthermore, from the joint posterior distribution, the \emph{posterior based systematic error} can be defined as $\Delta_P \BF{\lambda} = \BF{\lambda}_{\mathrm{MP}}-\bftr{\lambda}$, where $\BF{\lambda}_{\mathrm{MP}}$ is the maximum posterior point.
The two estimates of systematic error $\Delta_L \BF{\lambda}$ and $\Delta_P \BF{\lambda}$ will coincide when the priors are flat or they are a multivariate Gaussian centered at the maximum likelihood point.
When assessing biases induced by priors, using $\Delta_P \BF{\lambda}$ is more appropriate.
In this work, we are interested in biases due to waveform mismodeling, so we mainly focus on $\Delta_L \BF{\lambda}$.
For the sake of completeness, in Appendix~\ref{sec:additional-figures}, we show a comparison between $\Delta_L \BF{\lambda}$ and $\Delta_P \BF{\lambda}$ for a subset of the analyses done in Sec.~\ref{sec:bias_prec}.

 \subsection{Statistical significance of systematic biases} 
\label{subsec:distinguishability}

Having introduced relevant definitions and methods for computing systematic biases (along with their geometric interpretation) in Sec.~\ref{subsec:definitions}, we move on to discuss the statistical significance of systematic biases by relating it to various measures that are used in GW data analysis.
The main question we address here is the following:  given that $h_M$ is an approximation of $h_S$, when will the systematic error become ``significant''?
We define a \textit{significant systematic bias} in the measurement of a given parameter by the condition that the systematic error on that parameter must be larger than the statistical error.

Several avenues have been explored to assess systematic and statistical error in Bayesian inference with GW data~\cite{Flanagan:1997kp,Cutler:2007mi,Favata:2013rwa,Lindblom:2008cm,Moore:2021eok,Hu:2022rjq,Chatziioannou:2017tdw,Chatziioannou:2014bma,Favata:2021vhw,Toubiana:2024car,Kalaghatgi:2019log,Apostolatos:1995,Vallisneri:2012qq,Cornish:2011ys,Purrer:2019jcp}.
These works have used the following methods:
(i) mismatch criterion for assessing the accuracy of $h_M$ compared to $h_S$,  
(ii) loss in SNR predicted by the fitting factor $FF$ when using $h_M$,
(iii) Bayes factor between $h_S$ and $h_M$ to quantify model selection.
The mismatch criterion is equivalent to comparing the systematic and statistical errors, which we will review below and explicitly relate to other criteria used commonly in the literature, while also adding a criterion based on the \emph{effective cycles}~\cite{Sampson:2014qqa}.

We begin with the mismatch criterion~\cite{Chatziioannou:2017tdw}, which quantifies the accuracy of $h_M$ with respect to $h_S$ at a given SNR $\rho$, because one can show that it implies that the systematic error will be contained within the $1\sigma$ statistical error. Mathematically, the mismatch criterion states that model $h_M$ will not introduce a significant systematic bias to $1\sigma$ if 
\begin{align}
    1-\mathcal{F}(\bftr{\lambda},\bftr{\theta}) < \dfrac{N_{M}}{2 \rho^2} +  (1-FF) \label{eqn:mismatch_criterion}\,,
\end{align}
where recall that $N_M$ is the number of parameters in $h_M$, and ${\cal{F}}$ is the faithfulness or match.
Let us now unpack the different terms in~\cref{eqn:mismatch_criterion}.
The mismatch $1-\mathcal{F} (\bftr{\lambda},\bftr{\theta})$ between the signal and the model evaluated at the injected parameters $h_M(\bftr{\lambda})$ can be related to the distance between this model and the data, $\Delta h (\bftr{\lambda},\bftr{\theta})$ in Eq.~\eqref{eq:Deltah-def}, via~\cite{Flanagan:1997kp} 
\begin{align}
\begin{split}
    1-\mathcal{F} (\bftr{\lambda},\bftr{\theta}) &\approx \dfrac{\norm{\Delta h (\bftr{\lambda},\bftr{\theta})}^2}{2 \rho^2}+ \mathcal{O} \left( 1- \dfrac{\rho_M(\bfml{\lambda})}{\rho_S(\bftr{\theta})} \right), 
    \label{eqn:match_deltah}    
\end{split}
\end{align}
where we have neglected amplitude corrections, as we did in~\cref{eqn:FF_hperp}. 
The first term on the right-hand side of~\cref{eqn:mismatch_criterion} decreases with increasing $\rho$, while the second term is SNR-invariant, as it depends on the fitting factor given by~\cref{eqn:FF_hperp}.
Note that $FF=1$ is only true when $\Delta h_{\perp} = 0$, but when $1-FF \ll 1$, one can neglect the second term on the right-hand side of~\cref{eqn:mismatch_criterion} to recover the standard mismatch criterion of Ref.~\cite{Chatziioannou:2017tdw}.
Doing so makes the mismatch criterion more pessimistic, and thus, it is useful for setting accuracy goals in waveform modeling~\cite{blake_2018,blake_2019,Arredondo:2024nsl}.

Why does the mismatch criterion of ~\cref{eqn:mismatch_criterion} imply that a significant systematic bias will not be incurred in parameter estimation? 
To answer this question, we first need to express the log-likelihood at the maximum likelihood point in terms of the systematic error and the mismatch.
Using~\cref{eqn:log-likelihood-max1}, the Pythagorean theorem below~\cref{eqn:h_parallel}, as well as~\cref{eqn:mismatch_criterion} for $||\Delta h||^2$ and~\cref{eqn:h_parallel} for $||\Delta h_{||}||^2$, we obtain
\begin{subequations}
\begin{align}
\begin{split}
\log \mathcal{L}(d| \bfml{\lambda},H_M) &= \log \mathcal{L} (d | \bftr{\lambda},H_M) \\
&+ \dfrac{1}{2} (\Delta_L \lambda^i) (\Delta_L \lambda^j) \Gamma^{ij} (\bftr{\lambda}),
\label{eqn:logL_max_LSA}
\end{split}\\
 \log \mathcal{L} (d | \bftr{\lambda},H_M) &= - \rho^2 [1-\mathcal{F}(\bftr{\theta},\bftr{\lambda})],
\end{align}   
\end{subequations}
where the first term of Eq.~\eqref{eqn:logL_max_LSA} is just the log-likelihood at the injected parameters, and the second term is the correction received from shifting the parameters. 
Geometrically, from~\cref{eqn:logL_max_LSA}, we see that in a small enough region of parameter space around $\bftr{\lambda}$, the Fisher matrix $\Gamma^{ij}$ plays the role of the \emph{metric} of the submanifold $\mathcal{T}$~\cite{Cornish:2011ys,Cutler:1994ys,Flanagan:1997kp}.
From~\cref{eqn:logL_max_LSA}, we see also that $\log \mathcal{L}(d| \bfml{\lambda},H_M)$ increases quadratically with $\rho$ due to both terms, since $\Gamma^{ij}$ is proportional to $\rho^2$. 
From~\cref{eqn:log-likelihood-max1} and ~\cref{eqn:FF_hperp}, we can recast~\cref{eqn:logL_max_LSA} as
\begin{align}
    FF = \mathcal{F} (\bftr{\lambda},\bftr{\theta}) +  \dfrac{1}{2 \rho^2} (\Delta_L \lambda^i) (\Delta_L \lambda^j) \Gamma^{ij} (\bftr{\lambda}). \label{eqn:FF_sys_error}
\end{align}
For concreteness, let us say here that a significant systematic error occurs when $\Delta_L \lambda^i > \delta \lambda^i$, where $\delta \lambda^i$ is the $1\sigma$ statistical error in measuring the parameter $\lambda^i$.
Then, using~\cref{eqn:FF_sys_error} we see that a significant systematic error occurs when
\begin{align}
    FF - \mathcal{F} (\bftr{\lambda},\bftr{\theta}) > \dfrac{1}{2 \rho^2} (\delta \lambda^i) (\delta \lambda^j) \Gamma^{ij} (\bftr{\lambda}). \label{eqn:FF_stat_err}
\end{align}
To further relate this condition to the mismatch criterion of~\cref{eqn:mismatch_criterion} we must use that the expectation value of $(\delta \lambda^i) (\delta \lambda^j)$ is by definition given by the   elements $C^{ij} (\bftr{\lambda})$ of the Fisher-covariance matrix, which is the inverse of the Fisher matrix~\cite{Cutler:1994ys}.
The contraction between the Fisher matrix and its inverse is just the dimension of the submanifold $\mathcal{T}$, which is the number of waveform parameters $N_M$.
We therefore see that Eq.~\eqref{eqn:FF_stat_err} becomes the mismatch criterion of~\cref{eqn:mismatch_criterion} (see also Appendix G of~\cite{Chatziioannou:2017tdw}).
Enforcing the mismatch criterion is then equivalent to requiring that the systematic error be contained within the $1\sigma$ statistical error.\footnote{As a side note, the mismatch criterion is equivalent to the indistinguishability criterion of Refs.~\cite{Flanagan:1997kp,Lindblom:2008cm}, which states that $h_S$ is indistinguishable from $h_M$ when $\norm{\Delta h_{||}}^2 < \epsilon^2$. Here, $\epsilon$ is some ``desired accuracy threshold'' needed for indistinguishability. When we set $\epsilon = \sqrt{N_M}$ and use~\cref{eqn:match_deltah,eqn:FF_hperp} with the triangle law, we recover~\cref{eqn:mismatch_criterion}.
}

We now extend the mismatch criterion by expressing it in terms of the effective cycles.
The effective cycles $\mathcal{N}_e$ between the signal and the model at the injected parameters~\cite{Sampson:2014qqa},
\begin{align}
\mathcal{N}_e &\equiv \min_{\Delta t, \Delta \phi} \dfrac{1}{\pi \rho} \sqrt{ \int \limits_{f_{\min}}^{f_{\max}} \dfrac{df}{S_n(f)} A_M^2(\bftr{\lambda} ; f) \left( \Delta \Phi (\bftr{\theta} ; f) \right)^2 }\,, \label{eqn:eff_cycles_def}
\end{align}
are a measure of how much ``dephasing'' there is between $h_S$ and $h_M$ when the waveform parameters are the injected ones and the waveform amplitudes are similar to each other, $A_S \approx A_M$. Recall here that $\Delta \Phi$ is the time- and phase-shifted difference between the phase of the signal and the waveform model (see~\cref{eq:DeltaPhiinDeltaPsi}). The effective cycles are defined with a normalization factor of $\pi$, so that $\mathcal{N}_e = 1$ when $\Delta \Phi = 2\pi$.
When the phase difference $\Delta \Psi$ (which also defines $\Delta \Phi \ll 1$ as we discuss below) is small and $A_S \approx A_M$, the effective cycles are related to the match via $1-\mathcal{F} \approx 2\pi^2 \mathcal{N}_e^2$, as shown in Ref.~\cite{Sampson:2014qqa}. Then, conservatively (i.e., neglecting the FF term),~\cref{eqn:mismatch_criterion} in terms of effective cycles becomes
\begin{align}
    \mathcal{N}_e \leq \mathcal{N}_e^{\rm upper} < \dfrac{\sqrt{N_{M}}}{2 \pi \rho}\,, \label{eqn:eff_cycles_criterion}
\end{align}
where the upper bound $\mathcal{N}_e^{\mathrm{upper}}$ on the effective cycles,
\begin{align}
    \mathcal{N}_e^{\mathrm{upper}} \equiv  \dfrac{1}{\pi \rho} \sqrt{ \int \limits_{f_{\min}}^{f_{\max}} \dfrac{df}{S_n(f)} A_M^2(\bftr{\lambda} ; f) \left( \Delta \Psi (\bftr{\theta} ; f) \right)^2 }\,,
\end{align}
is obtained when $\Delta t=0$ and $\Delta \phi=0$, which corresponds to $N_e = \mathcal{N}_e^{\mathrm{upper}}$. Therefore, we can now interpret~\cref{eqn:eff_cycles_criterion} as requiring an upper bound on the effective cycles of dephasing incurred by using the less accurate waveform, which is useful for waveform modeling.

Other extensions and improvements of the mismatch criterion have been studied in the literature. 
One such improvement arises when one realizes that~\cref{eqn:mismatch_criterion} is equivalent to $\log \left[ \mathcal{L} (d|\bfml{\lambda}, H_M)/ \mathcal{L} (d|\bftr{\lambda},H_M) \right]$ being smaller than the expectation value of the log-likelihood around the maximum (as we also argued above)~\cite{Toubiana:2024car}.
Using this, Ref.~\cite{Toubiana:2024car} improved the mismatch criterion by requiring that the difference $\log \left[ \mathcal{L} (d|\bfml{\lambda},H_M)/ \mathcal{L} (d|\bftr{\lambda},H_M) \right]$ be smaller than half the specified quantile of the $\chi^2(N_M)$ distribution, as opposed to a $1\sigma$ interval.
A similar discussion of using quantiles of the log-likelihood is given in~\cite{Baird:2012cu}. 
Another improvement arises when one realizes that the total number of parameters in the model $N_M$ may not be appropriate for use in ~\cref{eqn:mismatch_criterion}, because some parameters can be  strongly biased while others need not be biased at all~\cite{Purrer:2019jcp}. One can thus improve~\cref{eqn:mismatch_criterion} by replacing $N_M$ with the number of parameters that are expected to be strongly biased.  

 \subsection{Identifying a false GR deviation} 
\label{subsec:claiming-false-GR-deviation}

A significant systematic bias in parameter estimation is naturally a problem when inferring astrophysical parameters, but it can also be a serious problem when carrying out tests of GR. If nature is as described by GR, but a significant systematic bias is incurred in the inference of a non-GR deviation parameter because of waveform mismodeling, one may be misled into thinking that one has detected a GR deviation, when in reality one has not. Such a scenario is what we define as a \textit{false GR deviation}. The question then arises as to whether one can identify such a false GR deviation from the data. The answer to this question is intimately connected to model selection between $h_M$ and $h_S$ (recalling that the corresponding hypotheses are nested, $H_S \supseteq H_M$).

One of the tools often employed in model selection is the Bayes factor between $h_S$ and $h_M$, which is a measure of whether the data prefers one model over the other, given equal prior odds for each model. 
In the Laplace approximation to the evidence, which assumes that the region surrounding the maximum of the
posterior distribution is well approximated by a multivariate Gaussian~\cite{Cornish:2011ys,gregory_2005}, the Bayes factor in favor of the hypothesis $H_S$ is given by
\begin{align}
\log BF_{S,M} \approx (1-FF)\rho^2 + \Delta \log O_{S,M}, \label{eqn:BF_FF}   
\end{align}
with $\Delta \log O_{S,M}$ the log of the ratio of the Occam factor between hypotheses $H_S$ and $H_M$. 
As shown in Ref.~\cite{Sampson:2014qqa}, the effective cycles given by~\cref{eqn:eff_cycles_def} place an upper bound on the fitting factor and Bayes factor,
\begin{align}
    \log BF_{S,M} < 2 \pi^2 \mathcal{N}_e^2 \rho^2 + \Delta \log O_{S,M}\,, \label{eqn:BF_effective_cycles}
\end{align}
because $FF > {\cal{F}}$ and $1 - {\cal{F}} \approx 2 \pi^2 N_e^2$, as we mentioned in the last subsection. Since here $h_S$ differs from $h_M$ due to the approximate nature of the latter, the larger $\mathcal{N}_e$ or $\rho$, the larger the Bayes factor, and thus, the more the data prefers $h_S$ over $h_M$. 
We can obtain a better handle on $BF_{S,M}$ by expressing it in terms of the systematic error explicitly.
Using~\cref{eqn:FF_sys_error} in~\cref{eqn:BF_FF}, we obtain
\begin{align}
\begin{split}
 \log BF_{S,M} &\approx \rho^2 \left[ 1- \mathcal{F}(\bftr{\theta},\bftr{\lambda})\right]  + \Delta \log O_{S,M}\\
 &-  \dfrac{1}{2} (\Delta_L \lambda^i) (\Delta_L \lambda^j) \Gamma^{ij} (\bftr{\lambda}). \label{eqn:BF_sys_error_accuracy}    
\end{split}
\end{align}
Thus, setting an accuracy threshold on $\Delta_L \BF{\lambda}$ explicitly translates to a threshold on the Bayes factor between $h_M$ and $h_S$ via~\cref{eqn:BF_sys_error_accuracy}.

Given the above expressions for the Bayes factor, we can now address whether the data would prefer a GR model over a non-GR model, using a ppE model for a proxy of the latter. In the Laplace approximation, \cref{eqn:BF_FF} tells us that the Bayes factor $BF_{\mathrm{ppE,GR}}$ in favor of the ppE model over the GR model given the data is
\begin{subequations}
\begin{align}
        BF_{\mathrm{ppE,GR}} &\approx \dfrac{1}{\beta_{\max}-\beta_{\min}} \sqrt{2\pi \dfrac{\det \Gamma_{\mathrm{GR}}}{\det \Gamma_{\mathrm{ppE}}}} e^{\left[ \rho^2 \Delta FF \right]}, \label{eqn:BF_Laplace} \\
        \Delta FF &\equiv FF_{\mathrm{ppE}} - FF_{\mathrm{GR}}, 
\end{align}    
\end{subequations}%
where $\Gamma_{\mathrm{ppE}}^{ij}$ and $\Gamma_{\mathrm{GR}}^{ij}$ are the Fisher matrices when using the ppE and GR models respectively, $\beta_{\max}$ and $ \beta_{\min}$ are the maximum and minimum values of the uniform prior on the additional ppE parameter $\beta$, and $FF_{\mathrm{GR}}$ is the fitting factor of the GR model.
\Cref{eqn:BF_Laplace} corresponds to Eq. (24) of~\cite{Moore:2021eok}, and we make a more explicit comparison with their work in Sec.~\ref{subsec:toy}. Given the data, if $BF_{\mathrm{ppE,GR}} \gg 1$, one would conclude that the ppE model is greatly preferred over the GR model by the data; more precisely, in the Jeffreys' scale, if $BF_{\mathrm{ppE,GR}} > BF_{\rm thresh} \equiv 10$, the ppE model would be ``strongly'' favored over the GR model by the data~\cite{jeffreys1961theory}.  

Waveform mismodeling can throw a wrench into this model selection procedure because it is possible that the GR and the non-GR models are both bad representations of the data. Such is the case, for example, if one attempts to analyze the GW signal produced by a highly eccentric binary with GR or non-GR quasicircular waveforms. One could then be in a situation where the Bayes factor favors the non-GR model over the GR model due purely to mismodeling. Worse yet, one may even be in a situation in which the non-GR model can capture most of the SNR in the signal (because the non-GR parameters can absorb the mismodeling error), yielding a high $FF$, and thus, passing the residual SNR test. One would thus be misled into thinking one has detected a GR deviation, when in reality one has not. 

Given this, let us consider the different possible scenarios that may arise if there is mismodeling in GR and non-GR waveforms when carrying out GW tests of GR. For concreteness, let us assume that we have detected a GW signal that is described by GR and that we analyze it with both a GR model and a ppE model, both of which contain mismodeling errors in their GR sectors. Let us further assume we first carry out a parameter estimation study with both the GR and the ppE models, obtaining corner plots of the posterior probability distributions of all parameters of each model. We then carry out a model comparison study by computing the Bayes factor between the GR and the ppE models, and we compute the fitting factors for both models to determine whether their maximum likelihood realizations leave behind any significant amount of SNR. We now wish to determine whether the signal is consistent with GR or not.   

The first junction in the decision tree 
is whether there is significant systematic error in the inference of the ppE parameter, i.e., that the systematic error in the inference of the ppE parameter is larger than the statistical error. If this occurs, we would incorrectly recover a posterior for the ppE parameter that excludes GR to a high statistical significance. We shall refer to this scenario as \textit{Possible Inference of a False GR Deviation}. Since significant systematic error is directly connected to the Bayes factor through~\cref{eqn:BF_sys_error_accuracy}, there is a critical Bayes factor $BF_{{\rm{ppE,GR}}}^{\rm{crit}}$ above which we are in this scenario. When the Bayes factor is below this critical value, on the other hand, there will not be any significant systematic error in the inference of the ppE parameter, and we are thus in the case of \textit{Definite Inference of No GR Deviation}.

Let us now assume we are in the Possible Inference of a False GR Deviation scenario and consider the next two junctions of the decision tree. One of these junctions is whether the data prefers the ppE model over the GR model. This can be assessed by determining whether the Bayes factor is not just larger than the critical value (for there to be significant systematic bias), but also larger than some Jeffreys' threshold $BF_{\rm thresh}$ above which the data would ``strongly'' prefer the ppE model over the GR model, i.e., is $BF_{\rm thresh} < BF_{{\rm{ppE,GR}}} > BF_{{\rm{ppE,GR}}}^{\rm{crit}}$? If the answer is no, then the data has no strong preference between the ppE and the GR models. 

The second junction is whether the ppE and the GR models can accurately fit the signal or not. This can be assessed by determining whether the fitting factor of the ppE and the GR models is above some heuristic threshold $FF_{\rm{ppE,GR}}^{\rm dist}$. We will define this threshold such that 
\begin{align}
1 - FF_{\rm{ppE,GR}}^{\rm dist} = N_{\rm{ppE,GR}}/(2 \rho^2)\,, \label{eqn:FF_criteria}    
\end{align}
so that the amount of SNR lost is smaller than the dimensionality of the parameter space. The second junction is then whether the SNR of the residual between the data and the maximum likelihood ppE or GR models is sufficiently small, i.e., is $FF_{\rm ppE,GR} > FF_{\rm ppE,GR}^{\rm dist}$? If the answer is no for both models, then they are both bad fits to the signal.  

Given these two junctions of the decision tree, we therefore have the following 4 subscenarios (shown in~\cref{fig:stealth-overt}) within the Possible Inference of a False GR Deviation scenario:
\begin{itemize}[leftmargin=*]
    \setlength \itemsep{0.5em}
    \item \textbf{Strong Inference of No GR Deviation}: $BF_{\mathrm{ppE,GR}}^{\mathrm{crit}}< BF_{\mathrm{ppE,GR}}< BF_{\rm thresh}$ and $FF_{\mathrm{ppE}} < FF_{\mathrm{ppE}}^{\mathrm{dist}}$, top-middle region (shaded in blue) in Fig.~\ref{fig:stealth-overt}. In this region, the ppE model is not a good fit to the signal, and it is not preferred strongly over the GR model by the data. Both the residual test and the model selection test would fail. Therefore, even though parameter estimation would indicate a GR deviation (due to inferring a nonzero ppE parameter), such a deviation can be easily attributed to mismodeling instead and the GR deviation could be easily identified as false.
    \item \textbf{Weak Inference of No GR Deviation I}: $BF_{\mathrm{ppE,GR}}^{\mathrm{crit}}< BF_{\mathrm{ppE,GR}}< BF_{\rm thresh}$ and $FF_{\mathrm{ppE}} > FF_{\mathrm{ppE}}^{\mathrm{dist}}$, bottom-middle region (shaded in yellow) in Fig.~\ref{fig:stealth-overt}. In this region, the ppE model is a good fit to the signal, but it is not preferred strongly over the GR model by the data. The residual test would thus be passed, but the model selection test would fail. Therefore, even though parameter estimation would indicate a GR deviation (due to inferring a nonzero ppE parameter), such a deviation can be attributed to mismodeling instead, and the GR deviation could be positively identified as false.
    \item \textbf{Weak Inference of No GR Deviation II}: $BF_{\mathrm{ppE,GR}}^{\mathrm{crit}}< BF_{\mathrm{ppE,GR}} > BF_{\rm thresh}$ and $FF_{\mathrm{ppE}} < FF_{\mathrm{ppE}}^{\mathrm{dist}}$, top-right region (shaded in yellow) in Fig.~\ref{fig:stealth-overt}. In this region, the ppE model is not a good fit to the signal, but it is strongly preferred over the GR model by the data. The model selection test would thus be passed, but the residual test would fail. Therefore, even though parameter estimation would indicate a GR deviation (due to inferring a nonzero ppE parameter), such a deviation can be attributed to mismodeling instead and the GR deviation could be identified as false.
     \item \textbf{Incorrect Inference of GR Deviation}: $BF_{\mathrm{ppE,GR}}^{\mathrm{crit}}< BF_{\mathrm{ppE,GR}} > BF_{\rm thresh}$ and $FF_{\mathrm{ppE}} > FF_{\mathrm{ppE}}^{\mathrm{dist}}$, bottom-right region  (shaded in red) in Fig.~\ref{fig:stealth-overt}. In this region, the ppE model is a good fit to the signal, and it is strongly preferred over the GR model by the data. Both the model selection and the residual tests would thus be passed. Therefore, parameter estimation would indicate a GR deviation (due to inferring a nonzero ppE parameter), and one would conclude \textit{incorrectly} that one has detected a GR deviation (i.e., one would have identified a ``false positive''), when in reality the ppE model has captured a GR effect in the signal that was not properly captured by the ppE model.
\end{itemize}

Clearly, even when there is a significant systematic bias on the inference of a ppE parameter that may indicate a GR deviation, there are still several regions where further analysis would reveal that there is actually no GR deviation in the signal, and instead, the ppE inference is due to mismodeling error. In the Weak Inference of No GR Deviation regions (in yellow in Fig.~\ref{fig:stealth-overt}), either the Bayes factor test (region I) or the residual test (region II) would fail, and thus, indicate that there is no GR deviation. In the Strong Inference of No GR Deviation region (blue in Fig.~\ref{fig:stealth-overt}), both the Bayes factor and the residual tests would fail, clearly indicating that the signal does not contain a GR deviation. However, in the Incorrect Inference of GR Deviation region, both the Bayes factor and the residual tests would pass, leading to a false positive identification of a GR deviation. Such false positives, fortunately, depend on the SNR, because the louder the signal, the lower $FF_{\rm{ppE}}^{\rm{dist}}$, which scales as $1/\rho^2$. Therefore, even if one signal at a given SNR were to lead to a false positive, future signals at higher SNRs would exclude this possibility, slowly shrinking the red region.  

Let us close this section by placing our results in context. Previous work had assessed the systematic biases induced because of analyzing a non-GR signal with a GR model~\cite{Cornish:2011ys,Vallisneri:2012qq,Vallisneri:2013rc}.
One of the main takeaways from these analyses is that a non-GR signal can be ``stealth'' biased when the parameters of the GR model used to analyze this signal can absorb the non-GR effect in the signal, resulting in no significant loss of SNR.
Our analysis shows that the reverse can happen as well: neglected GR effects in the signal (like eccentricity, spin precession, astrophysical environments, etc.) can be absorbed by ppE parameters, resulting in a different kind of ``stealth bias,'' which could incorrectly signal the presence of a GR deviation. 
We have thus extended the characterization of systematic biases for assessing false GR deviations.

\section{Systematic biases in ppE tests due to waveform inaccuracy: a toy model
} \label{subsec:toy}

Having established the general criteria for assessing the statistical significance of systematic biases in ppE tests, we now use a toy model to explicitly quantify the biases and their significance.
Consider the following toy model that illustrates how ppE tests can become biased due to inaccurate GR waveforms.
Let the GW model that can represent the signal be $h(A_0,\mathcal{M},\kappa_1,\kappa_2,\beta ; f) = A_0 f^{-7/6} e^{i \Psi(\mathcal{M},\kappa_1,\kappa_2,\beta ; f)}$, with amplitude $A_0$ and phase
\begin{align}
    \Psi(\mathcal{M},\kappa_1,\kappa_2,\beta ; f) = \dfrac{3}{128}u^{-5} + \kappa_1 u^k + \kappa_2 u^{k+2} + \beta u^b,
\end{align}
where $u = (\pi \mathcal{M} f)^{1/3}$ is the orbital velocity, and $\mathcal{M}=(m_1 m_2)^{3/5}/(m_1+m_2)^{1/5}$ is the chirp mass.
In this simple model, $\beta$ represents a single ppE deviation from GR, while $\kappa_1$ and $\kappa_2$ represent any parametric corrections to the GR waveform that are sourced by a particular physical effect within GR. 
For example, these can be corrections due to eccentricity, spin, finite size, or environmental effects.
For brevity, we refer to this as the ``$\kappa$-effect''.
We have considered the simplest case where the GR phase with a vanishing  $\kappa$-effect is just the leading-order PN point-particle term~\cite{Blanchet:2013haa} . 
For simplicity, we have also chosen to parametrize the amplitude in such a way that $A_0$ is uncorrelated with the other waveform parameters.

The signal and model hypothesis can all be represented with different realizations of the waveform model. 
The hypothesis $H_S$ corresponds to when the signal $s$ is given by $s = h(A_{0,\mathrm{tr}},\tr{\mathcal{M}},\kappa_{1,\mathrm{tr}},\kappa_{2,\mathrm{tr}},0 ; f)$. 
We will perform two sets of parameter recoveries with approximate models $h_{\mathrm{ppE}}$ (hypothesis $H_{\mathrm{ppE}}$) and $h_{\mathrm{GR}}$ (hypothesis $H_{\mathrm{GR}}$). 
The ppE hypothesis corresponds to recoveries with the model  $h_{\mathrm{ppE}} (A_0,\mathcal{M},\beta;f) \equiv h(A_0,\mathcal{M},0,0,\beta ; f)$, where the $\kappa$ GR corrections are set to zero, and the number of parameters is clearly $N_{\mathrm{ppE}} = 2$; the intent of this model is to search for ppE deviations (i.e., posterior support at $\beta \neq 0$) in the GW data.
The GR hypothesis corresponds to recoveries with the model $h_{\mathrm{GR}} (A_0,\mathcal{M},\beta;f) \equiv h(A_0,\mathcal{M},0,0,0 ; f)$, and the number of parameters is $N_{\mathrm{GR}} = 1$; the intent of this model is to estimate the astrophysical parameters $(A_0,{\cal{M}})$ from the GW data.
Note that the hypotheses for each model are all nested within one another, since $H_S \supseteq H_{\mathrm{ppE}} \supseteq H_{\mathrm{GR}}$.

Since $s \neq h_{\mathrm{ppE}}$ even when $\beta = 0$, the approximate nature of the recovery model will result in systematic errors when inferring the existence of a ppE deviation. 
In Sec.~\ref{subsec:toy-ppE-error}, we first present our analytic calculations for the systematic errors in $\mathcal{M}$ and $\beta$. 
We allow for different injected 
PN orders of the $\kappa$-effect, and different PN orders of the ppE recovery. 
We then assess the statistical significance in Sec.~\ref{subsec:toy-stat-sig} using the tools introduced in Sec.~\ref{subsec:distinguishability}, and also explicitly quantify the regimes of systematic bias.
In Sec.~\ref{subsec:toy-numerics}, we present a numerical example to illustrate and further quantify the systematic biases.

\subsection{Systematic errors using LSA\label{subsec:toy-ppE-error}}

Using~\cref{eqn:systematic_error_explicit_main_LO_expanded}, we compute the systematic errors in the ppE and GR models.
Given that we have considered the amplitude of the signal and waveform model to be the same, and given that there is no correlation between $A_0$ and other waveform parameters, there will be no bias in the measurement of $A_0$ at $\mathcal{O}(\Delta \Psi)$ (this is consistent with Appendix A of Ref.~\cite{Dhani:2024jja}, which we rederive in Appendix~\ref{sec:LSA} for completeness).
On the other hand, the inference of $\mathcal{M}$ and $\beta$ will generally be biased at leading order in $\Delta \Psi$. 
We can therefore simplify the recovery model by setting $A_0 = A_{0,\mathrm{tr}}$, and only compute the systematic errors on $\mathcal{M}$ and $\beta$.

For the ppE recovery, we find
\allowdisplaybreaks[4]
\begin{widetext}
\begin{subequations}
\begin{align}
\begin{split}
    \Delta_L^{\mathrm{ppE}} \mathcal{M} &= -\dfrac{128}{5} u_{0,\mathrm{tr}}^{5+k} \mathcal{M}_{\mathrm{tr}} \Big( \kappa_{1,\mathrm{tr}} \dfrac{\left[I(12-k)I(7-2b) - I(7-b-k)I(12-b) \right]}{\det \hat{\Gamma}(b)} \\
    & + \kappa_{2,\mathrm{tr}}  u_{0,\mathrm{tr}}^{2} \dfrac{\left[ - I(7-2b) I(10-k) + I(12-b) I(5-b-k) \right]}{\det \hat{\Gamma}(b)} \Big), \label{eqn:Delta_chirpM_ppE_0PN}
\end{split}\\
\begin{split}
    \Delta_L^{\mathrm{ppE}} \beta &= u_{0,\mathrm{tr}}^{k-b} \Big( \kappa_{1,\mathrm{tr}} \dfrac{\left [ I(17)I(7-b-k)-I(12-b)I(12-k) \right]}{\det \hat{\Gamma}(b)} \\
    &+ \kappa_{2,\mathrm{tr}} u_{0,\mathrm{tr}}^{2} \dfrac{\left[I(17)I(5-b-k) - I(12-b)I(10-k) \right]}{\det \hat{\Gamma}(b)} \Big), \label{eqn:Delta_beta_ppE_0PN}  
\end{split}
\end{align}\label{eqn:systematic_error_ppE_toy}        
\end{subequations}
\end{widetext}
where $ \det \hat{\Gamma}(b) = I(17)I(7-2b)-I^2(12-b)$ is the determinant of the rescaled Fisher matrix corresponding to the ppE model, $u_{0,\mathrm{tr}} = (\pi \mathcal{M}f_0)^{1/3}$ with $f_0$ a characteristic frequency, and $I(p)$ defined as
\begin{align}
    I(p) = 4 \int \limits_{f_{\min}}^{f_{\max}} \dfrac{dx}{S_n(f_0 x)} x^{-p/3},
\end{align}
where $x=f/f_0$ and $p$ is an integer~\cite{Poisson:1995ef}.
Note that $\det \hat{\Gamma}(b) >0$ or that the Fisher matrix is positive definite for the log-likelihood to correspond to a local maximum.
However, $\det \hat{\Gamma}(b)$ vanishes for $b=-5$, which corresponds to a 0PN deviation. 
The reason for this is the complete degeneracy between $\beta$ and $\mathcal{M}$ in this toy model when $b = -5$. 
Such full (or partial) degeneracies affect the measurability of the ppE deviations even when doing a full Bayesian analysis, as first shown in Ref.~\cite{Cornish:2011ys}. 
Given this, we will exclude the $b=-5$ case for the calculations that follow.

To obtain some insight into~\cref{eqn:Delta_chirpM_ppE_0PN,eqn:Delta_beta_ppE_0PN}, let us consider the special case $k=b$, which corresponds to testing for ppE 
deviations that enter the phase at the same PN order as the neglected $\kappa-$effect. 
Setting $k=b$ in~\cref{eqn:Delta_chirpM_ppE_0PN,eqn:Delta_beta_ppE_0PN}, we find
\begin{widetext}
\begin{subequations}
\begin{align}
    \Delta_L^{\mathrm{ppE}} \mathcal{M} &= -\dfrac{128}{5} u_{0,\mathrm{tr}}^{7+k} \mathcal{M}_{\mathrm{tr}} \kappa_{2,\mathrm{tr}} \dfrac{\left[ I (7-2k)I(10-k) - I(5-2k)I(12-k)\right]}{ \det \hat{\Gamma}(k)}, \\
    \Delta_L^{\mathrm{ppE}} \beta &= \kappa_{1,\mathrm{tr}} + \kappa_{2,\mathrm{tr}} u_{0,\mathrm{tr}}^{2} \dfrac{I(17)I(5-2k)-I(10-k)I(12-k)}{\det \hat{\Gamma}(k)}, \label{eqn:Delta-beta_ppE_nequalsb}
\end{align}        
\end{subequations}
\end{widetext}
where we see that $\Delta_L^{\mathrm{ppE}} \mathcal{M}$ is sourced by the higher PN $\kappa_2$ term, while $\Delta_L^{\mathrm{ppE}} \beta$ is sourced simply by $\kappa_{1,\mathrm{tr}}$.
We can easily interpret this by setting $\kappa_{2,\mathrm{tr}} = 0$.
In this limit, the $\kappa$ term in the signal is entirely degenerate with the ppE term in the recovery model, and therefore,
$\Delta_L^{\mathrm{ppE}} \mathcal{M} =0$, while $\Delta_L^{\mathrm{ppE}} \beta =\kappa_{1,\mathrm{tr}}$. In other workds, the transformation $\beta \rightarrow \kappa_{1,\mathrm{tr}}$ absorbs the systematic error from $\mathcal{M}$. 
Observe that, in this case, $FF=1$ and there is no loss in SNR because $\Delta h_{\perp}$ is identically zero.
Therefore, the ppE parameter can completely absorb the phase correction by a simple transformation when the $\kappa$-effect (consisting of a single term) enters at the same PN order as the ppE deviation.

In reality, the neglected phase correction will be a PN series instead of a single term. 
As shown in~\cref{eqn:Delta-beta_ppE_nequalsb}, even when $b=k$, the ppE parameter will not absorb all of the systematic error induced by the neglected $\kappa$-effect. This was also pointed out in Ref.~\cite{Vallisneri:2012qq} in the context of detecting modified gravity effects.
When additional ppE parameters are included, parameter estimation can still be carried out through the use of ``asymptotic priors'' on the ppE deviations, as shown in Ref.~\cite{Perkins:2022fh}.
From~\cref{eqn:Delta_chirpM_ppE_0PN,eqn:Delta_beta_ppE_0PN}, we find that correlations between ppE and GR parameters have a strong impact on the systematic errors.
Recently, the authors of Ref.~\cite{Kejriwal:2023djc} explored the role of strong correlations and degeneracies between ppE and GR parameters in the context of testing GR and measuring environmental effects. Their findings are consistent with our toy example.
As mentioned in Sec.~\ref{sec:introduction}, by including additional phenomenological parameters that can also ``absorb'' the neglected $\kappa$-effect, the biases can be mitigated by sampling on the phenomenological parameters and marginalizing over them~\cite{Owen:2023mid,Read:2023hkv}.
In fact, we can view the ppE deviations as phenomenological fitting terms of the GR model.\footnote{A similar reinterpretation is possible for the problem of systematic errors due to calibration uncertainties: see Ref.~\cite{Lindblom:2008cm}.}
In this reinterpretation, the absorption of the $\kappa$-effect into the ppE parameter precisely mitigates the biases in the GR parameter (in our toy model, the chirp mass).
In this sense, the toy model provides analytical understanding of the marginalization procedure used for mitigating systematic biases.

While every ppE deviation can in principle acquire systematic error due to waveform inaccuracy, the significance of the systematic bias will vary. 
Based on Sec.~\ref{subsec:distinguishability}, we now address the statistical significance of the biases in different ppE parameters (that enter at different PN orders) for a given $\kappa$-effect.
Since the $\kappa_2$ term is a subleading PN correction to $\kappa_1$, it is sufficient to set $\kappa_{2,\mathrm{tr}} = 0$ to answer these questions.
\Cref{eqn:systematic_error_ppE_toy} then simplifies to
\begin{subequations}
\begin{align}
    \Delta_L^{\mathrm{ppE}} \mathcal{M} &= \dfrac{128}{5} \tr{\mathcal{M}} \kappa_{1,\mathrm{tr}} u_{0,\mathrm{tr}}^{k+5} \dfrac{I_{\mathcal{M}} (b,k)}{\det \hat{\Gamma}(b)}, \label{eqn:systematic-error-ppE-kappa1-a}\\ 
    \Delta_L^{\mathrm{ppE}} \beta &= \kappa_{1,\mathrm{tr}} u_{0,\mathrm{tr}}^{k-b} \dfrac{I_{\beta}(b,k)}{ \det \hat{\Gamma} (b) }, \label{eqn:systematic-error-ppE-kappa1-b}
\end{align}
\end{subequations}
where we have introduced the shorthand 
$I_{\beta}(b,k) = I(17)I(7-b-k) - I (12-b)I(12-k)$, and $I_{\mathcal{M}}(b,k) = I(12-b)I(7-b-k) - I(7-2b) I (12-k)$. 
As we discussed in Sec.~\ref{subsec:distinguishability}, we see that $ \Delta_L^{\mathrm{ppE}} \beta$ is proportional to the effective cycles induced by the $\kappa_1$ term, given by
\begin{align}
    \mathcal{N}_{e,\kappa} &=\dfrac{ |\kappa_{1,\mathrm{tr}}| u_{0,\mathrm{tr}}^{k}}{2\pi} \sqrt{ \dfrac{I(7-2k)}{I(7)}}\,. \label{eqn:eff-cycles-kappa}
\end{align}
As expected, the effective cycles decreases as $k$ increases because $u_0 < 1$, which is consistent with the fact that higher PN order terms contribute less to a model than lower PN order terms.
Since we simplified the waveform models by not including $t_c$ and $\phi_c$, we also have that $\mathcal{N}_{e,\kappa}=\mathcal{N}_{e,\kappa}^{\mathrm{upper}}$.

The match $\mathcal{F}_{\kappa}$ between the signal and the model $h_{\rm GR}$ evaluated at the injected parameters is given by
\begin{align}
    \mathcal{F}_{\kappa} = 1- \kappa^2_{1,\mathrm{tr}} u_{0,\mathrm{tr}}^{2k} \dfrac{I(7-2k)}{2 I(7)}, \label{eqn:match_kappa}
\end{align}
where we have used the fact that $\mathcal{F}_{\kappa} = 1-2\pi^2 \mathcal{N}_{e,\kappa}^2$ to the lowest nontrivial order in $\kappa_{1,\mathrm{tr}}$.
When $\kappa_{1,\mathrm{tr}} \rightarrow 0$, there is a perfect match between the signal and the $h_{\rm GR}$ model.
Since the ppE model reduces to the GR model at the injected parameters, the match between the signal and the $h_{\rm ppE}$ model is also given by~\cref{eqn:match_kappa}.

\subsection{Statistical significance of biases \label{subsec:toy-stat-sig}}
We now address the significance of the biases by comparing the systematic error when estimating $\beta$ (presented above) to the statistical error.
The statistical errors on $\mathcal{M}$ and $\beta$ are obtained from $C_{\mathrm{ppE}}(\bftr{\lambda})$, the Fisher-covariance matrix of the ppE model evaluated at the true parameters. 
By the Cramer-Rao bound~\cite{Cutler:1994ys}, a lower limit on the variance in the estimation of any parameter (which we will use as a measure of the statistical error) is simply given by the square root of the diagonal elements of the covariance matrix, 
$\sigma_{\lambda^{A}} \geq \sqrt{C^{AA}_{\mathrm{ppE}} (\bftr{\lambda})}$. 
More concretely, 
\begin{subequations}
\begin{align}
    \sigma_{\mathcal{M}} &= \dfrac{128}{5 \rho} u_{0,\mathrm{tr}}^5 \tr{\mathcal{M}} \sqrt{\dfrac{I(7)I(7-2b)}{ \det \hat{\Gamma}(b)}}, \label{eqn:stat-ppE-1} \\
    \sigma_{\beta} &= \dfrac{u_{0,\mathrm{tr}}^{-b}}{\rho}\sqrt{\dfrac{I(7)I(17)}{ \det \hat{\Gamma}(b)}}, \label{eqn:stat-ppE-2} \\
    \sigma_{\mathcal{M} \beta} &= \dfrac{8\sqrt{2}}{\sqrt{5} \rho} \sqrt{\tr{\mathcal{M}} u_{0,\mathrm{tr}}^{(5-b)}} \sqrt{\dfrac{I(7)I(12-b)}{\det \hat{\Gamma}(b) }}, \label{eqn:stat-ppE-3}   
\end{align}    
\end{subequations}
where $\sigma_{\mathcal{M}}$ and $\sigma_{\beta}$ are the statistical errors in measuring $\mathcal{M}$ and $\beta$, respectively, $\sigma_{\mathcal{M} \beta}$ is the statistical correlation between $\mathcal{M}$ and $\beta$, and $\rho$ is the SNR of the signal. 
As expected, the statistical errors are inversely proportional to $\rho$. 
Furthermore, for a fixed $\rho$ and $u_{0,\mathrm{tr}}$, the statistical errors on $\beta$ increase with increasing $b$.
We see this by noting that the statistical error on a parameter is inversely proportional to the phasing induced by the parameter at the characteristic frequency.
At a given $f_0$ (typically chosen to be in the early inspiral), higher PN ppE deviations will typically contribute fewer cycles than lower PN terms.
In other words, higher PN deviations are typically harder to measure (see e.g.~\cite{Cutler:1994ys}).

Consequently, the critical SNR where the systematic error becomes comparable to the statistical error, denoted by $\rho_{\mathrm{crit}}$, will be different for each ppE deviation. 
We estimate $\rho_{\mathrm{crit}}$ from $|\Delta_L^{\mathrm{ppE}} \beta | \approx 1.645 \sigma_{\beta}$, corresponding to the $90\%$ credible interval, which results in
\begin{align}
     \rho_{\mathrm{crit}} \approx 1.645 \dfrac{1}{\left| \kappa_{1,\mathrm{tr}}\right| u_{0,\mathrm{tr}}^{k}} \dfrac{\sqrt{I(7)I(17) \det \hat{\Gamma}(b) }}{\left| I_{\beta}(b,k) \right| }.\label{eqn:rho-crot}
\end{align}    
Observe that $\rho_{\mathrm{crit}}$ decreases with increasing $\kappa_{1,\mathrm{tr}}$.
Similarly, the higher the PN order of the $\kappa$ effect (i.e., the larger the $k$), the smaller the $u_{0,\mathrm{tr}}^{k}$ term, and the larger the critical SNR at which systematic errors become important. In other words, for larger dephasing due to the neglected $\kappa$-effect (either because $\kappa_{1,\mathrm{tr}}$ is large or $k$ is small), a smaller critical SNR is sufficient to result in significant systematic error, which is rather intuitive. 

To completely assess the bias, we further need to calculate the fitting factors of the ppE and GR models along with the Bayes factor in favor of the ppE model over the GR model.
To do so, we need the systematic error in $\mathcal{M}$ when using the $h_{\rm GR}$ model.
With a reasoning similar to the one leading to~\cref{eqn:systematic-error-ppE-kappa1-a}, we obtain
\begin{align}
\begin{split}
  \Delta_L^{\mathrm{GR}} \mathcal{M}  &= -\dfrac{128}{5} u_{0,\mathrm{tr}}^{5+k} \tr{\mathcal{M}} \kappa_{1,\mathrm{tr}} \dfrac{I(12-k)}{I(17)}. \label{eqn:sys_error_chirp_GR}
\end{split}
\end{align} 
Given~\cref{eqn:systematic-error-ppE-kappa1-a,eqn:systematic-error-ppE-kappa1-b,eqn:sys_error_chirp_GR}, we calculate $FF_{\mathrm{ppE}}$ and $FF_{\mathrm{GR}}$ using~\cref{eqn:FF_sys_error,eqn:match_kappa}, and consistently expand to the lowest nontrivial order in $\kappa_{1,\mathrm{tr}}$ to obtain
\begin{subequations}
\begin{align}
    FF_{\rm ppE} &= 1 - \dfrac{\kappa_{1,\mathrm{tr}}^2 u_{0,\mathrm{tr}}^{2k}}{2} \left[\dfrac{I(7-2k)}{I(7)} -  \dfrac{I_{\mathcal{M \beta}}(b,k)}{I(7) \det \hat{\Gamma}(b)} \right], \label{eqn:ff_ppE_kappa}\\
    FF_{\rm GR} & = 1- \dfrac{\kappa_{1,\mathrm{tr}}^2 u_{0,\mathrm{tr}}^{2k}}{2} \dfrac{\det \hat{\Gamma}(k)}{I(17)I(7)}, \label{eqn:ff_GR_kappa} \\
    I_{\mathcal{M} \beta}(b,k) &= I_{\beta}(b,k) I(7-b-k) - I_{\mathcal{M}}(b,k) I(12-k).
\end{align}    
\end{subequations}
Using the Laplace approximation of the Bayes factor given by~\cref{eqn:BF_Laplace}, the systematic errors for the ppE and GR recovery given by~\cref{eqn:systematic-error-ppE-kappa1-a,eqn:systematic-error-ppE-kappa1-b,eqn:sys_error_chirp_GR}, and the fitting factors given by~\cref{eqn:ff_ppE_kappa,eqn:ff_GR_kappa}, we obtain $BF_{\mathrm{ppE,GR}}$ as
\begin{subequations}
\begin{align}
\begin{split}
  BF_{\mathrm{ppE,GR}} & \approx \dfrac{\sqrt{2\pi}\sigma_{\beta}}{\beta_{\max}-\beta_{\min}} \exp \left[ \dfrac{1}{2}\left(\dfrac{\Delta_L^{\mathrm{ppE}} \beta}{\sigma_{\beta}} \right)^2\right] \label{eqn:BF_toy_example_general},
\end{split}\\
\begin{split}
 \left(\dfrac{\Delta_L^{\mathrm{ppE}} \beta}{\sigma_{\beta}} \right)^2 & = \rho^2 \kappa_{1,\mathrm{tr}}^2 u_{0,\mathrm{tr}}^{2k} \left(\dfrac{I^2_{\beta}(b,k)}{I(7)I(17) \det \hat{\Gamma}(b)} \right), \label{eqn:BF_toy_example_specific}    
\end{split}\\
\sigma_{\beta}^2& = \dfrac{\det \Gamma_{\mathrm{GR}}}{\det \Gamma_{\mathrm{ppE}}}. \label{eqn:BF_toy_example_c}
\end{align}\label{eqn:BF_toy_example}%
\end{subequations}
Clearly, the prior volume enters the denominator of the Bayes factor, implying that the larger the prior volume, the more the GR model will be preferred. Of course, for a ppE model that is to represent a \textit{small} deformation from GR, the width of the prior on $\beta$ cannot be arbitrarily large, since, if that were the case, the ppE corrections would dominate over the GR part of the ppE model (see~\cite{Perkins:2022fh} for a detailed discussion).   

\begin{table*}[ht!]
    \centering
    \setlength{\tabcolsep}{0.7em} %
\renewcommand{\arraystretch}{2.5}
\resizebox{\textwidth}{!}{%
    \begin{tabular}{p{2.75cm}|c|p{2.8cm}|c}
    \hline
    \hline
       \centering Regime of significant systematic bias  & \centering Quantitative description & 
 \centering Interpretation & Graphical Representation \\
       \hline
      Strong Inference of No GR Deviation   & \makecell{\\ $FF_{\rm ppE} < FF_{\rm ppE}^{\rm dist}$ \& $BF_{\rm ppE,GR}<BF_{\rm ppE, GR}^{\rm thresh}$, \\ $\Delta_L^{\rm ppE} \beta/\sigma_{\beta}>(\Delta_L^{\rm ppE} \beta/\sigma_{\beta})_{\rm dist}$ \& $\Delta_L^{\rm ppE} \beta/\sigma_{\beta}<(\Delta_L^{\rm ppE} \beta/\sigma_{\beta})_{\rm thresh}$, \\
      $\mathcal{N}_{e,\kappa}>\mathcal{N}_{e,\kappa}^{\rm dist}$ \& $\mathcal{N}_{e,\kappa}<\mathcal{N}_{e,\kappa}^{\rm thresh}$}  & \vspace{-0.9cm}Model selection test fails and residual SNR test fails. & \adjustbox{valign=c}{\includegraphics[width=8.5cm, height=1cm]{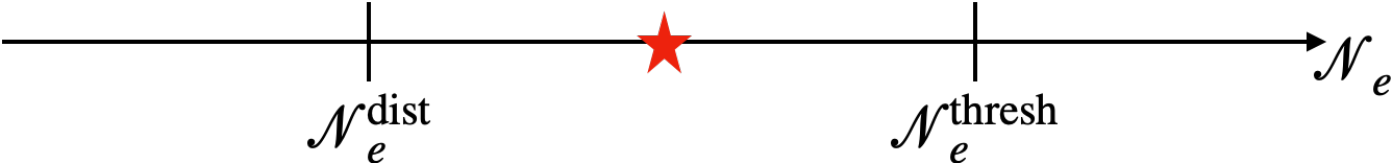}} \\
      \hline
      Weak Inference of No GR Deviation I  & \makecell{\\ $FF_{\rm ppE} > FF_{\rm ppE}^{\rm dist}$ \& $BF_{\rm ppE,GR}<BF_{\rm ppE, GR}^{\rm thresh}$ , \\ $\Delta_L^{\rm ppE} \beta/\sigma_{\beta}<(\Delta_L^{\rm ppE} \beta/\sigma_{\beta})_{\rm dist}$ \& $\Delta_L^{\rm ppE} \beta/\sigma_{\beta}<(\Delta_L^{\rm ppE} \beta/\sigma_{\beta})_{\rm thresh}$,\\
      $\mathcal{N}_{e,\kappa}<\mathcal{N}_{e,\kappa}^{\rm dist}$ \& $\mathcal{N}_{e,\kappa}<\mathcal{N}_{e,\kappa}^{\rm thresh}$}  & \vspace{-0.9cm}Model selection test fails and residual SNR test passes. & \adjustbox{valign=c}{\includegraphics[width=8.5cm, height=1cm]{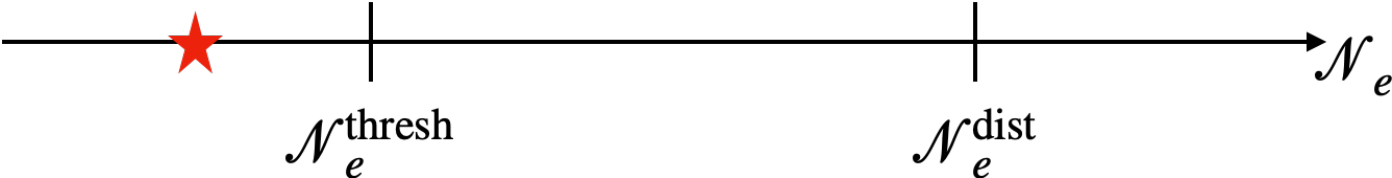}} \\
      \hline  
      Weak Inference of No GR Deviation II  & \makecell{\\ $FF_{\rm ppE} < FF_{\rm ppE}^{\rm dist}$ \& $BF_{\rm ppE,GR}>BF_{\rm ppE, GR}^{\rm thresh}$, \\ $\Delta_L^{\rm ppE} \beta/\sigma_{\beta}>(\Delta_L^{\rm ppE} \beta/\sigma_{\beta})_{\rm dist}$ \& $\Delta_L^{\rm ppE} \beta/\sigma_{\beta}>(\Delta_L^{\rm ppE} \beta/\sigma_{\beta})_{\rm thresh}$,\\
      $\mathcal{N}_{e,\kappa}>\mathcal{N}_{e,\kappa}^{\rm dist}$ \& $\mathcal{N}_{e,\kappa}>\mathcal{N}_{e,\kappa}^{\rm thresh}$}  & \vspace{-0.9cm}Model selection test passes and residual SNR test fails. & \adjustbox{valign=c}{\includegraphics[width=8.5cm, height=1cm]{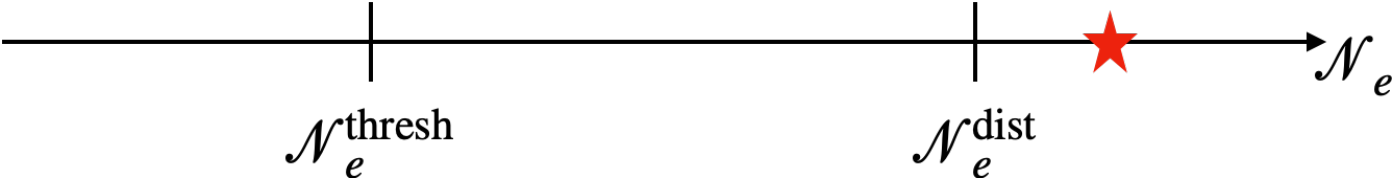}}  \\
      \hline    
      Incorrect Inference of GR Deviation   & \makecell{\\ $FF_{\rm ppE} > FF_{\rm ppE}^{\rm dist}$ \& $BF_{\rm ppE,GR}>BF_{\rm ppE, GR}^{\rm thresh}$, \\ $\Delta_L^{\rm ppE} \beta/\sigma_{\beta}<(\Delta_L^{\rm ppE} \beta/\sigma_{\beta})_{\rm dist}$ \& $\Delta_L^{\rm ppE} \beta/\sigma_{\beta}>(\Delta_L^{\rm ppE} \beta/\sigma_{\beta})_{\rm thresh}$,\\
      $\mathcal{N}_{e,\kappa}<\mathcal{N}_{e,\kappa}^{\rm dist}$ \& $\mathcal{N}_{e,\kappa}>\mathcal{N}_{e,\kappa}^{\rm thresh}$}  & \vspace{-0.9cm}Model selection test passes and residual SNR test passes. & \adjustbox{valign=c}{\includegraphics[width=8.5cm, height=1cm]{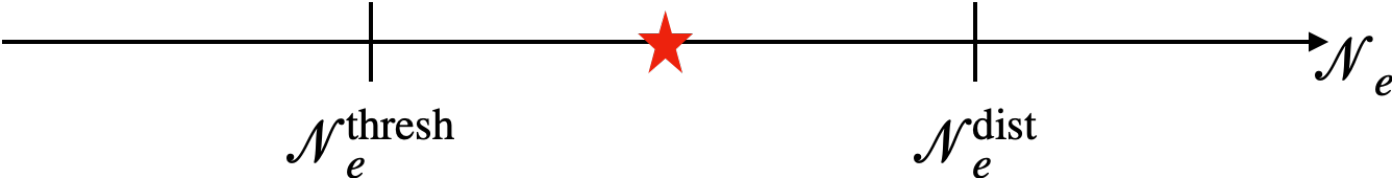}}
      \\
      \hline
    \end{tabular}
    }
    \caption{Characterization of biases into the different subregimes of Possible Inference of a False GR Deviation.}
    \label{tab:classification_toy_table}
\end{table*}

Observe that $\log BF_{\mathrm{ppE,GR}}$ has a quadratic dependence on $\kappa_{1,\mathrm{tr}}$ arising from the LSA, consistent with the relationship with the effective cycles given in~\cref{eqn:BF_effective_cycles,eqn:eff-cycles-kappa}.
The dependence of $ BF_{\mathrm{ppE,GR}}$ on $\rho$ has two contributions: one that is exponential in $\rho^2$, which arises from the ratio of maximum likelihoods between ppE and GR models, and another that is proportional to $1/\rho$, which arises from the $\sigma_{\beta}$ term in~\cref{eqn:BF_toy_example_specific}. 
In other words, the Occam penalty between the ppE and GR models scales inversely with the SNR due to the additional ppE parameter (see also~\cite{Moore:2021eok}).
Note that~\cref{eqn:BF_toy_example_c} is the same as~\cref{eqn:stat-ppE-2}, except that in the former case, we have expressed the statistical error in $\beta$ in a more general form.
The exponential growth of the likelihood ratio with $\rho^2$ will always compensate the decay caused by the Occam factor.
In other words, the ppE model will eventually become preferred with increasing SNR, even though both models are not faithful to the signal.

When the systematic error in the ppE parameter is sufficiently smaller than its statistical error, there is no significant bias in ppE tests of GR. 
Concretely, when the systematic error is smaller than the statistical error at $90\%$ confidence, we have that
\begin{align}
    BF_{\mathrm{ppE,GR}} < \dfrac{9.698 \sigma_{\beta}}{\beta_{\max}-\beta_{\min}} \equiv BF_{\mathrm{ppE,GR}}^{\mathrm{crit}}, \label{eqn:BF_crit}
\end{align}
where $ BF_{\mathrm{ppE,GR}}^{\mathrm{crit}}$ is the critical Bayes factor.
The regime of Definite Inference of No GR Deviation discussed in Sec.~\ref{subsec:distinguishability} is thus given by~\cref{eqn:BF_crit}, which is equivalent to the conditions $\rho < \rho_{\mathrm{crit}}$ or $(\Delta_L^{\rm ppE} \beta/ \sigma_{\beta}) < 1.645$.

At the critical SNR where the systematic error is comparable to the statistical error
and when $\sigma_{\beta} < (\beta_{\max} - \beta_{\min})/10$, we have that {$BF^{\rm crit}_{\rm ppE,GR} < 1$.} 
Thus, it is possible that even when the systematic error exceeds the $90\%$ statistical error, the ppE model need not be favored, which highlights the importance of requiring the Bayes factor to exceed the threshold for strong model preference.
When $BF_{\mathrm{ppE,GR}} > BF_{\mathrm{ppE,GR}}^{\mathrm{crit}}$ or $(\Delta_L^{\rm ppE} \beta/\sigma_{\beta}) > 1.645$, the bias in the ppE parameter becomes significant and can fall into the subregimes of Possible Inference of a False GR Deviation (see~ Sec.~\ref{subsec:claiming-false-GR-deviation}) depending on both $BF_{\mathrm{ppE,GR}}$ and $FF_{\mathrm{ppE}}$.
For simplicity, we adopt~\cref{eqn:FF_criteria} with $1-FF_{\rm ppE}^{\rm dist} = N_M/(2\rho^2)$ as the distinguishability threshold for assessing the fitting factor or, equivalently, the SNR loss (residual test).
A less stringent criterion would be~\cref{eqn:mismatch_criterion} or the improved criterion of Ref.~\cite{Toubiana:2024car}, but we use~\cref{eqn:FF_criteria} as a pessimistic criterion to illustrate the significance of the biases.

To classify the biases into each subregime of Possible Inference of a False GR Deviation, we will determine corresponding inequalities that $(\Delta_L^{\mathrm{ppE}} \beta/ \sigma_{\beta})$ must satisfy.
We do so by first translating $FF_{\rm ppE}^{\rm dist}$ to the bound $(\Delta_L^{\mathrm{ppE}} \beta/ \sigma_{\beta})_{\rm dist}$, and $BF_{\rm thresh}$ to the bound $(\Delta_L^{\mathrm{ppE}} \beta/ \sigma_{\beta})_{\rm thresh}$.
When $FF_{\rm ppE}$ is larger (smaller) than $FF_{\rm ppE}^{\rm dist}$, the bound $(\Delta_L^{\mathrm{ppE}} \beta/ \sigma_{\beta})_{\rm dist}$ is an upper (lower) bound on $(\Delta_L^{\mathrm{ppE}} \beta/ \sigma_{\beta})$;
when $BF_{\rm ppE} $ is larger (smaller) than $ BF_{\rm thresh}$, the bound $(\Delta_L^{\mathrm{ppE}} \beta/ \sigma_{\beta})_{\rm thresh}$ is a lower (upper) bound on $(\Delta_L^{\mathrm{ppE}} \beta/ \sigma_{\beta})$. 

The bound $(\Delta_L^{\mathrm{ppE}} \beta/ \sigma_{\beta})_{\mathrm{dist}}$ is found from $FF_{\mathrm{ppE}} = FF^{\mathrm{dist}}_{\mathrm{ppE}} $, which results in
\begin{align}
\begin{split}
\left( \dfrac{\Delta_L^{\mathrm{ppE}} \beta}{\sigma_{\beta}} \right)^2_{\mathrm{dist}} &=\dfrac{ N_M I^2_{\beta}(b,k)}{I(17) [\det \hat{\Gamma}(b) I(7-2k)-I_{\mathcal{M}\beta}(b,k)]}. \label{eqn:sig-stat-dist}
\end{split}
\end{align}
Meanwhile, the bound $(\Delta_L^{\mathrm{ppE}} \beta/ \sigma_{\beta})_{\rm thresh}$ is found from $BF_{\rm ppE} = BF_{\rm thresh}$, which results in 
\begin{align}
\left( \dfrac{\Delta_L^{\mathrm{ppE}} \beta}{\sigma_{\beta}} \right)_{\rm thresh} = \left[ 2 \log \left( \dfrac{BF_{\rm thresh}}{\sqrt{2\pi}} \dfrac{\beta_{\max} - \beta_{\min}}{\sigma_{\beta}}\right) \right]^{1/2}.\label{eqn:sig-stat-thresh}
\end{align}
Given $\left( \Delta_L^{\rm ppE} \beta/ \sigma_{\beta} \right)_{\rm dist/thresh}$, using~\cref{eqn:eff-cycles-kappa,eqn:systematic-error-ppE-kappa1-b,eqn:stat-ppE-2}, the corresponding bounds on the effective cycles $\mathcal{N}_{e,\kappa}^{\rm dist/thresh}$ are given by
\begin{align}
\begin{split}
\mathcal{N}_{e,\kappa}^{\rm dist / thresh} &= \left| \left(\dfrac{\Delta_L^{\rm ppE} \beta}{\sigma_{\beta}} \right)_{\rm dist / thresh} \right| \\
& \times \dfrac{\sqrt{I(17)I(7-2k) \det \hat{\Gamma}(b)}}{2\pi \rho |I_{\beta}(b,k)|}\,,\label{eqn:eff_cycles_thresh}
\end{split}
\end{align} 
or equivalently
\begin{align}
\mathcal{N}_{e,\kappa}^{\rm dist} &= \frac{\sqrt{N_M}}{2 \pi \rho} {\cal{I}}_{\rm dist}(b,k)\,,  
\\
\mathcal{N}_{e,\kappa}^{\rm thresh} &= \frac{{\cal{I}}_{\rm thresh}(b,k)}{2 \pi \rho} \left\{2 \log\left[\frac{BF_{\rm thresh}}{\sqrt{2 \pi} {\cal{O}}_p} \right] \right\}^{1/2}
\end{align} 
where 
\begin{align}
    {\cal{I}}_{\rm dist}(b,k) &\equiv \left | 1 - \frac{I_{{\cal{M}} \beta}(b,k)}{I(7-2 k) {\rm det} \hat{\Gamma}(b)} \right|^{-1/2}\,,
    \\
    {\cal{I}}_{\rm thresh}(b,k) &\equiv \frac{\left[I(17) I(7-2k) {\rm det}\hat{\Gamma}(b) \right]^{1/2}}{|I_\beta(b,k)|}\,,
\end{align}
and recall that the Occam penalty is defined here as ${\cal{O}}_p = \sigma_\beta/V_\beta = \sigma_\beta/(\beta_{\rm max} - \beta_{\rm min})$. 
The functions ${\cal{I}}_{\rm dist}(b,k)$ and $ {\cal{I}}_{\rm thresh}(b,k)$ are typically in the range $\sim 3 $--$17$ and $\sim 2 $--$5$ respectively; of course, ${\cal{I}}_{\rm dist}(b,k)$ diverges when the ppE parameter can exactly absorb the $\kappa$-effect (with $FF_{\rm ppE}=1$), which occurs when $b = k$.
With~\cref{eqn:sig-stat-dist,eqn:sig-stat-thresh,eqn:eff_cycles_thresh} in hand, the subregimes of Possible Inference of a False GR Deviation can be characterized.
We tabulate the characterization in~\cref{tab:classification_toy_table} by quantifying the set of inequalities that must be satisfied for each subregime of significant systematic bias.
Further details on how the characterization can change from one subregime to another is discussed in Appendix~\ref{app:bias_char_toy}.

\subsection{Numerical example \label{subsec:toy-numerics}}

For completeness and to further highlight the statistical significance and properties of the biases, we look at a fiducial system with 
$m_{1,\mathrm{tr}} = m_{2,\mathrm{tr}} = 10 M_\odot$, implying
$\tr{\mathcal{M}} \simeq 8.7$, 
and setting  $f_0=f_{\min} = 20\mathrm{Hz}$, $f_{\max} = 200 \mathrm{Hz}$, and $\rho=\{30,150\}$.
We use the O5 sensitivity curve~\cite{KAGRA:2013rdx} and work with a single detector.
We choose  a $\kappa$-effect that enters at 1PN order, implying that $k=-3$, and we choose a fiducial value of $\kappa_{1,\mathrm{tr}} = 7.10 \times 10^{-4}$, so that 
$\kappa_{1} u^{k}$ is smaller by an order of magnitude compared to 1 radian throughout the inspiral (i.e., at both $f_{\min}$ and $f_{\max}$).
We also choose a prior on $\beta$ (given by~\cref{eqn:relative_Newtonian_ppe_bound} in Sec.~\ref{sec:bayesian_framework})
such that the ppE term does not exceed the leading GR term in magnitude.
The injected signal induces a mismatch of 
$1-\mathcal{F}_{\kappa} \approx 5.70 \times 10^{-3}$, or $\mathcal{N}_{e,\kappa}\approx 0.0169$ effective cycles of dephasing between the signal and the GR model evaluated
at $\bftr{\lambda}$.
For $\rho=150$, this mismatch exceeds the distinguishability bound $N_{\mathrm{ppE}}/(2\rho^2) = 4.44 \times 10^{-5}$ by nearly two orders of magnitude, signaling that there can be potentially significant biases.
Note that the mismatch between the signal and the ppE model is the same as that with the GR model when evaluating the ppE model at $\bftr{\lambda}$, 
since the ppE model then reduces to the GR model.

With a GR recovery, we incur a systematic error in the chirp mass of $\Delta_L^{\mathrm{GR}} \mathcal{M} = -4.83 \times 10^{-4}$, resulting in a loss of fitting factor of
$1-FF_{\mathrm{GR}} \approx 3.78 \times 10^{-4}$. 
At $\rho=150$, compared to the threshold $N_{\mathrm{GR}}/(2\rho^2) = 2.22 \times 10^{-5}$, the GR recovery is indeed distinguishable from the signal because $1-FF_{\mathrm{GR}} > N_{\mathrm{GR}}/(2\rho^2)$.
In~\cref{tab:toy} we consider ppE recoveries with three selected values of $b$ and we present the systematic errors $\Delta_L^{\mathrm{ppE}} \beta$ and $\Delta_L^{\mathrm{ppE}} \log \mathcal{M}$, the critical SNR $\rho_{\mathrm{crit}}$, the loss of fitting factor $1-FF_{\mathrm{ppE}}$, the statistical errors $\sigma_{\beta}$ and $\sigma_{\log \mathcal{M}}$, the Bayes factor $BF_{\mathrm{ppE,GR}}$, and the effective cycles bounds $\mathcal{N}_{e,\kappa}^{\rm dist}$ and $\mathcal{N}_{e,\kappa}^{\rm thresh}$, along with the characterization of the bias.

\begin{table}[ht]
    \centering
     \setlength{\tabcolsep}{0.7em} %
\renewcommand{\arraystretch}{1.5}
\resizebox{\columnwidth}{!}{%
    \begin{tabular}{c||c|c|c}
    \hline
       $b$  &  $-7$ & $-3$ & $-1$ \\
    \hline 
       $\Delta_L^{\mathrm{ppE}} \beta$  & $-5.93 \times 10^{-7}$ & $\kappa_{1,\mathrm{tr}}$ & $9.81 \times 10^{-3}$ \\
    \hline
    $\Delta_L \log \mathcal{M}$ & $-1.09 \times 10^{-3}$ & $0$ & $-2.92 \times 10^{-4}$ \\
    \hline
    $\rho_{\mathrm{crit}}$ & $62.4$ & $58.9$ & $60.4$\\
    \hline
    $1-FF_{\mathrm{ppE}}$ & $4.22 \times 10^{-5}$ & $0$ & $1.90 \times 10^{-5}$ \\
    \hline
    \hline
    \multicolumn{4}{c}{$\rho=30$}\\
    \hline
    $\sigma_{\beta}$ & $7.50 \times 10^{-7}$ & $8.48 \times 10^{-4}$ & $1.20 \times 10^{-2}$ \\ 
    \hline
    $\sigma_{\mathcal{\log M}}$ & $7.77 \times 10^{-4}$ & $5.98 \times 10^{-4}$ & $2.81 \times 10^{-4}$ \\
    \hline
    $BF_{\mathrm{ppE,GR}}$ & $8.27 \times 10^{-3}$ & $9.79 \times 10^{-3}$ & $2.07 \times 10^{-2}$ \\
    \hline
    $\mathcal{N}_{e,\kappa}^{\rm thresh}$ & $0.0827$ & $0.0774$ & $0.0751$ \\
    \hline
    $\mathcal{N}_{e,\kappa}^{\rm dist}$ & $0.123$ & $\infty$ & $0.1836$ \\
    \hline
    Bias region & Weak I & Weak I & Weak I \\
    \hline
    \hline    
    \multicolumn{4}{c}{$\rho=150$}\\
    \hline
    $\sigma_{\beta}$ & $1.50 \times 10^{-7}$ & $1.70 \times 10^{-4}$ & $2.40 \times 10^{-3}$ \\ 
    \hline
    $\sigma_{\mathcal{\log M}}$ & $1.55 \times 10^{-4}$ & $1.20 \times 10^{-4}$ & $5.62 \times 10^{-5}$ \\
    \hline
    $BF_{\mathrm{ppE,GR}}$ & $2.99$ & $8.80$ & $12.3$ \\
    \hline
    $\mathcal{N}_{e,\kappa}^{\rm thresh}$ & $0.0182$ & $0.0171 $ & $0.0167$ \\
    \hline
    $\mathcal{N}_{e,\kappa}^{\rm dist}$ & $0.0246$ & $\infty$ & $0.0367$ \\
     \hline
    Bias region & Weak I & Weak I/Incorrect & Weak II \\
    \hline
    \hline
    \end{tabular}
    }
    \caption{
    Statistical significance of biases in ppE tests for the toy example.
    The three columns correspond to ppE deviations at $-1$PN, 1PN, and 2PN order, or $b=\{ -7, -3, -1 \}$.
    The injected signal has an SNR of $\rho = 30$ (top) and $150$ (bottom), and it has $\kappa_{1,\mathrm{tr}} = 7.10 \times 10^{-4}$ and ${\cal{N}}_{e,\kappa} \approx 0.0169$.
    When $\rho=150$, observe that even though $\rho > 2 \rho_{\mathrm{crit}}$ (significant systematic error), the Bayes factor strongly favors the ppE model only for $b=-3$ and $b=-1$.
    The statistical errors in $\beta$ increase with increasing $b$, while the statistical errors in $\mathcal{M}$ decrease.
    The systematic error in $\beta$ increases with increasing $b$, and so does the Bayes factor in favor of the ppE model.
    }
    \label{tab:toy}
\end{table}

\begin{figure}[t]
    \centering
    \includegraphics[width=0.49\textwidth]{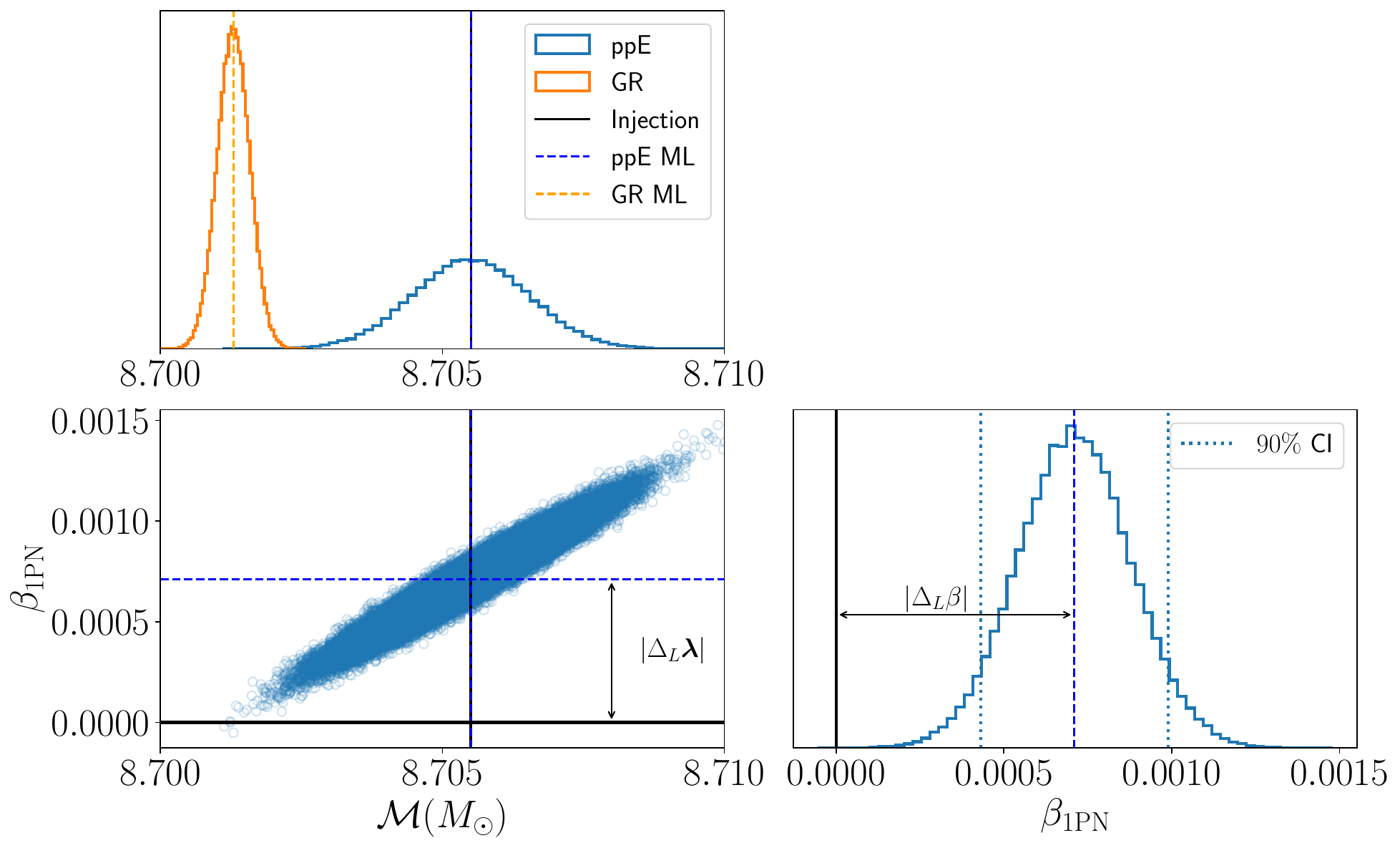}
    \caption{Systematic biases in the ppE recovery (with 1PN deviation) and the GR recovery for the toy example.
    The SNR of the signal is 150 and the neglected $\kappa$-effect enters at 1PN order with $\kappa_{1,\mathrm{tr}} = 7.10 \times 10^{-4}$. 
    In the top panel we show the marginalized posteriors on $\mathcal{M}$ when using the ppE model (blue histogram) and the GR model (orange histogram) along with their respective maximum likelihood values (dashed lines).
    The injected value of $\mathcal{M}$ coincides with the maximum likelihood value when using the ppE model.
    In the bottom right panel we show the the marginalized posterior on $\beta$ and the $90\%$ credible region, corresponding to a statistical error of $1.645 \sigma_{\beta}$ (see~\cref{eqn:stat-ppE-3}). 
    Observe that the systematic error (annotated) is larger than the statistical error.
    }
    \label{fig:corner_toy}
\end{figure}

For $\rho=30$, the systematic error in the ppE deviation is comparable to or smaller than the corresponding statistical error, consistent with $\rho < \rho_{\rm crit}$.
Meanwhile, for $\rho=150$, in each ppE recovery shown in~\cref{tab:toy}, the systematic error is larger than the statistical error, 
which is also indicated by $\rho > 2 \rho_{\mathrm{crit}}$.
As expected, the 1PN ppE recovery ($k=-3$) completely fits the signal having $FF_{\mathrm{ppE}}=1$, while being significantly biased compared to the statistical error.

To further illustrate this absorption of the systematic error by the 1PN ppE deviation, in~\cref{fig:corner_toy} we show the marginalized posteriors of $\mathcal{M}$ and $\beta$. 
For contrast, we also show the marginalized posterior of $\mathcal{M}$ in the GR recovery.
To obtain the posteriors, we used flat priors and drew from a normal distribution centered around the predicted $\bfml{\lambda}$ (see~\cref{eqn:systematic-error-ppE-kappa1-b,eqn:sys_error_chirp_GR}) with the Fisher-covariance\footnote{We invert the Fisher matrix analytically and then evaluate it numerically. We have checked that this is consistent with evaluating the Fisher matrix numerically and then inverting via singular value decomposition~\cite{numerical_recipes} to obtain the Fisher-covariance matrix.} matrix (see~\cref{eqn:stat-ppE-1,eqn:stat-ppE-2,eqn:stat-ppE-3}). 
From~\cref{fig:corner_toy}, we see that the ppE deviation completely absorbs all of the systematic error. 
As we go to higher PN order ppE deviations, the systematic error in $\beta$ increases, as does the statistical error in measuring $\beta$.  
Interestingly, the statistical error in $\mathcal{M}$ decreases as the PN order of the ppE deviation increases.
The systematic error in $\mathcal{M}$ for the 2PN ppE model is an order of magnitude smaller than in the $-1$PN ppE model.
As shown in~\cref{eqn:systematic-error-ppE-kappa1-a,eqn:systematic-error-ppE-kappa1-b}, these properties of the errors are due to the correlations between $\mathcal{M}$ and $\beta$.

\begin{table*}[ht!]
    \centering
    \setlength{\tabcolsep}{0.7em} %
\renewcommand{\arraystretch}{1.5}
\resizebox{\textwidth}{!}{%
    \begin{tabular}{c|c|c|c}
    \hline \hline
       Physical parameters  &  Interpretation & Prior & Injected value  \\
       \hline \hline
       RA  & Right Ascension -- Longitude on celestial 2-sphere & $\mathrm{RA} \in \mathcal{U} (0, 2\pi)$ & $1.3750$ \\
        \hline
       DEC  & Declination -- Latitutde on celestial 2-sphere & $\sin \mathrm{DEC} \in \mathcal{U} ( -1, 1 )$ & $-1.2108$ \\ 
       \hline
       $\psi$  & Polarization angle & $\psi \in \mathcal{U}( 0, 2\pi)$ & $2.6590$ \\ 
       \hline
       $\iota$  & Inclination angle & $\cos \iota \in \mathcal{U}( -1, 1)$ & varied (see text) \\ 
       \hline
       $\phi_{\mathrm{ref}}$  & Phase at reference frequency $f_{\mathrm{ref}}$ (50Hz) & $\phi_{\mathrm{ref}} \in \mathcal{U}( 0, 2\pi)$ & $1.3$ \\ 
       \hline
       $t_c$  & Time of coalescence & $t_c \in \mathcal{U}( t_c^{\mathrm{inj}} - 0.2, t_c^{\mathrm{inj}} + 0.2)$ & $T_{\mathrm{signal}}-2$ \\ 
       \hline       
       $D_L$  & Luminosity distance to source & Uniform in volume (see text) & varied (see text) \\ 
       \hline
       $m_1$  & Mass of primary & Uniform in $m_1$ (see text) & varied (see text) \\ 
       \hline  
       $m_2$  & Mass of secondary & Uniform in $m_2$ (see text) & varied (see text) \\ 
       \hline
       $\{\chi_{1x},\chi_{1y} \}$  & Spins of primary in the orbital plane & -- & $\chi_{1x}$ varied (see text), $\chi_{1y} = 0$ \\ 
       \hline
       $\{\chi_{2x},\chi_{2y} \}$  & Spins of secondary in the orbital plane & -- & $\chi_{1x}=0$, $\chi_{1y} = 0$ \\ 
       \hline
       $\chi_{1z}$  & Spin of primary along orbital angular momentum & Aligned spin prior (see text) & $0.1$ \\ 
       \hline 
       $\chi_{2z}$  & Spin of secondary along orbital angular momentum & Aligned spin prior (see text) & $-0.15$ \\ 
       \hline
       $\betappe{n}$  & ppE deviation that enters at $n$PN order &  Uniform in $\betappe{n}$ (see text) & $0$ \\ 
       \hline 
    \end{tabular}
    }
    \caption{Physical parameters of the waveform models with their interpretation, prior, and injected values.
    The parameters that are varied for the injection are explained in Sec.~\ref{subsec:inj-models}. 
    The parameters whose prior is not listed (`--') are not part of the recovery and are not sampled for the main analysis.}
    \label{tab:param-table}
\end{table*}

When $\rho=30$, the ppE recovery is strongly disfavored for all the ppE deviations.
The loss in SNR is also not significant (as can be seen from the effective cycles listed in~\cref{tab:toy}), implying that the biases are all characterized as a Weak Inference of No GR Deviation I.
For $\rho=150$, we observe that the Bayes factor strongly favors the
2PN recovery (over the GR recovery), in comparison to how the $-1$PN and 1PN order recoveries are favored.
Specifically, the Bayes factor decreases when going to lower PN order, with the least significant support for the $-1$PN recovery.
The $-1$PN recovery is clearly an example of a Weak Inference of No GR Deviation I.
In the 1PN recovery, the bias is characterized as a Weak Inference of No GR Deviation, but it is effectively on the borderline with an Incorrect Inference of a GR Deviation, owing to the Bayes factor being very close to the threshold.
The loss in SNR is significant for the 2PN recovery, which, when combined with the strong preference for the ppE model, results in a Weak Inference of No GR Deviation II.

We conclude this section by comparing our results with those in the literature, and then by briefly discussing caveats in using the LSA and Laplace approximation for assessing systematic errors and distinguishability.
Our~\cref{eqn:BF_toy_example_general} is essentially a rederivation of Eq. (24) in~\cite{Moore:2021eok}, but we have obtained our result without neglecting correlations between ppE and GR parameters. 
When neglecting correlations, our~\cref{eqn:BF_toy_example_general} reduces to that of~\cite{Moore:2021eok}.
Our result in~\cref{eqn:sig-stat-thresh} 
for how $(\Delta_L^{\rm ppE} \beta/\sigma_{\beta})_{\rm thresh}$ or the Bayes factor given by~\cref{eqn:BF_toy_example_specific} scales with SNR is effectively reflected in Fig.~2 of~\cite{Moore:2021eok}, where the authors have also considered the impact of stacking multiple events.
Another difference with~\cite{Moore:2021eok} is that we have set the prior odds ratio between the ppE and GR models to unity.
While we derived $BF_{\mathrm{ppE,GR}}$ for the simple case where there is a single GR term in the phase of $h_{\mathrm{GR}}$, the result holds even when the GR phase consists of additional terms beyond the leading PN order.
The reason is that~\cref{eqn:BF_toy_example_general} does not manifestly depend on the recovery performed with $h_{\mathrm{GR}}$, which is the result from computing $BF_{\mathrm{ppE,GR}}$ using the \emph{Savage-Dickey ratio}~\cite{Chatziioannou:2014bma,Moore:2021eok} between the models $h_{\mathrm{ppE}}$ and $h_{\mathrm{GR}}$.

Typically, the regime of validity of the LSA to compute $\Delta_L \lambda^i $ is when $\rho \gg 1$ \emph{and} $\Delta \Psi \ll 1$.
The validity regime defined by $\rho \gg 1$  is well studied~\cite{Vallisneri:2007ev}, but similar studies need to be performed to determine the validity regime defined by $\Delta \Psi \ll 1$.
Higher order corrections to~\cref{eqn:systematic_error_explicit_main_LO_expanded} include $\mathcal{O}(\Delta \Psi)^2, \mathcal{O}(\Delta \Psi \Delta A)$ terms that arise from expanding the difference between the waveform and signal (we compute these in Appendix~\ref{sec:LSA}), as well as next-to-leading order corrections $\mathcal{O}((\Delta_L \lambda^i)^2)$ to~\cref{eqn:systematic_error_explicit_main_LO_unexpanded} that arise from expanding the waveform about the maximum likelihood point. 
These corrections have to be systematically computed in order to assess the validity of using the LSA for computing systematic error.
This is currently beyond the scope of this project and we leave it for future work.
To characterize the regimes of the biases, we also used the Laplace approximation to LO and neglected higher order corrections. 
For assessing biases for more realistic models, where bimodalities are a common feature, the Laplace approximation and LSA will likely break down. 
Despite these limitations, the results obtained from the simplified toy model provide an analytic understanding into the characterization of false GR deviations induced by waveform inaccuracy.

\section{Injection--recovery setup and Bayesian parameter estimation} \label{sec:bayesian_framework}

The methods and statistical criteria reviewed in Sec.~\ref{sec:general_problem} set up the background for understanding the general problem of systematic errors due to waveform inaccuracies.
The toy example in Sec.~\ref{subsec:toy} provides some analytic insight into how ppE tests become biased, explicitly showing how the biases can be classified as belonging to different regimes of significant systematic bias: Strong Inference of No GR Deviation, Weak Inference of No GR Deviation I \& II, and Incorrect Inference of GR Deviation.
With this understanding, we now focus on addressing potential systematic biases in ppE tests due to neglecting spin precession and higher modes.
We analyze the biases induced by spin precession and higher modes independently by injecting signals with either spin precession or higher modes, and recovering with waveform models that contain neither.

We first discuss the details of the injections in Sec.~\ref{subsec:inj-models}, where we list the parameters that will be varied for both the spin precession and higher modes analyses. 
In Sec.~\ref{subsec:rec-models} we discuss the details of the recovery models and explain which ppE deviations we consider in the ppE model.
In Sec.~\ref{subsec:priors} we motivate the priors for the parameters, and the parameters we sample on when performing Bayesian parameter estimation.
We use both MCMC and nested sampling methods for sampling the posterior. 
Specifically, we use both the GW-Analysis-Tools (\texttt{GWAT}) pipeline developed in~\cite{Perkins:2021mhb}, and the more widely used \texttt{BILBY}~\cite{bilby_paper} pipeline.
In Sec.~\ref{subsec:posterior} we explain the details of the samplers used in each pipeline, along with the details of computing the systematic and statistical errors for assessing the biases.
In~\cref{tab:param-table} we list the physical parameters of the models along with their interpretation, priors, and injected values.

\subsection{Injection models}\label{subsec:inj-models}
We specialize to GW signals observable by the LVK detectors (Hanford, Livingston, Virgo, and KAGRA) operating at design sensitivity with O5 sensitivity curves~\cite{KAGRA:2013rdx} for each detector.
We inject noise-less data described by $d=s$ into each detector with $f_{\min} = 20 \mathrm{Hz}$ and $f_{\max} = 2048 \mathrm{Hz}$ (or a sampling frequency of $4096\mathrm{Hz}$) and we set the duration of the signal\footnote{For frequency-domain data analysis, the duration is used to set the linear spacing in frequency $\Delta f = 1/T_{\mathrm{signal}}$ when sampling the waveform.} to $T_{\mathrm{signal}}=32$s.
As mentioned in Sec.~\ref{sec:general_problem}, we set the noise $n=0$ because we wish to isolate systematic error induced by waveform inaccuracy from errors induced by a specific noise realization, but we do include a noise spectral density to account for the average Gaussian noise of the detectors. In particular, this means we do not include any type of glitches in the noise model, which could by themselves generate a different type of systematic error that could also be confused with a GR deviation~\cite{Gupta:2024gun}. 

\subsubsection{Injections with spin precession} \label{subsub:inj-prec}
To investigate biases induced by spin precession, we inject using the \texttt{IMRPhenomPv2} model~\cite{Hannam:2013oca,Schmidt:2012rh,Khan:2018fmp}, resulting in a signal $s=h_S=h^{\mathrm{PhenomPv2}}$ in each detector.
This (precessing, dominant $\ell =2$ mode) waveform model depends on all 15 parameters listed in~\cref{tab:param-table}, with the ppE deviation set to zero. 
Specifically, the waveform model depends on $\BF{\lambda} = \{\mathrm{RA}, \mathrm{DEC}, \psi, \iota, \phi_{\mathrm{ref}}, t_c, D_L, m_1, m_2, \chi_{1z}, \chi_{2z} \}$ and $\BF{\kappa} = \{ \chi_{1x}, \chi_{1y}, \chi_{2x}, \chi_{2y} \}$. 
Recall that $\vec{S}_{1} = m_{1}^2 \{ \chi_{1x} ,\chi_{1y}, \chi_{1z} \}$ is the spin of the primary in the orbital basis (and likewise for the secondary).
Note that the nonprecessing, quasicircular dominant mode waveform model \texttt{IMRPhenomD} shares the parameters $\BF{\lambda}$ with \texttt{IMRPhenomPv2}, while $\BF{\kappa}$ are the additional parameters of the precessing model that are not needed in the \texttt{IMRPhenomD} model.

The effects of spin precession~\cite{Apostolatos:1994,Apostolatos:1995,Kesden:2014sla,Gerosa:2015tea,Chatziioannou:2017tdw} on the waveform strongly depend on the spins in the orbital plane $\vec{S}_{1 \perp} = m_1^2 \{\chi_{1x} , \chi_{1y}\}$ and $\vec{S}_{2 \perp} = m_2^2 \{\chi_{2x} , \chi_{2y}\}$. 
In particular, precession is essentially dependent on the mass-weighted combination $\chi_p$, defined as~\cite{Hannam:2013oca,Schmidt:2012rh,Khan:2018fmp} 
\begin{align}
\chi_p \equiv \dfrac{\max \left(A_{1} S_{1 \perp}, A_{2} S_{2 \perp}\right)}{A_{1} m_{1}^{2}},
\end{align}
where $A_{1}=2+\frac{3 m_{2}}{2 m_{1}}, A_{2}=2+\frac{3 m_{1}}{2 m_{2}}$, $S_{1\perp} = \mathrm{mag}(\vec{S}_{1\perp})$, and $S_{2\perp} = \mathrm{mag}(\vec{S}_{2\perp})$.
The effects of precession become stronger with increasing $\chi_p$.
Further, the inclination of the orbital plane with respect to the line of sight (i.e.~the inclination angle $\iota$) and the masses (or duration of the signal) also affect the impact of precession on the waveform~\cite{Apostolatos:1994,Apostolatos:1995,Trifiro:2015zda,Chatziioannou:2017tdw,Gangardt:2021lic,Gangardt:2022ltd}. 

To analyze biases due to spin precession, we therefore vary the injected $\chi_p$, $\iota$, and $\{m_1,m_2\}$.
For simplicity, we always inject with $\chi_{1y}=0,\chi_{2x}=0,\chi_{2z}=0$.
Doing so, we simply have that $\chi_p = \chi_{1x}$, and we vary $\chi_{1x}$ over 3 values that result in 3 $\chi_p$ injections, namely $\chi_p^{\mathrm{inj}} = \{ 0,0.45,0.9\}$.
For all injections, we set the spin components along the orbital angular momentum vector to be $\chi_{1z}^{\mathrm{inj}} = 0.1$ and $\chi_{2z}^{\mathrm{inj}} = - 0.15$.
We choose $\iota^{\mathrm{inj}} = \{ 0.05, \pi/4, \pi/2 \}$, corresponding to $\{$ ``face-on'', ``mid-inc'', ``edge-on''$\}$ systems, respectively.
We do not choose $\iota^{\mathrm{inj}}$ to be exactly zero when studying face-on systems, because we primarily use \texttt{GWAT} (see Sec.~\ref{subsec:posterior} and Sec.~\ref{subsec:priors}) for the spin precession analysis;
injecting with  $\iota^{\mathrm{inj}} = 0$ using \texttt{GWAT} can result in edge issues related to the prior.\footnote{Further, it can also result in a highly ill-conditioned Fisher matrix, which affects the proposal jumps along the eigenvectors of the Fisher matrix deployed in \texttt{GWAT} (see Sec.~\ref{subsec:posterior}).}
The spins and inclination are also defined at a reference frequency of $f_{\mathrm{ref}} = 50$Hz.
We choose two different total masses $m_{\rm tot} =\{ 20\msun,40\msun\}$ with a mass ratio of ${q\equiv m_2/m_1} = 2/3$, 
where by definition ${m_2<m_1}$, so $0 < q < 1$. 
To analyze the statistical significance of the biases, we also choose two fiducial SNRs, $\rho = \{ 30, 60 \}$. 
When changing the SNR, we restrict the recovery to the GR model and the ppE model with a 1PN deviation.
We obtain the desired SNR by rescaling the luminosity distance $D_L$.
For the extrinsic parameters, $\{\mathrm{RA}, \mathrm{DEC},\psi, \phi_{\mathrm{ref}}, t_c\}$, we use the injected values listed in~\cref{tab:param-table}.

\subsubsection{Injections with higher modes}\label{subsub:inj-hm}
To investigate biases induced by higher modes, we inject using the \texttt{IMRPhenomHM} model~\cite{London:2017bcn}, resulting in a signal $s=h_S=h^{\mathrm{PhenomHM}}$ in each detector.
The (nonprecessing, quasicircular) higher-mode waveform model contains the $(\ell, m) = \{ (2,2), (2,1), (3,3), (3,2), (4,4), (4,3) \}$ modes, and depends on the waveform parameters $\BF{\lambda} = \{\mathrm{RA}, \mathrm{DEC}, \psi, \iota, \phi_{\mathrm{ref}}, t_c, D_L, m_1, m_2, \chi_{1z}, \chi_{2z} \}$, with the ppE deviation set to zero.
Note that \texttt{IMRPhenomHM} does not introduce new waveform parameters compared to the \texttt{IMRPhenomD} model.

The effects of higher modes increases with increasingly asymmetric masses, and they are more pronounced for edge-on systems. To study biases in the ppE recovery due to neglect of higher modes, we vary the mass ratio $q = m_2/m_1$ (where recall that $0 < q \leq 1$), the inclination angle $\iota$, the total mass $m_{\rm tot}$, the network SNR, and the PN order of the ppE recovery. 
For the mass ratio, we inject with $q^{\mathrm{inj}} = \{ 1/7, 2/3, 9/10\}$ to contrast the effects of a highly-asymmetric and a highly-symmetric system.
Our choice of $q^{\mathrm{inj}} =1/7$ is motivated by the validity of the \texttt{IMRPhenomHM} model, which has been well tested up to a mass ratio of $1/8$ (see~\cite{London:2017bcn,Kalaghatgi:2019log}).
We consider three fiducial face-on, mid-inc, and edge-on injections, corresponding to $\iota^{\mathrm{inj}} = \{0, \pi/4, \pi/2 \}$.
We primarily use the \texttt{BILBY} pipeline for the higher-mode analysis, and therefore, we can choose $\iota^{\mathrm{inj}}=0$ without edge issues.
We again choose two fiducial total masses $m^{\mathrm{inj}}_{\rm tot} = \{20\msun, 40\msun \}$ and  SNRs $\rho =\{30,60 \}$, just as we did in~\cref{subsub:inj-prec}.
For the remaining parameters of \texttt{IMRPhenomHM}, we inject using the values listed in~\cref{tab:param-table}.

\subsection{Recovery with approximate GR and ppE models}\label{subsec:rec-models}

For a given injection, we perform two sets of parameter recoveries: one with $h_M = h^{\mathrm{PhenomD}}$, and another with $h_M = h^{\mathrm{PhenomD+ppE}}$. 
The $h^{\mathrm{PhenomD}}$ waveform depends on the parameters $\BF{\lambda} = \{ \mathrm{RA}, \mathrm{DEC}, \psi, \iota, \phi_{\mathrm{ref}}, t_c, D_L, m_1, m_2, \chi_{1z}, \chi_{2z} \} $~\cite{Khan:2015jqa}. 
The $h^{\mathrm{PhenomD+ppE}}$ waveform contains, in addition, a single ppE deviation $\delta \Psi_{\rm ppE}=\betappe{(5+b)} u^b$ added to the inspiral phase of the \texttt{IMRPhenomD} model at $(5+b) \mathrm{PN}$ order,\footnote{Continuity and differentiability are enforced at the junctions between the inspiral and intermediate stage, and at the intermediate and merger-ringdown stage, as prescribed in the \texttt{IMRPhenomD} model. The ppE modification to the inspiral phase naturally changes the continuity and differentiability constants, therefore propagating modifications in the inspiral stage to the other stages.}
so it contains an additional ppE parameter $\betappe{5+b}$, where $b$ is the ppE index.
For each recovery of the waveform parameters, we fix $b$ to certain values, discussed below.
We do not consider ppE deviations in the amplitude because typically parametric tests of GR with corrections to the phase are more sensitive
(in general, phase corrections are more sensitive than amplitude corrections because of the cross term $\rho^2_{MS}$ in the log-likelihood given by~\cref{eqn:log-likelihood})~\cite{Perkins:2022fh}.
In other words, biases in the inspiral ppE tests are sufficient for characterizing potential false GR deviations due to waveform systematics.
We briefly review each waveform model in Appendix~\ref{sec:waveform_review}. 

For each injection, the ppE recovery is performed by testing for deviations at $-1$PN, 1PN, and 2PN order, or $b=\{ -7, -3, -1 \}$.
We consider these ppE deviations based on the mapping to specific modified theories of gravity~\cite{Yunes_2009} and their distinct observational signatures.
A deviation at $-1$PN order maps to theories where dipole radiation is activated by violating the strong equivalence principle, due to either a scalar or a vector field (such as Brans-Dicke gravity~\cite{Will:2014kxa}, scalar-Gauss-Bonnet gravity~\cite{Boulware:1985,GROSS198741,Kanti:1995vq}, or Einstein-AEther theory~\cite{Jacobson:2007veq}).
A deviation at 1PN maps to theories with a massive graviton~\cite{Will:2014kxa}, which modifies the dispersion relation in the propagation of GWs.
A deviation at 2PN order maps to theories where magnetic dipole radiation is activated, due to both the strong equivalence principle and parity invariance being violated (such as in dynamical Chern-Simons gravity~\cite{Alexander:2009tp}).
We do not consider a 0PN deviation because of the strong correlation with the chirp mass, which can introduce sampling biases.
In Appendix~\ref{sec:review_ppE} we review the ppE framework with more examples of theories for each ppE deviation we consider in this work.

\subsection{Priors and sampling parameters}\label{subsec:priors}
For the ppE model, we sample the posterior distribution $\mathcal{P}(\BF{\lambda}|d, H_{\mathrm{ppE}})$ (and similarly for the GR model), on all parameters of the waveform model. 
The parameter vector is 12 dimensional for the ppE model (under the hypothesis $H_{\mathrm{ppE}}$) and 11 dimensional for the GR model (under the hypothesis $H_{\mathrm{GR}}$). 
We use different sampling parameters when analyzing \texttt{IMRPhenomPv2} injections and \texttt{IMRPhenomHM} injections.
The choice of sampling parameters is tied to the sampling method of the data analysis pipelines, \texttt{GWAT} and \texttt{BILBY}.
We discuss the specifics of the sampler settings in Sec.~\ref{subsec:posterior} below.

In \texttt{GWAT}, we use an agnostic prior that is uniform in volume $\propto D_L^3$ and sample on $\log D_L$ instead of $D_L$.
With the transformation, the log-prior acquires a contribution from the log-Jacobian of the transformation, which is proportional to $3 \log D_L$. 
For all the recoveries, we set $\min(D_L) = 50 \mathrm{Mpc}$ and $\max(D_L) = 1200 \mathrm{Mpc}$ in the prior, so that the injected distance (that was rescaled to obtain the fiducial SNRs) is included in the prior range.
Our priors on the component masses $m_1$ and $m_2$ is agnostic and uniform.
However, we sample on $\{ \log \mathcal{M}, \eta \}$ instead of $\{ m_1, m_2 \}$, where $\eta = m_1 m_2 / m_{\rm tot}^2$ is the symmetric mass ratio, $\mathcal{M} = \eta^{3/5} m_{\rm tot}$ is the chirp mass, and $m_{\rm tot} = m_1 + m_2$ is the total mass. 
Sampling on the transformed mass variables results in a contribution to the log-prior from the log-Jacobian of the transformation\footnote{The transformation becomes singular at $\eta = 1/4$, which is in practice not an issue as long as we do not inject very close to the edge of the prior.} which is proportional to $\log \left[\mathcal{M}^2 /(\eta^{6/5}\sqrt{1-4\eta})  \right]$.
We set the range on $m_1$ to $\min (m_{1}) = 2 \msun$ and $\max (m_{1}) = 40 \msun$ (and the same for $m_2$). 
For $\{\chi_{1z} , \chi_{2z}\}$, we use an equivalent of the aligned spin prior detailed in~\cite{Lange:2018pyp}.

We assume an agnostic uniform prior on $\betappe{(5+b)}$ with bounds determined by requiring that the magnitude of the ppE correction be smaller than the leading GR phase throughout the inspiral~\cite{Perkins:2022fh}.
In general, this translates to
\begin{align}
    \left| \betappe{(5+b)}\right|< \dfrac{3}{128} (\pi \mathcal{M}f)^{-(5+b)/3}. \label{eqn:relative_Newtonian_ppe_bound}
\end{align}
To evaluate the bound in~\cref{eqn:relative_Newtonian_ppe_bound}, for $5+b>0$, we conservatively choose $f=f_{\mathrm{ISCO}} = (6^{3/2}\pi \max(m_{\rm tot}))^{-1}$ and $\mathcal{M} = \max(\mathcal{M})$, where $\max(m_{\rm tot})$ and $\max(\mathcal{M})$ correspond to the maximum values of the total mass and chirp mass respectively, given the priors on $m_1$ and $m_2$. 
For $5+b<0$, we instead choose $f=f_{\min}$ and $\mathcal{M}=\min(\mathcal{M})$ to evaluate the bound, where $\min(\mathcal{M})$ is the minimum value of the chirp mass, given the priors on $m_1$ and $m_2$.
For the remaining (extrinsic) parameters of the waveform model, namely $\{\mathrm{RA}, \mathrm{DEC}, \iota, \psi, \phi_{\mathrm{ref}}, t_c \}$, we use the priors given in~\cref{tab:param-table}.

When using \texttt{BILBY}, we adopt the same set of priors used in \texttt{GWAT} with the following changes. We sample on $D_L$ (instead of $\log D_L$), using the \texttt{UniformInSourceFrame} prior, and we sample on $\{\mathcal{M},q \}$ (instead of $\{ \mathcal{M} , \eta \}$) for the mass variables.
We find that these differences in the sampling do not affect the biases observed in the chirp mass or the ppE deviation, which we demonstrate in Appendix~\ref{sec:convergence}. 

\subsection{Posterior sampling, and computing systematic and statistical errors}\label{subsec:posterior}
For sampling the posterior, we use both the \texttt{GWAT} and \texttt{BILBY} pipelines. 
GWAT is a GW focused data analysis pipeline that builds off of \texttt{BayesShip} (general purpose MCMC sampler), both of which were developed in~\cite{Perkins:2021mhb}.
We utilize the Parallel Tempering Markov Chain Monte Carlo (PTMCMC) routine in \texttt{GWAT} that is based on~\cite{Vousden:2016} (see also~\cite{gregory_2005,Littenberg:2009bm,Littenberg:2010gf} for a review of PTMCMC).
In Appendix~\ref{sec:convergence}, we provide details of the sampler settings, robustness, and convergence checks we performed.

\begin{figure*}[ht!]
    \centering
    \includegraphics[width=0.75\textwidth]{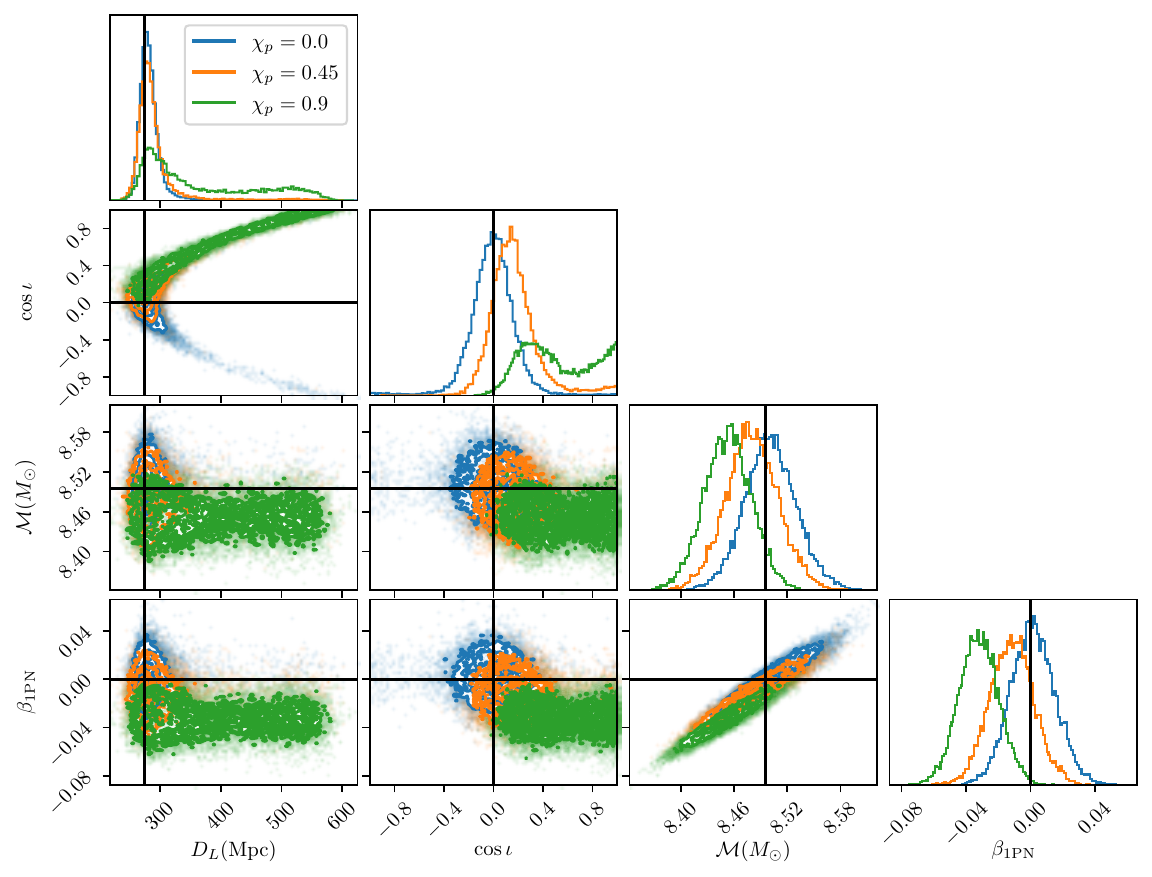}
    \caption{Corner plot of the marginalized 2d and 1d posteriors of $\{ D_L, \cos \iota, \mathcal{M}, \betappe{1} \}$. 
The posteriors show precession-induced systematic biases in the recovery of these parameters for an edge-on $(12,8)\msun$ injection with varying $\chi_p$ at an SNR of 30.
The black lines indicate the injected values, while
the blue, orange and green posteriors correspond to $\chi_p = \{ 0,0.45,0.9 \}$, respectively.
Observe that there are strong biases in the recovery of the intrinsic parameters $\{ \mathcal{M}, \betappe{1} \}$, and there are bimodal features in the extrinsic parameters $\{ D_L, \cos \iota \}$, which are induced by the neglected amplitude precession effects.
Unsurprisingly, there are strong correlations between $D_L$ and $\cos \iota$, as there are between $\mathcal{M}$ and $\betappe{1}$.
The bias in chirp mass propagates onto the non-GR sector and contaminates the parametrized test of GR, suggesting the detection of a GR deviation in the high $\chi_p$ injection case, when in reality there is no GR deviation in the signal.}
\label{fig:corner_extrinsic_intrinsic_edgeOn_precession}
\end{figure*}

For the injections with spin precession, we primarily use \texttt{GWAT} for sampling the posterior.
To ensure robustness in the injection-recovery studies that reveal systematic biases, we compare results from the \texttt{GWAT} pipeline and the \texttt{BILBY} pipeline, along with other convergence and sanity checks, which are detailed in Appendix~\ref{sec:convergence}.
For the injections with higher modes, we use the \texttt{BILBY} pipeline as the waveform library in the \texttt{GWAT} pipeline does not currently contain \texttt{IMRPhenomHM}.
In Appendix~\ref{sec:convergence}, we also describe the convergence and robustness checks we performed when using \texttt{BILBY} for parameter estimation.

For the injections with higher modes, we use the \texttt{BILBY} pipeline as the waveform library in the \texttt{GWAT} pipeline does not currently contain \texttt{IMRPhenomHM}.
While we do not need \texttt{IMRPhenomHM} for computing biases when using \texttt{IMRPhenomD}+\texttt{ppE}, we perform sanity checks by injecting and recovering with the base \texttt{IMRPhenomHM} model.

With the recovered posterior, we determine the maximum value of the likelihood using $\max[\mathcal{L} (d| \BF{\lambda},H_{M})]$ and the corresponding $\bfml{\lambda}$ using $\mathrm{argmax}[\mathcal{L} (d| \BF{\lambda},H_{M})]$. 
With $\bfml{\lambda}$ in hand, we compute the likelihood-based systematic error $\Delta_L \BF{\lambda} = \bfml{\lambda} - \bftr{\lambda}$, defined back in Sec.~\ref{subsec:definitions}.
We ensure that our estimation of the maximum posterior is robust by comparing $\log \mathcal{L} (d| \bfml{\lambda},H_M)$ with the predicted value~\cite{Toubiana:2023cwr} of $\overline{\log \mathcal{L} (d | \BF{\lambda},H_M)} + N_M/2$, where $\overline{\log \mathcal{L} (d | \BF{\lambda},H_M)}$ is the sample mean of the log-likelihood distribution.\footnote{This is equivalent to computing the variance of the log-likelihood distribution~\cite{Veitch:2014wba}.}
We find that the differences are within the sampling error and that a direct estimation of $\log \mathcal{L} (d| \bfml{\lambda},H_M)$ using $\max[\mathcal{L} (d| \BF{\lambda},H_{M})]$ suffices for our analyses.

We also determine the maximum value of the posterior using $\max[\mathcal{P} ( \BF{\lambda}| d, H_{M})]$ and the corresponding $\BF{\lambda}_{\mathrm{MP}}$ using $\mathrm{argmax}[\mathcal{P} ( \BF{\lambda}| d, H_{M})]$.
With $\BF{\lambda}_{\mathrm{MP}}$, we determine the statistical error $\Sigma_{\lambda^i}$ corresponding to the $90\%$ confidence interval of the marginalized posteriors $\mathcal{P}(\lambda^i|d,H_M)$ (for a given index $i$ with other parameters marginalized over) in each parameter as follows.
For a given parameter $\lambda^i$, we determine the quantile $Q^{\mathrm{MP}}_{\lambda^i}$ of the $\mathcal{P}(\lambda^i | d)$ corresponding to $\BF{\lambda}_{\mathrm{MP}}$ and check where it lies relative to the $5\%$ quantile $Q_{\lambda^i}^{5\%}$ and the $95\%$ quantile $Q_{\lambda^i}^{95\%}$.
We then compute
\begin{align}
  \Sigma_{\lambda^i} \equiv
    \begin{cases}
        Q_{\lambda^i}^{90\%} - Q_{\lambda^i}^{0\%} \quad Q^{\mathrm{MP}}_{\lambda^i} \leq Q_{\lambda^i}^{5\%} \\
        \dfrac{Q_{\lambda^i}^{95\%} - Q_{\lambda^i}^{5\%}}{2}, \quad   Q_{\lambda^i}^{5\%}< Q^{\mathrm{MP}}_{\lambda^i} < Q_{\lambda^i}^{95\%}  \\
        Q_{\lambda^i}^{100\%} - Q_{\lambda^i}^{10\%}, \quad Q^{\mathrm{MP}}_{\lambda^i} \geq Q_{\lambda^i}^{95\%}
    \end{cases}, \label{eqn:stat-error-def}
\end{align}
which defines the statistical error in a fully Bayesian setting, generalizing the definition used for the toy example in Sec.~\ref{subsec:toy}.
\Cref{eqn:stat-error-def} is used to account for cases where the marginalized posterior can be one-sided, similar to what was done in~\cite{Owen:2023mid}.
The statistical errors obtained from~\cref{eqn:stat-error-def} are also comparable to those obtained from the highest posterior density estimate of \texttt{PESummary}~\cite{Hoy:2020vys}.
We then define the normalized (likelihood-based) systematic error as:
\begin{align}
\bar{\Delta}_L \lambda^i \equiv \dfrac{\Delta_L \lambda^i}{\Sigma_{\lambda^i}}, \label{eqn:normalized-sys-error}   
\end{align}
which is used to assess the significance of the systematic biases.
With our injection-recovery setup and Bayesian parameter estimation methods laid out, we move on to present the results for the biases due to spin precession in Sec.~\ref{sec:bias_prec} and higher modes in Sec.~\ref{sec:bias_HM}.

\section{Systematic bias in ppE tests due to neglecting spin precession} \label{sec:bias_prec}

In this section, we present our results on the biases induced by neglecting spin precession and specifically address the following questions:
\begin{itemize}
    \setlength \itemsep{0.5em}
    \item[(A)] How do different waveform parameters of the recovery model become biased with varying $\chi_p^{\mathrm{inj}}$?
    \item[(B)] For a given precessing signal, which ppE test is most biased when precession is neglected, and how do the biases change with $\iota^{\mathrm{inj}}$ and $\chi_p^{\mathrm{inj}}$? 
    \item[(C)] How do the biases in ppE tests change with increasing SNR and total mass independently?
\end{itemize}
Below, we answer each of these questions systematically, using the injection-recovery setup outlined in Sec.~\ref{sec:bayesian_framework}.
We present all the results for the spin precession induced biases based on the analysis done with \texttt{GWAT}. We show a comparison against \texttt{BILBY} in Appendix~\ref{sec:convergence}. 
\subsection{Dependence of biases on injected \texorpdfstring{$\chi_p$}{chip}}
For the edge-on $(12,8)\msun$ injection, we first study how the biases in the recovered parameters depend on $\chi_p^{\mathrm{inj}}$ (focusing on the 1PN ppE test).
In~\cref{fig:corner_extrinsic_intrinsic_edgeOn_precession}  we show 
the marginalized 2d and 1d posteriors for the subspace of parameters $\{ D_L, \cos \iota, \mathcal{M}, \betappe{1} \}$.
The injected values are denoted by black lines, while the blue, orange, and green histograms correspond to $\chi_p^{\mathrm{inj}} = \{ 0,0.45,0.9 \}$ respectively.
When $\chi_p^{\mathrm{inj}} = 0$, there are no biases because the \texttt{IMRPhenomPv2} model reduces to the \texttt{IMRPhenomD} model.
As expected, there are very strong correlations between $D_L$ and $\cos \iota$, both of which enter the amplitude of the waveform.
The strong correlations between $\mathcal{M}$ and $\betappe{1}$ are also expected~\cite{Cornish:2011ys} and can be understood from the Fisher analysis of the toy example in Sec.~\ref{subsec:toy}. 
For $\chi_p^{\mathrm{inj}}=0.45$, we see that there are mild biases in the parameters, as the marginalized posteriors begin to move away from the injected values.

The biases are strongest when $\chi_p^{\mathrm{inj}} = 0.9$, as expected, and there are two set of features in the posteriors that we emphasize below.
First, the marginalized posteriors on $\mathcal{M}$ and $\betappe{1}$ become significantly biased. 
In the marginalized distribution of $\betappe{1}$, the GR value of $\betappe{1}^{\mathrm{inj}}=0$ has negligible posterior support, suggesting the detection of a GR deviation (when in reality there is none in the signal).
The biases are driven by the correlation between $\mathcal{M}$ and $\betappe{1}$, with both $\mathcal{M}$ and $\betappe{1}$ biased to lower values relative to the injection.
Second, the marginalized posterior on $\cos \iota$ becomes notably bimodal.
We see that the bimodality appears with increasing $\chi_p^{\mathrm{inj}}$, suggesting that the neglected precessional modulations in the amplitude of the recovery model result in an ambiguous inference of the inclination.
To further illustrate the bimodality of $\cos \iota$, we also show the parameter recovery using $h^{\mathrm{PhenomD}}$ in~\cref{fig:corner_GR_ppE_compare} of Appendix~\ref{sec:additional-figures}. 

While the bias in $\betappe{1}$ strongly suggests that we have detected a GR deviation, the bimodal inclination recovery instead hints to the presence of inaccuracies in the waveform model, implying a false GR deviation.
Since we know that the signal is a spin precessing injection, this point might appear pedantic, but in practice we do not know what effects the signal contains.
When we do not know what the true signal is, looking for biases in \textit{all} parameters may show a clearer picture of whether the GR deviation is due to waveform inaccuracies or not.

\begin{figure}[t]
\centering
\includegraphics[width=0.4\textwidth]{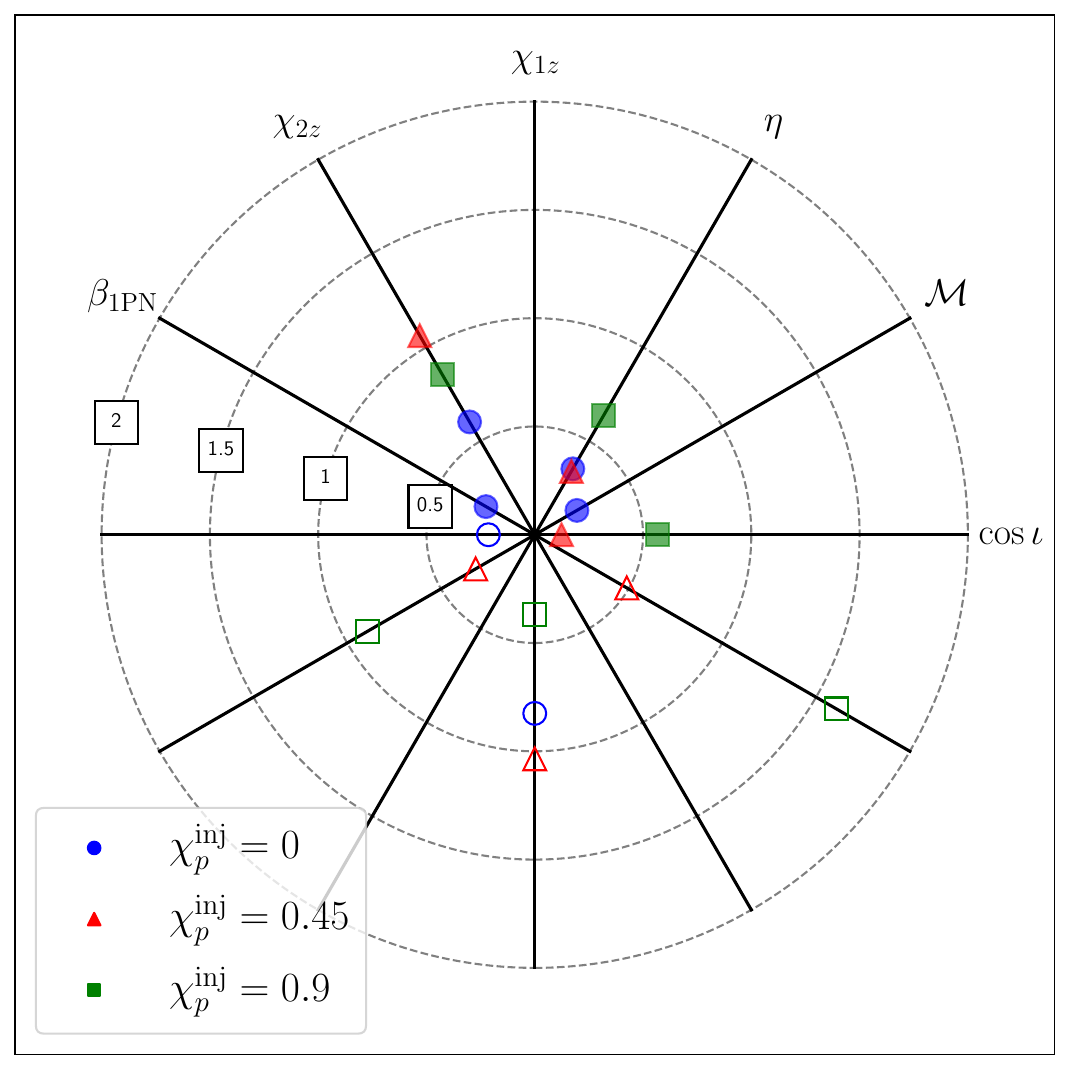}
\caption{Spoke-wheel plot depicting the normalized systematic error $\bar{\Delta}_L \lambda^i$ in parameters $\{\cos \iota, \mathcal{M}, \eta, \chi_{1z}, \chi_{2z}, \betappe{1} \}$.
The biases are incurred from neglecting precession when performing parameter recovery against edge-on injections with $(12,8)\msun$ systems at SNR of 30 that have $\chi_p^{\mathrm{inj}} = \{ 0, 0.45,0.9 \} $ (blue circles, red triangles and green squares, respectively).
The ratio $\bar{\Delta}_L \lambda^i$ is plotted along the spokes for each parameter, with filled markers indicating $\bar{\Delta}_L \lambda^i>0$, and unfilled markers indicating $\bar{\Delta}_L \lambda^i<0$.
The dotted circles at a fixed radius indicate when $|\bar{\Delta}_L \lambda^i|$ crosses a certain threshold, with $|\bar{\Delta}_L \lambda^i|=1$ indicating the threshold when systematic error equals the statistical error in magnitude.
The normalized systematic errors $\bar{\Delta}_L \betappe{1}$ and $\bar{\Delta}_L \mathcal{M}$ clearly grow as $\chi_p^{\mathrm{inj}}$ is increased.
}
\label{fig:wheel_plot_chip}
\end{figure}

\begin{figure*}[ht!]
\centering
\includegraphics[width=0.75\textwidth]{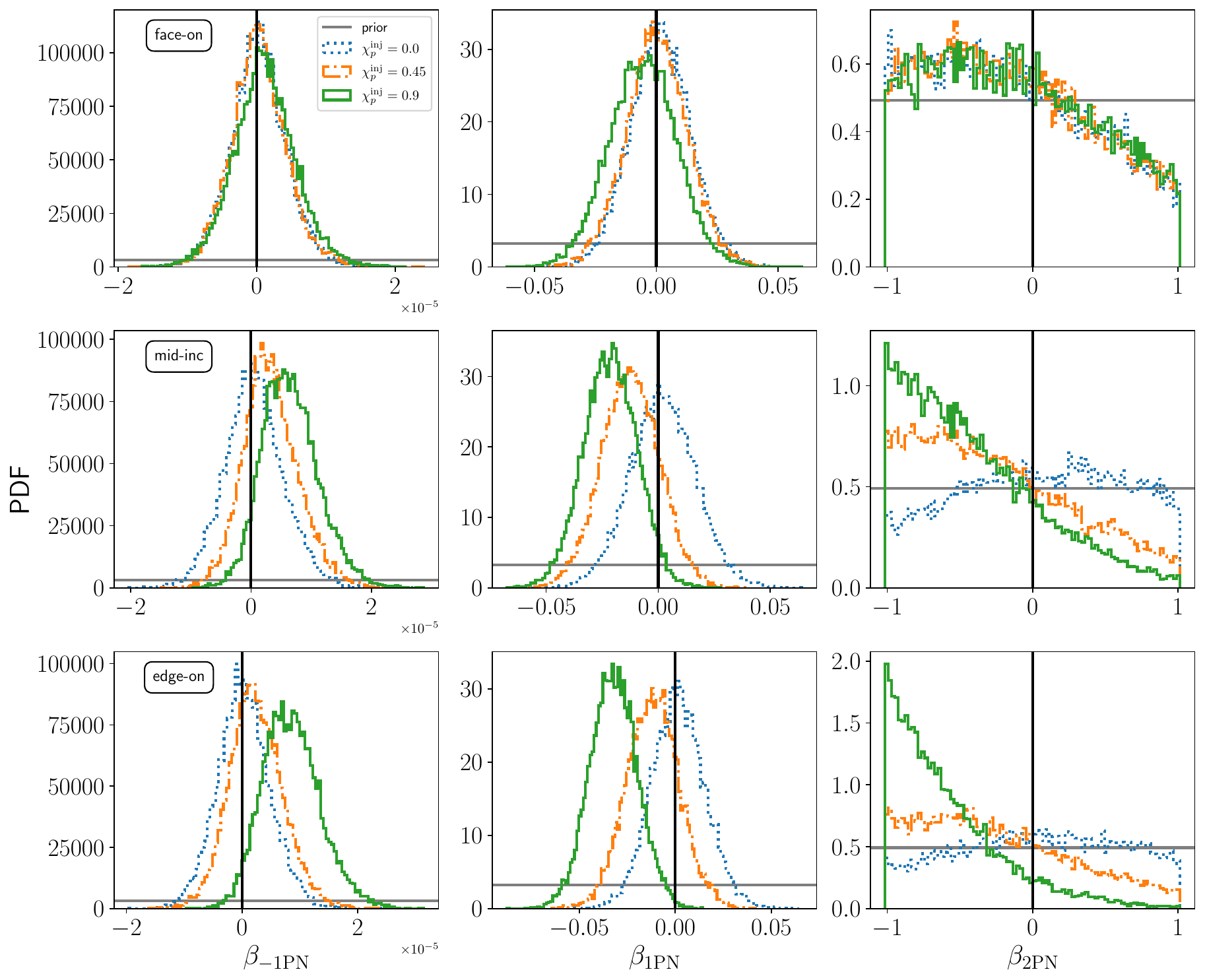}
\caption{Recovered marginalized posteriors on ppE parameters for 
$(12,8)\msun$ injections with varying $\chi_p$ and inclination at SNR of 30.
The rows represent different inclinations $ \iota^{\mathrm{inj}} = \{ 0.05,\pi/4,\pi/2 \}$, while the columns represent different ppE deviations at $-1$PN, 1PN, and 2PN orders.
The blue, orange, and green histograms correspond to $\chi_p=\{0,0.45,0.9\}$ respectively.
The vertical black line represents the injected values and the horizontal gray line represents the priors.
Note that each ppE deviation is recovered independently.
Observe that the biases become stronger with increasing inclination and $\chi_p$ of the injection.
All ppE deviations are significantly biased for edge-on systems with $\chi_p=0.9$.
}
\label{fig:chi_p_iota_dependence}
\end{figure*}

In~\cref{fig:wheel_plot_chip}, we use a ``spoke-wheel'' plot to summarize the normalized systematic error $\bar{\Delta}_L \lambda^i$ for the parameters $\{ \cos \iota, \mathcal{M}, \eta, \chi_{1z}, \chi_{2z}, \betappe{1} \}$. The spoke-wheel plot is based on the radar plot of~\cite{matplotlib_radar_chart} and it is a polar representation of Fig.~5 in~\cite{Owen:2023mid}.
In the spoke-wheel plot, the normalized systematic error $\bar{\Delta}_L \lambda^i$ lies only along the spokes of the wheel, with different angles representing different parameters, and circles of different radii representing different magnitudes of $\bar{\Delta}_L \lambda^i$.
We see a clear trend in the growth of $| \bar{\Delta}_L \lambda^i |$ for $\{ \cos \iota, \mathcal{M}, \betappe{1} \}$ with increasing $\chi_p^{\mathrm{inj}}$, consistent with~\cref{fig:corner_extrinsic_intrinsic_edgeOn_precession}.
While $\eta$ becomes biased more at $\chi_p^{\mathrm{inj}}=0.9$ than at $\chi_p^{\mathrm{inj}}=\{ 0, 0.45\}$, part of the bias comes from the prior. 
The prior on $\eta$ is highly peaked at $\eta = 1/4$ due to the divergent contribution resulting from the Jacobian.
We also see this dependence on the prior when estimating the systematic error using $\Delta_L \eta$ as opposed to $\Delta_P \eta$ (see Appendix~\ref{sec:additional-figures} for more details).
Regarding inferences on the aligned spins, observe that there is no clear trend in $\bar{\Delta}_L \chi_{1z}$ or $\bar{\Delta}_L \chi_{2z}$.
The aligned spin components enter the phase of $h^{\mathrm{PhenomD}}$ as a mass-weighted combination $\chi_{\mathrm{eff}}$ 
,
which is better measured than the individual components~\cite{Miller:2024sui,Callister:2022qwb}. 
Furthermore, the choice of spin priors can affect the inference of the measured spins~\cite{Miller:2024sui,Callister:2022qwb}, which also explains why we cannot extract any clear trends in the biases of $\chi_{1z}$ or $\chi_{2z}$.

\subsection{Dependence of biases on injected inclination and PN order of ppE deviation}

We now study the biases in the recovery of $\betappe{-1}, \betappe{1}$, and $\betappe{2}$ (one deviation at a time) for the $(12,8) \msun$ system injected with $\iota^{\mathrm{inj}} = \{ 0.05, \pi/4, \pi/2 \}$ and $\chi_p^{\mathrm{inj}} = \{0,0.45,0.9 \}$.
The biases are shown in~\cref{fig:chi_p_iota_dependence}, where the columns correspond to a given ppE deviation, the rows correspond to each $\iota^{\mathrm{inj}}$, the blue, orange and green colored histograms correspond to each $\chi_p^{\mathrm{inj}}$, and the priors are indicated with gray lines.
Observe that the biases in all three ppE deviations grow with both $\iota^{\mathrm{inj}}$ and $\chi_p^{\mathrm{inj}}$, and, as expected, they are largest when the injection is edge-on and highly precessing.
Observe also that when $\chi_p^{\mathrm{inj}} = 0.9$ and $\iota^{\mathrm{inj}} = \{ \pi/4, \pi/2\}$, the 2PN deviation peaks at the edge of the prior.
We checked that using a wider prior on the 2PN ppE deviation allows for the posterior to peak within the prior range, but close to its edge.
However, if the signal contained a true GR deviation, using $h^{\mathrm{PhenomD+ppE}}$ would have resulted in a posterior that peaks within the prior range (see~\cref{eqn:relative_Newtonian_ppe_bound}), given that the ppE deviation has to be  negligibly smaller than the leading GR term.
Since precession effects need not be negligibly small like deviations from GR, the posterior of a ppE deviation peaking strongly (railing) at the edge of the prior can indicate a Potential Inference of a GR Deviation.

\begin{figure*}[ht!]
\centering
\includegraphics[width=0.9\textwidth]{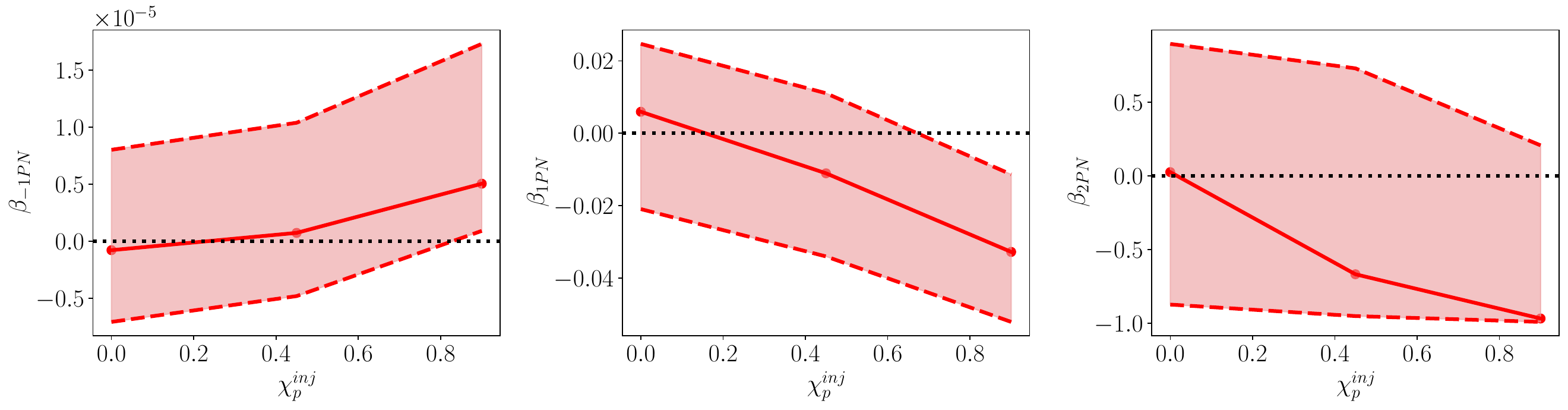}
\caption{Precession induced systematic biases in $\{\beta_{\mathrm{-1PN}}, \beta_{\mathrm{1PN}}, \beta_{\mathrm{2PN}} \}$ tests for edge-on $(12,8)\msun$ injections with varying $\chi_p$ at SNR of 30. 
The solid lines represent the maximum likelihood value of the ppE parameter with the filled region between the dashed lines representing the $90\%$ credible region.
Observe that the bias in ppE parameters increases with increasing $\chi_p$.
}
\label{fig:bias_line_beta_chip}
\end{figure*}

From the panels of~\cref{fig:chi_p_iota_dependence}, we observe that the 1PN deviation is more biased than the $-1$PN and 2PN deviations, which is visibly clear for the edge-on case, when precession effects are most dominant.
This ``preference'' for the 1PN deviation can be made more explicit by determining, for each ppE deviation, 
when the injected value is excluded from the $90\%$ confidence interval.
In~\cref{fig:bias_line_beta_chip}, for the edge-on case, we plot $\Delta_L \betappe{(5+b)}$ with the $90\%$ credible region (determined using~\cref{eqn:stat-error-def}) against $\chi_p^{\mathrm{inj}}$.
When either end of the confidence interval begins to exit the injected value $\betappe{(5+b)}^{\mathrm{inj}} = 0$ (shown by the black dotted line), the biases are strongest.
We note that the precise crossing of the edge of the credible region with the $\betappe{(5+b)}^{\mathrm{inj}} = 0$ line depends on the definition of the credible region, and it can be determined more precisely with more values of $\chi_p^{\mathrm{inj}}$.
Notwithstanding such nuances, it is clear from~\cref{fig:bias_line_beta_chip} that the 1PN test is the most biased when the system is highly precessing. 

\begin{figure*}[ht!]
\centering
\includegraphics[width=0.8\textwidth]{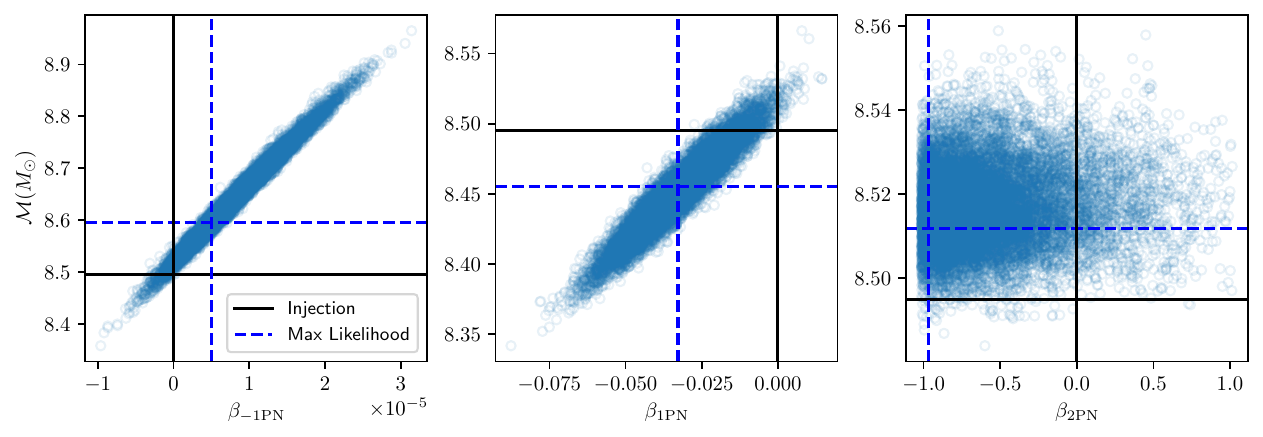}
\caption{Bias in $\mathcal{M}$ and $\betappe{(5+b)}$ (for $b=\{ -7, -3, -1\}$) corresponding to ppE recoveries with $-1$PN, 1PN, and 2PN deviations.
The black solid lines correspond to the injection and the blue dashed lines correspond to the maximum likelihood point.
Observe that the correlation between $\mathcal{M}$ and $\betappe{(5+b)}$ weakens with increasing $b$. 
Furthermore, observe that the systematic and statistical errors in $\mathcal{M}$ decrease with increasing $b$, while the opposite holds true for the ppE deviation.
}
\label{fig:chirpM_beta_bias_diff_ppE}
\end{figure*}

\begin{figure*}[ht!]
\centering
\includegraphics[width=0.9\textwidth]{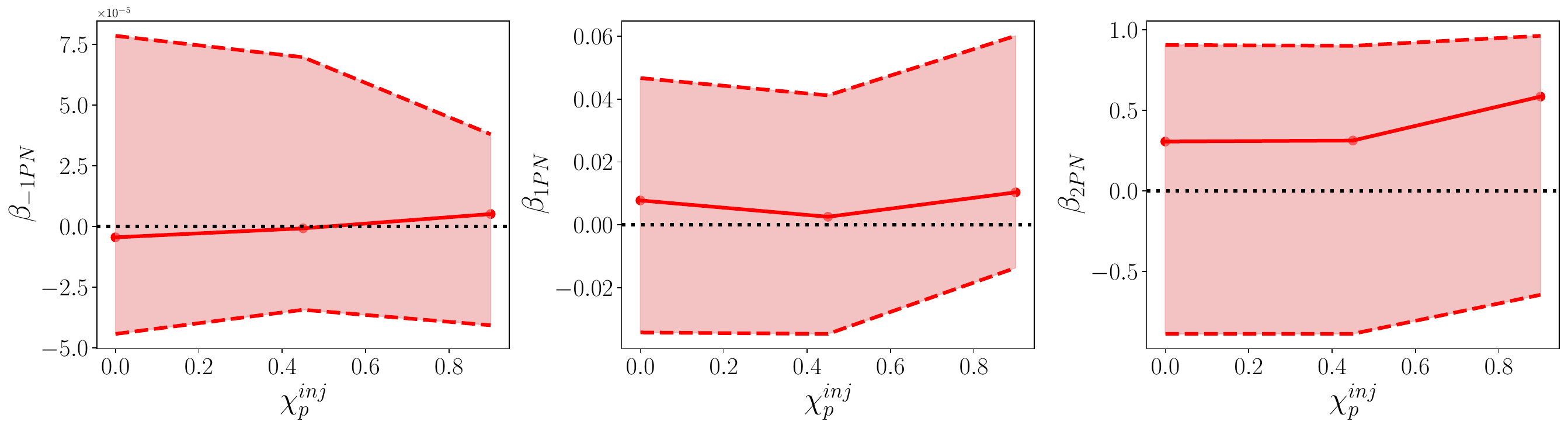}
\caption{Precession-induced systematic biases in $\{\beta_{\mathrm{-1PN}}, \beta_{\mathrm{1PN}}, \beta_{\mathrm{2PN}} \}$ tests for edge-on, $(24,16)\msun$ injections, with varying $\chi_p$ at an SNR of 30.
The solid lines represent the maximum likelihood value of the ppE parameter with the filled region between the dashed lines representing the $90\%$ credible region.
Observe that the biases are much less significant when compared to the $(12,8)\msun$ case, even at $\chi_p^{\mathrm{inj}} = 0.9$.
}
\label{fig:mass_dependence}
\end{figure*}

We have established from both~\cref{fig:corner_extrinsic_intrinsic_edgeOn_precession} and the toy model in Sec.~\ref{subsec:toy} that the biases in the ppE deviations depend strongly on the correlation with the chirp mass.
We illustrate this further in~\cref{fig:chirpM_beta_bias_diff_ppE}, where we show the 2d subspace of $\betappe{(5+b)}$ and $\mathcal{M}$, given the edge-on $(12,8)\msun$ injection with $\chi_p^{\mathrm{inj}}=0.9$.
Observe that the statistical error in $\mathcal{M}$ decreases with increasing PN order of the ppE recovery, while the statistical error in $\betappe{(5+b)}$ increases with increasing PN order (or increasing $b$), just as we saw using the Fisher analysis for the toy model in Sec.~\ref{subsec:toy}. 
Given that a $-1$PN and 1PN ppE deviation contribute more cycles than a 2PN ppE deviation, the posteriors on $\betappe{-1}$ and $\betappe{1}$ are far more constrained than the posterior on $\betappe{2}$, consistent with the green histograms in the bottom row of~\cref{fig:chi_p_iota_dependence}.
The correlations between $\mathcal{M}$ and $\betappe{(5+b)}$ become weaker with increasing $b$, which makes $\mathcal{M}$ more constrained while $\betappe{(5+b)}$ broadens (see Appendix~\ref{sec:additional-figures} for more details).
When recovering with the 2PN ppE deviation, given the correlation with chirp mass and how tightly the chirp mass is constrained, together with the prior range on the 2PN ppE deviation given by~\cref{eqn:relative_Newtonian_ppe_bound}, the systematic error is directed more toward the 2PN ppE deviation rather than the chirp mass. 
Both the railing of the 2PN ppE deviation and the lack of significant bias in the chirp mass are shown clearly in the third panel of~\cref{fig:chirpM_beta_bias_diff_ppE}.
Consequently, we have explicitly shown that the extent of the biases (due to spin precession in this case) in the ppE deviations is driven by the correlation with the chirp mass, which we had predicted using the LSA performed with the toy model in Sec.~\ref{subsec:toy}.

\subsection{Dependence of biases on injected total mass and network SNR}

We now look at how the biases in the ppE deviations depend on the total mass and network SNR of the injection.
In~\cref{fig:mass_dependence}, we show the equivalent of~\cref{fig:bias_line_beta_chip}, with a $(24,16)\msun$ edge-on injection instead of a $(12,8)\msun$ edge-on injection.
Since we found significant biases in the $(12,8)\msun$ case, we can meaningfully compare how the biases change when the total mass is doubled.
From~\cref{fig:mass_dependence}, we observe that there are no significant biases in any of the ppE deviations, even when $\chi_p^{\mathrm{inj}} = 0.9$.
This is because the signal corresponding to the $(24,16) \msun$ source has a shorter duration than the $(12,8) \msun$ source.
Consequently, at the same SNR, the heavier source will have larger statistical errors, which is reflected in the wider $90\%$ confidence interval of $\betappe{(5+b)}$ observed in~\cref{fig:mass_dependence} when compared against~\cref{fig:bias_line_beta_chip}. 
The heavier source contributes fewer precession cycles as well, resulting in a smaller systematic error.
We see both of these features in~\cref{fig:mass_dependence} and, therefore, at a given SNR, the significance of the bias is reduced when the source is heavier.
\begin{figure}[h!]
\centering
\includegraphics[width=0.45\textwidth]{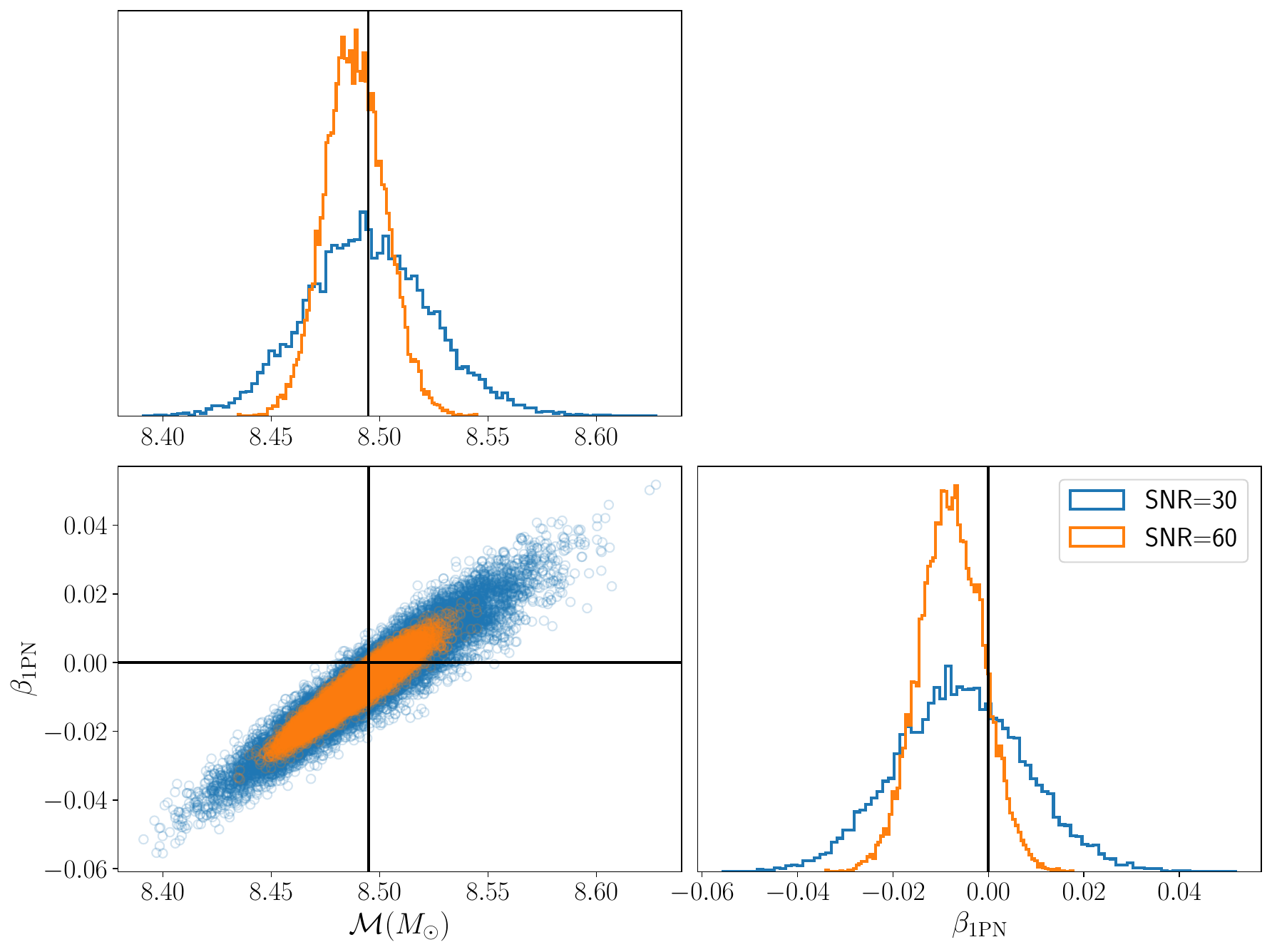}
\caption{Comparison of precession-induced systematic biases at SNR of 30 and 60. 
Shown are the marginalized 2d posteriors of $\{ \mathcal{M}, \betappe{1}\}$ recovered for an edge-on $(12,8)\msun$ injections with $\chi_p^{\mathrm{inj}}=0.9$.
The blue and orange histograms correspond to SNR of 30 and 60 respectively, while the black line corresponds to the injection.
Observe that the louder signal results in a more significant bias. 
}
\label{fig:corner_plot_snr}
\end{figure}

When the SNR of the source is doubled (by rescaling the distance) to $\rho = 60$, the systematic biases are more significant.
As we explained in Sec.~\ref{subsec:definitions} and Sec.~\ref{subsec:toy}, this is because the systematic error is SNR-scale invariant, while the statistical error is inversely proportional to the SNR.
In~\cref{fig:corner_plot_snr}, we show the dependence on SNR of the marginalized posteriors on $\mathcal{M}$ and $\betappe{1}$ for the $(12,8)\msun$, face-on injection with $\chi_p^{\mathrm{inj}}=0.9$.
We saw from~\cref{fig:chi_p_iota_dependence} that for the $\rho = 30$ case, the bias in $\betappe{1}$ was not significant, as the precession effects are subdued when the binary is face-on.
However, when $\rho = 60$, the bias is clearly more significant, with smaller support at the injected value.
As expected, the distributions simply become tighter, but are centered around the same point, reflecting that the systematic error does not change when the SNR is scaled by a constant.
Further, we see that even when the binary is face-on, a highly precessing source with an SNR greater than 60 can become significantly biased.

\subsection{What drives the biases?}
\label{subsec:stat-sig-drive-prec}

To better understand what drives the biases, we look at the contributions from the phase modulations (PM) and amplitude modulations (AM) of the injection.
Our analysis is motivated by the early fitting factor calculations done in~\cite{Apostolatos:1995} for spin-precession induced systematic errors.
We perform our analysis in the following way. 
We take the injected polarizations $h^{+,\times}_S (\bftr{\theta} ; f) = A^{+,\times}_S (\bftr{\theta} ; f )\exp [i \Psi^{+,\times}_S(\bftr{\theta} ;  f)]$ and construct their amplitudes $A^{+,\times}_S(\bftr{\theta} ; f )$ and phases $\Psi^{+,\times}_S ( \bftr{\theta} ; f )$ from
\begin{subequations}
\begin{align}
    A^{+,\times}_S(\bftr{\theta} ; f ) &= \left| h^{+,\times}_S (\bftr{\theta} ; f )\right|,\\
    \Psi^{+,\times}_S(\bftr{\theta} ; f ) &= \mathrm{arctan} \left[ \dfrac{\mathrm{Im} [h^{+,\times}_S(\bftr{\theta} ; f )]}{\mathrm{Re} [h^{+,\times}_S(\bftr{\theta} ; f )]} \right].
\end{align}    
\end{subequations}
Given the approximate model at the injected parameters, we take the polarizations $h^{+,\times}_M (\bftr{\lambda} ; f)$ and construct the amplitude $A^{+,\times}_M (\bftr{\lambda} ; f)$ and phase $\Psi^{+,\times}_M (\bftr{\lambda} ; f)$ for each polarization. 
We then construct a ``PM only'' injection using
\begin{align}
  h_{S,PM}^{+,\times}(\bftr{\theta} ; f) = A^{+,\times}_M (\bftr{\lambda} ; f)\exp[ i \Psi^{+,\times}_S(\bftr{\theta} ; f)],
\end{align}
and the corresponding ``AM only'' injection using
\begin{align}
  h_{S,AM}^{+,\times}(\bftr{\theta} ; f) = A^{+,\times}_S (\bftr{\theta} ; f)\exp[ i \Psi^{+,\times}_M(\bftr{\lambda} ; f)].
\end{align}

\begin{figure}[t]
\centering
\includegraphics[width=0.4\textwidth]{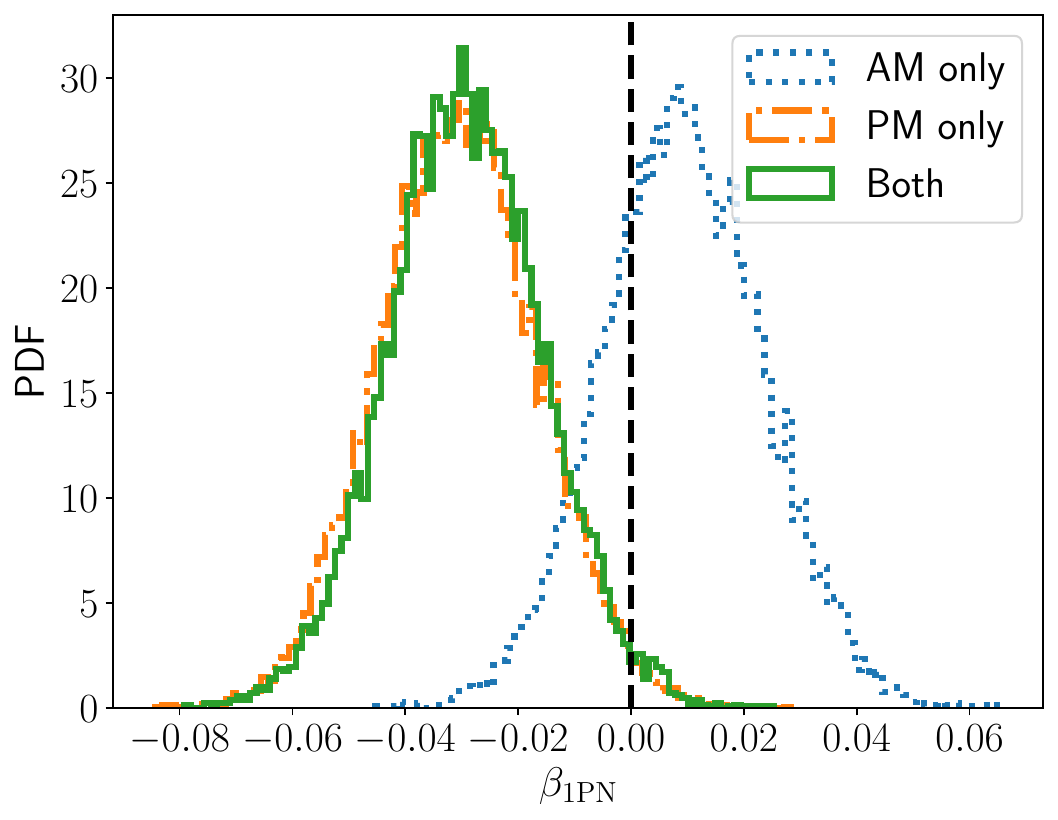}
\caption{PM and AM contribution to the spin precession induced bias for the $(12,8) \msun$ edge-on injection at an SNR of 30.
Shown are the marginalized 1d posteriors on $\betappe{1}$ when injecting with only PM (orange histogram), only AM (blue histogram), and both PM and AM (green histogram).
Observe that the injection with only PM results in a bias that is statistically consistent with including both AM and PM.
Furthermore, note that the injection with only AM does not result in significant bias, showing that PM drives the bias in $\betappe{1}$.}
\label{fig:beta_precession_PM_AM}
\end{figure}

With this in hand, we then inject either $h_{S,PM}^{+,\times}$ or $h_{S,AM}^{+,\times}$ and recover them using $h_M$ to analyze the biases induced by the phase or amplitude modulations.
Such an analysis is particularly helpful in understanding which parameters are biased due to PM and which due to AM.
In~\cref{fig:beta_precession_PM_AM} we show the marginalized posterior for $\betappe{1}$ when injecting with a signal from a $(12,8) \msun$ system with $\iota^{\mathrm{inj}}=\pi/2$ and $\chi_p^{\mathrm{inj}}=\pi/2$ at an SNR of 30.

As discussed in Sec.~\ref{subsec:stat-sig-bayesian-prec}, the incurred bias in $\betappe{1}$ is significant. 
From~\cref{fig:beta_precession_PM_AM}, we find that turning on only the AM results in insignificant bias for $\betappe{1}$.
On the other hand, turning on only the PM results in a posterior that is statistically consistent with having both AM and PM.
Our result is not surprising, as it is well known in GW data analysis that phase corrections are more significant than amplitude corrections for parameter estimation.

Having shown that the PM is more relevant than AM for the ppE bias, why is it that the 1PN deviation is significantly more biased than $-1$PN and 2PN when it comes to spin precession?
As first pointed out in Refs.~\cite{Apostolatos:1994,Apostolatos:1995}, the secular part of the PM due to spin precession corrects the nonprecessing phase starting at 1PN order.
This is clearly seen when looking at the Euler angles $\phi_z (f)$ and $\zeta(f)$ that modulate the nonprecessing waveform in the \texttt{IMRPhenomPv2} model (reviewed in Appendix~\ref{sec:waveform_review}). 
The angle $\phi_z(f)$ is commonly referred to as the precession angle, and to leading PN order, it scales as $f^{-1}$, which is 1PN relative to the leading term in the nonprecessing phase.
As shown in Ref.~\cite{Apostolatos:1995}, the effects of precession can be absorbed by adding a phenomenological term at 1PN order to a nonprecessing waveform.
If the phenomenological term is reinterpreted as a ppE deviation and the $\kappa$-effect refers to precession, the argument in Ref.~\cite{Apostolatos:1995} is essentially what we showed with our toy model in Sec.~\ref{subsec:toy}.

As $\phi_z(f)$ and $\zeta(f)$ have subleading PN corrections, a single ppE deviation that enters at 1PN order cannot completely absorb the effects of precession. 
However, as we showed in Sec.~\ref{subsec:toy}, the leading $\kappa$-effect will be absorbed by the ppE parameter when the neglected $\kappa$-effect enters at the same PN order as the ppE deviation.
Even though there are no spin precession contributions that enter at $-1$PN order, the $-1$PN ppE parameter is significantly biased due to its strong correlation with chirp mass and its small statistical error (see~\cref{fig:chirpM_beta_bias_diff_ppE}).
The $-1$PN ppE parameter is less significantly biased than the 1PN ppE parameter due to the absolute systematic error being much smaller for the former.
On the other hand, the 2PN ppE parameter incurs the larger share (with chirp mass) of the systematic error (when compared to the $-1$PN and 1PN ppE parameters), leading to the marginalized distribution of the 2PN ppE parameter peaking at the edge of the prior.
The systematic bias in the 2PN ppE parameter is not as significant as that of the 1PN ppE parameter due to the former's larger statistical error. 
The fact that the precession effects enter at 1PN order explains why the biases due to spin precession are strongest for the 1PN ppE parameter compared to the $-1$PN and 2PN ppE parameters.
Furthermore, it also explains why the biases are more significant for a lower total mass system: more precession cycles are accumulated at lower frequencies.

\subsection{Statistical significance of biases}
\label{subsec:stat-sig-bayesian-prec}

In~\cref{tab:stat-sig-prec-ppE} we show the statistical significance of spin-precession induced biases in both $\betappe{(5+b)}$ and $\mathcal{M}$ for the case of the $(12,8)\msun$ injection with $\iota^{\mathrm{inj}}= \pi/2$, $\chi_p^{\mathrm{inj}} = 0.9$, and $\rho = 30$.
This injection resulted in significant biases in all the ppE deviations, as shown in~\cref{fig:bias_line_beta_chip,fig:chi_p_iota_dependence}.
\Cref{tab:stat-sig-prec-ppE} is essentially the equivalent of~\cref{tab:toy} for the toy model of Sec.~\ref{subsec:toy}.
We do not list $\rho_{\mathrm{crit}}$ as it cannot be \emph{a priori} computed when we do a Bayesian MCMC analysis.
However, based on~\cref{tab:stat-sig-prec-ppE}, we can estimate what $\rho_{\mathrm{crit}}$ ought to be.
\begin{table}[ht!]
    \centering
     \setlength{\tabcolsep}{0.4em} %
\renewcommand{\arraystretch}{1.5}
\resizebox{\columnwidth}{!}{%
    \begin{tabular}{c||c|c|c}
    \hline
       $b$  &  $-7$ & $-3$ & $-1$ \\
    \hline 
       $\Delta_L \betappe{(5+b)}$  & $5.03 \times 10^{-6}
$ & $-3.28 \times 10^{-2}$ & $-0.966 $ \\
    \hline
    $\Sigma_{\beta}$ & $8.19 \times 10^{-6}$ & $2.03 \times 10^{-2}$ & $0.970$ \\ 
    \hline
    $\Delta_L^{\mathrm{ppE}} \log \mathcal{M}$ & $1.18\times 10^{-2}$ & $4.63 \times 10^{-3}$ & $1.99 \times 10^{-3}$ \\
    \hline
    $\Sigma^{\mathrm{ppE}}_{\mathcal{\log M}}$ & $1.44 \times 10^{-2}$ & $5.19 \times 10^{-3}$ & $1.79 \times 10^{-3}$ \\
    \hline
    $1-FF_{\mathrm{ppE}}$ & $3.34 \times 10^{-2}$ & $3.12 \times 10^{-2}$ & $3.31 \times 10^{-2}$ \\
    \hline
    $BF_{\mathrm{ppE,GR}}$ & $0.441$ & $5.41$ & $2.62$ \\
    \hline
    \end{tabular}
    }
    \caption{
    Statistical significance of spin precession induced biases in $\betappe{(5+b)}$ and $\mathcal{M}$ for the $(12,8) \msun$ edge-on injection with $\chi_p^{\mathrm{inj}} = 0.9$ at an SNR of 30.
    For comparison, $\Delta^{\mathrm{GR}}_L \log \mathcal{M} = 3.65 \times 10^{-3}$, $\Sigma^{\mathrm{GR}}_{\log \mathcal{M}} = 1.96 \times 10^{-3}$, and $1-FF_{\mathrm{GR}}=3.50 \times 10^{-2}$.
    }
\label{tab:stat-sig-prec-ppE}
\end{table}
Even for the spin precession injections analyzed with MCMC/nested sampling methods, several predictions of the toy model hold true.
First, due to the correlations between $\mathcal{M}$ and $\betappe{(5+b)}$ shown in~\cref{fig:chirpM_beta_bias_diff_ppE}, we see from~\cref{tab:stat-sig-prec-ppE} that the systematic and statistical errors in $\betappe{5+b}$ grows with increasing $b$, while the opposite happens for $\mathcal{M}$.
The magnitude of the systematic error in $\betappe{(5+b)}$ is comparable to, or larger than, the corresponding statistical errors.
This suggests that $\rho = 30$ is already close to the critical SNR $\rho_{\mathrm{crit}}$.

\begin{center}
\begin{table}[ht!]
\setlength{\tabcolsep}{0.7em} %
\renewcommand{\arraystretch}{1.7}
\begin{tabular}{ c|c|c|c} 
 \hline \hline
 \multicolumn{2}{c|}{Injection} & \multicolumn{2}{c}{1PN recovery}  \\
 \hline \hline 
 $\iota^{\mathrm{inj}}$ & $\chi_p^{\mathrm{inj}}$  & $\bar{\Delta}_L \beta_{1\mathrm{PN}}$ & $\log BF_{\mathrm{ppE,GR}}$ \\ 
 \hline
 \multirow{3}{4em}{\centering $0.05$} & $0$  & $-0.078$ & $-2.30$  \\ 
  & $0.45$ & $-0.24$  & $-2.28$ \\
 & $0.9$ & $-0.41$ & $-1.87$  \\
 \hline
 \multirow{3}{4em}{\centering $\pi/4$} & $0$  & $-0.016$ & $-2.79$ \\ 
  & $0.45$  & $-0.57$ & $-1.35$  \\
 & $0.9$ & $-1.43$  & $-1.17$ \\
 \hline
 \multirow{3}{4em}{\centering $\pi/2$} & $0$ & $0.26$  & $-2.43$ \\ 
  & $0.45$  & $-0.49$ & $-1.41$ \\
 & $0.9$  & $-1.61$ & $1.69$\\
 \hline
\end{tabular}
\caption{Statistical significance of spin-precession induced biases in the 1PN test for $(12,8)\msun$ injection with varying $\iota^{\mathrm{inj}}$ and $\chi_p^{\mathrm{inj}}$ at SNR of 30. 
Observe that the Bayes factor increases with both $\iota^{\mathrm{inj}}$ and $\chi_p^{\mathrm{inj}}$, with the largest value at $\iota^{\mathrm{inj}} = \pi/2$,$\chi_p^{\mathrm{inj}} = 0.9$.
The increase in Bayes factor is correlated with the increase in the ratio of systematic to statistical error.
This table is a companion to~\cref{tab:stat-sig-prec-ppE}.
}
\label{tab:stat-sig-prec-inc}
\end{table}
\end{center}

We compute the Bayes factors in favor of the ppE model by explicitly evaluating the model evidence for both the ppE and GR models. 
The results shown in~\cref{tab:stat-sig-prec-ppE} are from the analyses done using \texttt{GWAT}. 
The evidence is computed in \texttt{GWAT} using thermodynamic integration~\cite{gregory_2005,Littenberg:2009bm,Littenberg:2010gf,Vousden:2016}.
The calculation of the evidence is robust as we checked for convergence by increasing the number of temperature rungs, \texttt{ensembleSize}, and number of chains per temperature, \texttt{ensembleN}.
The Bayes factors in all three cases of $b= \{ -7, -3, -1 \}$ do not strongly favor the ppE model, with marginal preference shown in the 1PN case.
These values of the Bayes factor are qualitatively reflected in~\cref{fig:chi_p_iota_dependence} when one estimates them using the Savage-Dickey ratio.
Since there can be binning and sampling errors~\cite{Chatziioannou:2014bma} in the low likelihood regions, where the injection is, we do not use the Savage-Dickey ratio in computing the Bayes factor. 
We show how the Bayes factors change with $\iota^{\mathrm{inj}}$ and $\chi_p^{\mathrm{inj}}$ for the 1PN recovery in~\cref{tab:stat-sig-prec-inc}.
We find that the Bayes factor increases with both $\iota^{\mathrm{inj}}$ and $\chi_p^{\mathrm{inj}}$, and is correlated with the increase in the ratio of systematic to statistical error. 

To determine the regime of bias that these analyses land on, per the classification of~\cref{fig:stealth-overt}, we look at the fitting factor of the ppE model and the Bayes factor between the ppE and GR models.
Notably from~\cref{tab:stat-sig-prec-ppE}, $1-FF_{\mathrm{ppE}}$ is an order of magnitude larger than the fiducial FF threshold, given by $N_{\mathrm{ppE}}/(2\rho^2) = 6.67 \times 10^{-3}$, while $BF_{\rm ppE,GR}$ is below the fiducial BF threshold, $BF_{\rm thresh} = 10$. This implies that all analyses correspond to examples of biases in the Strong Inference of No GR Deviation subregime of Fig.~\ref{fig:stealth-overt}, since both the residual SNR and the model selection tests would not be passed. In particular, this implies that for the cases considered, it would be straightforward to identify the presence of systematic bias and avoid a false positive GR deviation inference.  

Having said this, several features of~\cref{tab:stat-sig-prec-ppE} should be discussed. First, the fitting factor is marginally larger for the 1PN than $-1$PN and 2PN cases.
Compared to the fitting factor of the GR model, each ppE model gives a slightly higher fitting factor.
Geometrically, we showed this behavior using the Laplace approximation in Sec.~\ref{subsec:distinguishability}, and we find that the Bayesian result is consistent: the inclusion of the ppE parameter reduces $\norm{\Delta h_{\perp}}$ compared to the GR value.
For each value of $b$, there is significant loss of SNR. 
Based on the analytical arguments in Sec.~\ref{subsec:claiming-false-GR-deviation} and Sec.~\ref{subsec:toy}, we expect that at higher SNR, the systematic bias will be classified as a Weak Inference of No GR Deviation II.
Increasing the signal SNR to 60, for the 1PN ppE recovery, we find that the Bayes factor increases to $BF_{\mathrm{ppE,GR}} \sim 80,822$, which greatly exceeds the threshold BF, resulting in a strong preference for the ppE model over the GR model. 
One can then clearly confirm that the systematic bias moves to the Weak Inference of No GR Deviation II subregime.
Owing to the appreciable loss of SNR, in none of the cases of neglecting spin precession, we find that the resulting systematic bias is a realization of an Incorrect Inference of GR Deviation. 
Therefore, even for a loud signal, we will not find a false positive inference of a GR deviation, as the ppE parameter cannot completely absorb the effects of spin precession.

\begin{figure*}[t]
\centering
\includegraphics[width=0.9\textwidth]{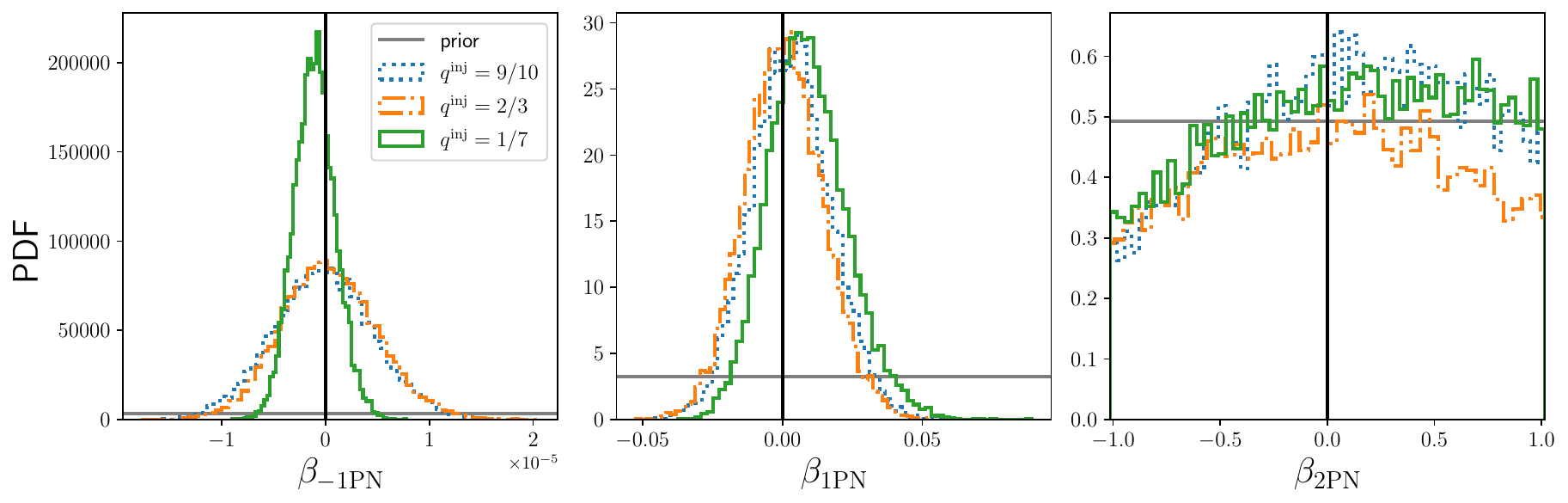}
\caption{Systematic biases in $-1$PN, 1PN, and 2PN ppE tests for edge-on injections with varying $q$ (with fixed total mass of $20 \msun$ at SNR of 30.
With decreasing $q$, the bias increases, which is most clearly observed in the $-1$PN deviation.
}
\label{fig:mass_ratio_dependence}
\end{figure*}

\subsection{Executive summary of spin-precession induced biases in ppE tests}
Below, we summarize the main results of the systematic biases caused by neglecting spin precession.
\begin{enumerate}[leftmargin=*]
    \setlength \itemsep{0.5em}
    \item 
Systematic biases in different ppE parameters:
    \begin{enumerate}[leftmargin=*]
        \setlength \itemsep{0.5em}
    	\item The 1PN ppE parameter is most significantly biased (compared to $-1$PN and 2PN ppE parameters) due to leading PN order precession effects modulating the phase at 1PN order (relative to the nonprecessing phase).
    	This is clearly observed when the signal is from an edge-on source.
    	\item When precession in the signal is significant (such as for $\chi_p = 0.9$), the 2PN ppE parameter gets significantly biased and rails at the edge of the prior.
        This is due to the weak correlation with the chirp mass, the tight constraint on measuring chirp mass, and the restricted prior range of the ppE deviation (requiring the ppE deviation to be suppressed compared to the leading GR term).
    \item The $-1$PN ppE parameter is biased even though the precession dynamics does not enter at $-1$PN order. 
    This is partly due to the fact that the $-1$PN ppE parameter is very well measured (small statistical error), owing to the many cycles contributed by this term.   
	\item The absolute systematic error in the ppE parameters increases with increasing PN order of the ppE term. 
    \end{enumerate}
	\item Dependence of systematic biases on source parameters and SNR:
	\begin{enumerate}[leftmargin=*]
     \setlength \itemsep{0.5em}
	\item Systematic biases in ppE parameters increase with inclination angle, orbital-plane spins, and SNR of the precessing signal. 
	\item Meanwhile, the systematic biases decrease with total mass (fewer GW cycles) for a fixed SNR.
	\end{enumerate}
	\item Statistical significance of systematic biases in ppE parameters:
	\begin{enumerate}[leftmargin=*]
     \setlength \itemsep{0.5em}
	\item For edge-on signals with $\chi_p = 0.9$ at an SNR of 30, the ppE model is not strongly favored over the GR model.
	The resulting systematic bias in each recovery is a Strong Inference of No GR Deviation.
	\item For edge-on signals with $\chi_p = 0.9$ at an SNR of 60, the ppE model is strongly favored over the GR model, resulting in a Weak Inference of No GR Deviation II.
	\item There is significant loss in SNR when recovering with the nonprecessing ppE model.
    Consequently, the systematic biases are not in the regime of Weak Inference of No GR Deviation I or Incorrect Inference of GR Deviation.
	\end{enumerate}
\end{enumerate}
This analysis shows how the use of FFs and BFs can help us classify what kind of systematic bias one may be encountering, and whether the inferences one draws about the validity of GR are robust. 

\section{Systematic bias in ppE tests due to neglecting higher modes} \label{sec:bias_HM}

In analogy with Sec.~\ref{sec:bias_prec}, in this section we present our results on the biases induced by neglecting higher modes. We specifically address the following questions using the injection-recovery setup outlined in Sec.~\ref{sec:bayesian_framework}:
\begin{itemize}
    \setlength \itemsep{0.5em}
    \item[(A)]  For a given signal with higher modes, which ppE test is most biased when higher modes are neglected, and how do the biases change with $\iota^{\mathrm{inj}}$ and $q^{\mathrm{inj}}$? 
    \item[(B)] How do the biases in ppE tests change with increasing SNR and total mass?
\end{itemize}
All the results for the higher-modes induced biases refer to the analysis done with \texttt{BILBY}. 

\subsection{Dependence of ppE biases on injected mass ratio, inclination angle, total mass and network SNR}

We first study the dependence on $q$ as the subdominant modes contribute more when $q \rightarrow 0$ (see~\cite{Blanchet:2013haa} for a review).
In~\cref{fig:mass_ratio_dependence}, we show the recovered $-1$PN, 1PN, and 2PN deviations for $q^{\rm inj} = \{9/10, 2/3,1/7 \}$ (with $m_{\rm tot, inj}=20 \msun$) at SNR of 30.
Observe that in all cases, the bias is truly minimal. 
As we saw in Sec.~\ref{sec:bias_prec}, the recovery of ppE deviations worsens with increasing PN order.
Notably, the $-1$PN ppE deviation is more sharply peaked for $q^{\rm inj} = 1/7$ than the other injections.
Because smaller mass ratio injections have longer signal duration, we can extract more information about the $-1$PN ppE deviation (which contributes at low frequencies) from the data.
In both the $-1$PN and 1PN ppE recoveries, the marginalized distributions move slightly away from the injection with decreasing mass ratio.
For the 2PN ppE recovery, not enough information can be extracted from the signals, with the marginalized posteriors being consistent with the priors.
We also checked that the correlations with chirp mass are consistent with what we found in Sec.~\ref{sec:bias_prec}.

\begin{figure}[t]
\centering
\includegraphics[width=0.45\textwidth]{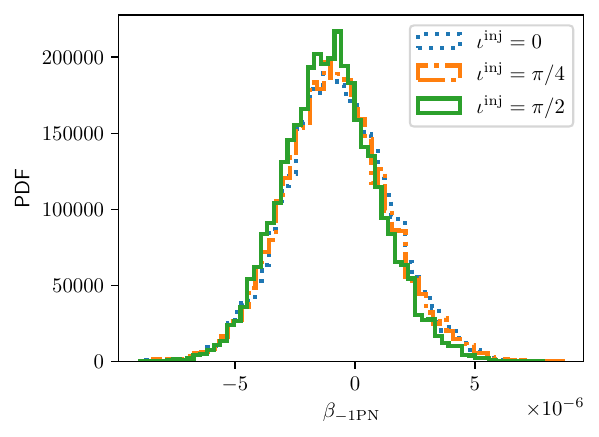}
\caption{Systematic biases in $-1$PN ppE deviation for $m_{\rm tot}=20,q=1/7$ injections with varying inclination at SNR of 30.
With decreasing $q$, the bias increases, which is most clearly observed in the $-1$PN deviation.
}
\label{fig:inc_dependence_HM}
\end{figure}

In~\cref{fig:inc_dependence_HM}, we vary the injected inclination and show the marginalized posteriors for the $-1$PN deviation for the $m_{\rm tot, inj}=20\msun,q_{\rm inj}=1/7$ system.
We find that the marginalized posteriors are consistent with one another for all inclinations.
In other words, the systematic bias is not very sensitive to the inclination. This is different from what we found in Sec.~\ref{sec:bias_prec} for spin-precession induced biases.
We confirmed that this trend holds also for the other ppE deviations and for the other mass ratio injections.

\begin{figure}[t]
\centering
\includegraphics[width=0.45\textwidth]{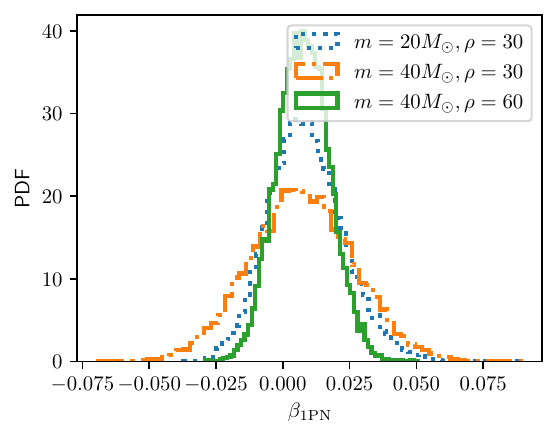}
\caption{Systematic biases in $1$PN ppE deviation for $m_{\rm tot}=\{20,40\}\msun$ injections at SNR of 30 and 60.
At a given SNR, increasing the total mass broadens the posterior; at a given total mass, increasing the SNR makes the posterior narrower.
}
\label{fig:mass_dependence_HM}
\end{figure}

In~\cref{fig:mass_dependence_HM}, we show the dependence of the ppE recovery on the injected total mass and SNR, for a fixed mass ratio $q^{\rm inj} = 1/7$.
For the $m_{\rm tot,inj}=20 \msun$ system, when the total mass is doubled, the resulting marginalized posterior for $\betappe{1}$ (and similarly for the other ppE deviations) widens, owing to the fewer cycles the signal spends in band.
However, the peaks of the distributions coincide with one another.
When the SNR of the injection of the $m_{\rm tot,inj}=40\msun$ system is doubled to $\rho_{\rm inj}=60$, the marginalized posterior for $\betappe{1}$ narrows, as the statistical error is inversely proportional to the SNR, as expected. Once more, in all cases considered, the systematic bias is minimal relative to what we found in the spin-precessing case.  

\subsection{What drives the biases?}

\begin{figure}[ht!]
\centering
\includegraphics[width=0.45\textwidth]{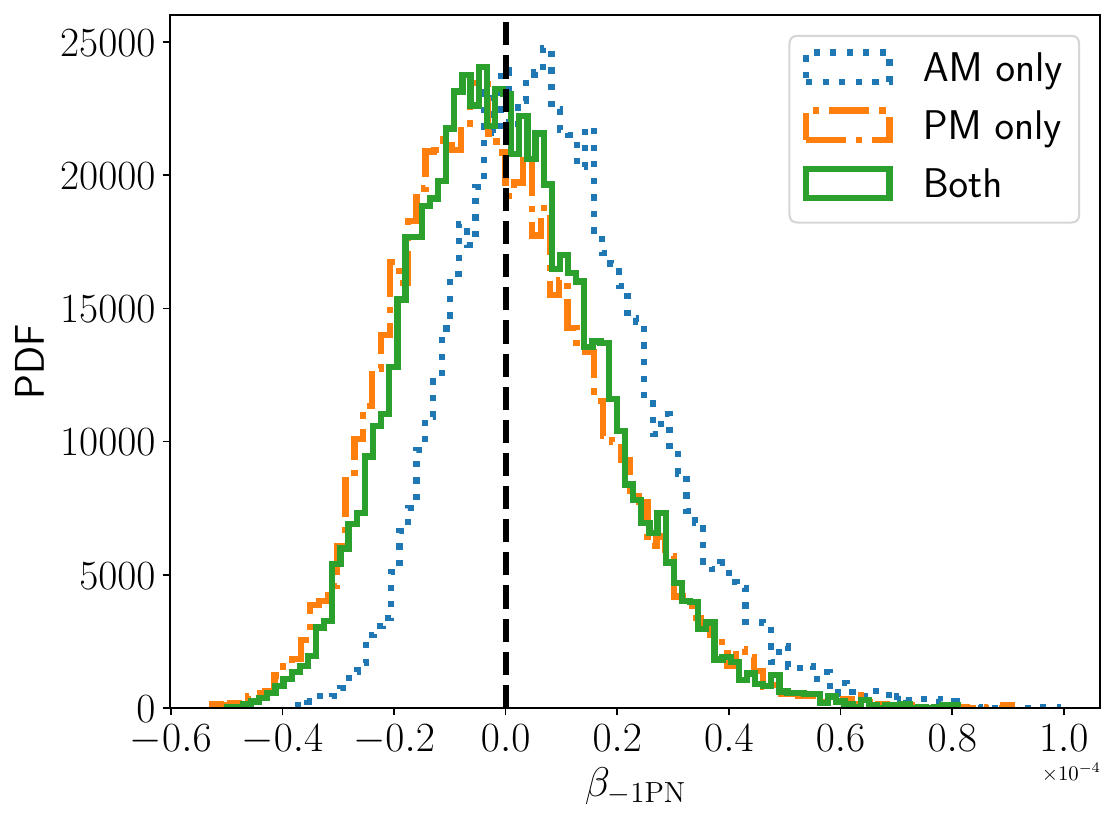}
\caption{Phase and amplitude modulation contributions to the recovery of the -1PN ppE deviation, given an injected higher modes signal with $m_{\rm tot}=40 \msun$ and $q=1/7$ at edge-on inclination.}
\label{fig:beta_higher-modes_PM_AM}
\end{figure}

As in Sec.~\ref{sec:bias_prec}, we determine the contribution (to the ppE recovery) of the phase and amplitude modulations present in the \texttt{IMRPhenomHM} signal.
Similar to what we found in Sec.~\ref{sec:bias_prec}, the phase modulations contribute more than the amplitude modulations.
We demonstrate this in~\cref{fig:beta_higher-modes_PM_AM} for the -1PN ppE recovery when the injection is from an edge-on system with $m_{\rm tot} = 40 \msun$ and $q=1/7$.
From~\cref{fig:beta_higher-modes_PM_AM}, we see that having only phase modulations in the higher modes injection produces a posterior that is nearly identical to the case where we have both phase and amplitude modulations, similar to what we found for spin-precessing injections.

To further explain the trends observed in the biases induced by the higher-mode injections, we turn to the structure of the subdominant mode contribution (reviewed briefly in Appendix~\ref{sec:waveform_review}) and contrast it with the structure of precessing signals.
For a waveform containing higher modes, in addition to the dominant mode, there are subdominant modes with $\ell > 2$ and $|m| \neq 2$. 
These modes have the same phase as the $(2,2)$ mode, but come with a different harmonic of the orbital frequency.
In other words, $\Phi_{\ell m } = m \Phi_{\rm orb}$, where $\Phi_{\ell m}$ is the phase of the $(\ell, m)$ mode and $\Phi_{\rm orb}$ is the orbital phase.
The dominant mode itself is not modified in the \texttt{IMRPhenomHM} waveform, and thus carries most of the information regarding the source.
Meanwhile, for a (phenomenological) precessing waveforms, the nonprecessing dominant mode is modulated in both amplitude and phase.
The dominant precessing mode is distinct from the dominant nonprecessing mode, and carries additional information about the spins.
The measurement of chirp mass for a precessing signal is thus affected by whether precession is included or not, while it is largely unaffected when higher modes are neglected.
Due to the correlations between chirp mass and the ppE deviations, the inference of a GR deviation is not as sensitive to the inclusion of higher modes, as it is for precession (at a given SNR).

The structure of the higher-mode waveform can also be used to understand the dependence of the biases on the inclination angle. 
Although the amplitude of the subdominant modes increases with the inclination, with the effects being strongest for edge-on systems, the changing inclination is primarily an amplitude modulation, and as such it does not contribute appreciably to the phase.
Since the ppE modifications are only present in the phase, the systematic bias is not very sensitive to the inclination, in contrast to what we found for spin-precession induced biases. 

In all cases, we found that the systematic bias induced by neglecting higher modes in the recovery ppE model is much smaller than that induced by spin precession. This implies that, for the systems and SNRs considered, biases induced in the higher-mode case lie in the Definite Inference of No GR Deviation subregime of~\cref{fig:stealth-overt}. This begs the question whether there are any SNRs or system properties for which biases due to the neglect of higher-modes could become a problem for inspiral tests of GR. Extrapolating the results of Fig.~\ref{fig:mass_dependence_HM} for the 1PN ppE test, we estimate that an SNR of $\approx 75$ (for an injection with $m_{\rm tot, inj}=40 M_\odot$ and $q_{\rm inj} = 1/7$) would be needed for GR to be excluded at 90\% confidence. 
Since an increase in SNR forces us to flow rightward and upward in the diagram of~\cref{fig:stealth-overt}, louder signals may land us in the Strong Inference of No GR Deviation or the Weak Inference of No GR Deviation II subregimes, but not in the more dangerous Incorrect Inference of GR deviation subregime (since, as explained above, the effect of higher modes cannot be mimicked by a ppE deformation).

\section{Conclusions} \label{sec:conclusion}

In this work, we studied biases in ppE tests due to waveform inaccuracies that lead to false GR deviations.

We quantified the ppE biases and introduced a generic classification of their nature.
Using the Bayes factor between the ppE and GR models, and the fitting factor of the ppE model (or SNR loss when using the ppE model), in Sec.~\ref{subsec:distinguishability}, we identified four regimes of significant systematic bias (see~\cref{fig:stealth-overt}): Strong Inference of No GR Deviation, Weak Inference of No GR Deviation I \& II, and Incorrect Inference of GR Deviation.

In Sec.~\ref{subsec:toy}, using the LSA and Laplace approximation within a toy model, we quantified these regimes in terms of the ratio of systematic to statistical error, as well as the effective cycles of the signal.
As a by-product, we explicitly connected different measures used in quantifying systematic biases.
Computing NLO corrections to the formulas predicted by the LSA can be useful for applying the LSA to a wider range of systematic bias studies. It may also help to explain the findings of Refs.~\cite{Dhani:2024jja,Kapil:2024zdn}.
Using improved criteria for distinguishability  (such as~\cite{Toubiana:2024car}) will also help with the classification criteria for the systematic biases.

We then specialized the general analysis of Sec.~\ref{sec:general_problem} to study the biases in ppE tests due to neglecting spin precession (Sec.~\ref{sec:bias_prec}) and higher modes (Sec.~\ref{sec:bias_HM}).
We investigated how the precession-induced biases change with inclination, total mass, $\chi_p$, and SNR of the injected signal, as well as with the PN order of the ppE parameter.
The bias in ppE parameters increases with increasing inclination, increasing $\chi_p$, and decreasing total mass.
We confirmed that the systematic error does not change with SNR, but the statistical significance of the bias increases due to the smaller statistical error at larger SNR.
The precession-induced bias in the ppE parameter and in the chirp mass, in turn, depends on the role of correlations between the ppE parameter and the chirp mass, which we predicted and explained in Sec.~\ref{subsec:toy}. 
We found the toy model predictions to hold true also in the full Bayesian analysis: the (precession-induced) systematic error in the ppE parameter increases with increasing PN order, while the systematic error in chirp mass decreases with decreasing PN order. 
When neglecting higher modes, we found that the ppE biases are not as large as when neglecting spin precession. This observation should be useful to set priorities in waveform modeling efforts.

In Sec.~\ref{subsec:stat-sig-bayesian-prec} and ~\cref{tab:stat-sig-prec-ppE}, we quantified the statistical significance of the systematic biases induced by spin precession. This is the full Bayesian version of the toy model results shown in~\cref{tab:toy}.
We found that at an SNR of 30, the biases are realizations of a Strong Inference of No GR Deviation. 
At a larger SNR of 60, the bias is instead characterized as a Weak Inference of No GR Deviation~II.
We found that phase modulations drive the biases more than amplitude modulations for spin-precessing injections.
Our analysis is the full Bayesian version of the fitting factor analysis done in~\cite{Apostolatos:1995}.

While we used \texttt{IMRPhenomPv2} to study the biases, the trends we found are not specific to this choice of the precessing waveform model.
This conclusion is corroborated by the analytic structure of precessing signals discussed in Sec.~\ref{subsec:stat-sig-drive-prec}.
Therefore, our results can be used as a baseline for assessing the biases due to other precessing waveform models, such as \texttt{IMRPhenomPv3} (or between the EOB and IMR precessing waveform models).
Our analytic arguments can also help understand potential ppE biases when using a less accurate precessing waveform -- e.g., the biases incurred in the ppE parameters when using \texttt{IMRPhenomPv2}, given a \texttt{IMRPhenomPv3} injection (the bias in the GR parameters has been studied in Ref.~\cite{Khan:2019kot}).
Possible future work can also include analyzing systematic errors in ppE tests when using NR injections containing both spin-precession and higher modes.
The priors we used in the analysis for the ppE parameters will also affect the statistical significance of the biases, especially with the Bayes factor calculation.
Using prior constraints on the ppE deviations will provide a more realistic assessment of the statistical significance of the biases.
Another limitation of this work is that we only considered single-event ppE tests. 
Quantifying the biases through a hierarchical analysis (see e.g.~\cite{Saini:2023rto}) can help to assess the statistical significance of the biases and to better classify them into the different regimes.

Finally, in this work we only considered ground-based detectors operating at O5 sensitivity.
While our analytic arguments will carry over to other detector networks and sensitivities, the significance and classification of the biases may not.
Next-generation detectors and LISA span a wider frequency window, so inspiraling sources will generally spend more cycles in band. 
This usually results in larger systematics and smaller statistical errors.
The typical sources that will be observed by ground-based detectors and space-based detectors will also be drastically different.
For example, the omission of higher modes is known to lead to large systematic errors for massive black hole mergers in LISA, even within GR~\cite{Pitte:2023ltw}.
The residual eccentricity produced by stellar and gaseous environments may induce large biases in tests of GR~\cite{Garg:2024qxq}, and precessional effects will only increase these systematic errors.
In the future we plan to extend our analysis of biases in ppE tests to next-generation ground-based and space-based detectors.
\section{Acknowledgements}
N.Y. acknowledges support from NASA grant GRANT NO. 80NSSC22K0806, R.S.C acknowledges support from the University of Illinois Graduate College Dissertation Completion Fellowship, and K.~P. acknowledges support from NSF Grant No. PHY-2244433. 
E.B. is supported by NSF Grants No. PHY-2207502, AST-2307146, PHY-090003 and PHY-20043, by NASA Grant No. 21-ATP21-0010, by the John Templeton Foundation Grant 62840, by the Simons Foundation, and by the Italian Ministry of Foreign Affairs and International Cooperation grant No.~PGR01167.
We thank Scott Perkins for helping with \texttt{GWAT} set up and for discussions in the early stage of the project.
We are grateful to Carl-Johan Haster, Abhishek Hegade, Anna Liu, Simone Mezzasoma, Caroline Owen, and Justin Ripley for helpful discussions.
\appendix

\section{Overview of systematic biases in GW inference}
\label{sec:lit_review}
Understanding systematic biases in GW inference is an active area of research.
There are several causes for these systematic biases, which we classified and depicted in~\cref{fig:systematics-overview}.
In this section, we expand on each cause of systematic bias ---inaccuracies in the modeling of detector noise, detector response (data characterization), waveform, astrophysical and nuclear physics effects, priors for Bayesian inference, and the underlying theory of gravity.

Systematic biases due to detector noise are caused by inaccurate noise models. 
GW detectors are prone to glitches that are transient, nonstationary processes~\cite{LIGOScientific:2016gtq}. 
Mischaracterizing the noise through inaccurate glitch subtraction can lead to imperfect parameter estimation and bias the inference. 
Recently, the authors of Ref.~\cite{Ghonge:2023ksb} showed the importance of incorporating a deglitching strategy into data analysis pipelines to mitigate biases caused by common detector glitches (see also~\cite{Heinzel:2023vkq}).
Deploying such deglitching techniques is essential in mitigating biases in parametrized tests of GR~\cite{Kwok:2021zny}.
Another recent study shows how to mitigate glitch systematics by deploying Gaussian process modeling of the underlying physics that generates the glitches, rather than the explicit glitch realization~\cite{Ashton:2022ztk}.
Glitch systematics have also been investigated for LISA and have been found to be significant~\cite{Spadaro:2023muy}.

Inaccuracies in data characterization can also introduce systematic errors.
For ground-based detectors, an important source of systematic errors is imperfect data calibration~\cite{LIGOScientific:2016xax}.
The raw data, consisting of a measured electrical signal, has to be processed and converted, via a response function, into the inferred GW signal, with observables such as the amplitude and phase.
Imperfections in the detector response function can then result in biased GW measurements of the amplitude and phase.
Under the assumption of stationary Gaussian noise, the calibration uncertainties follow a Gaussian process~\cite{cahillane:2017}.
Minimum accuracy requirements on the calibration errors for robust detection of GW signals were studied in~\cite{Lindblom:2008cm}. 
Physically motivated calibration models were used to assess potential biases in GW events from the first observing run~\cite{Payne:2020myg}. 
Closed-form expressions for the conditional likelihood on the calibration parameters, as well as the marginal likelihood integrated over calibration uncertainties, are provided in~\cite{Essick:2022vzl}.
Systematic biases can also potentially arise from the presence of overlapping GWs of distinct astrophysical origin.
For ground based detectors, the biases are most significant when the signals overlap in both time and frequency~\cite{Hourihane:2022doe}.
When unaccounted for, overlapping signals can also bias tests of GR~\cite{Hu:2022bji}.
With space-based detectors such as LISA, the signals can last for much longer, making the problem of overlapping signals more challenging~\cite{Cornish:2003vj}.
For LISA, gaps in the data can also cause biases, and mitigating these biases is essential for robust parameter estimation~\cite{Baghi:2019eqo,Dey:2021dem}.
Overcoming such data and noise related systematics has led to robust data analysis pipelines~\cite{gstlal_burst,Sachdev:2019vvd,Klimenko:2015ypf,Khan:2015jqa,Cornish:2014kda,Umstatter:2005su,Cornish:2005qw,Littenberg:2020bxy}. 

Exact waveform solutions in closed form do not exist for realistic compact binary mergers, so the parameters of the GW source are typically inferred using a parametric waveform model constructed using a wide range of techniques with input from post-Newtonian (PN) theory, black hole perturbation theory, and numerical relativity (NR).
Systematics due to potential inaccuracies in the construction of waveform models can be broken down into three parts (inspiral, merger, and ringdown) based on the distinct regimes of the waveform.
Typically, perturbation theory tools are used for the inspiral and ringdown stage of the GW, and NR simulations are used for the merger (see ~\cite{Yunes:2022ldq} for a pedagogical review).
Solutions to the GW from the distinct regimes are then ``stitched'' together to construct waveform templates.
Broadly, the waveform templates are constructed using phenomenological inspiral-merger-ringdown (IMR) models (e.g.,~\cite{Khan:2015jqa}), effective-one-body (EOB) models (e.g.,~\cite{Bohe:2016gbl}), and more recently NR-surrogate models~\cite{Varma:2018mmi,Blackman:2015pia}. 
For recent upgrades to these families of waveforms, including spin-precession, see~\cite{Varma:2019csw,Pratten:2020ceb,Ramos-Buades:2023ehm}, and for an overview of these waveform families, see e.g., Ref.~\cite{Isoyama:2020lls} and references therein. 
All of these methods are intrinsically perturbative (and suffer from truncation error) or numerical (and suffer from finite-difference errors, among others), which can induce systematic biases in parameter estimation.
Understanding these systematics is essential to mitigate biases in GW inference. 

State-of-the-art PN inspiral waveforms for quasicircular point-particle inspirals have been computed to 4.5PN order~\cite{Blanchet:2023bwj}.
Truncating this inspiral phase to a lower PN order can result in systematic errors~\cite{Owen:2023mid,Hoy:2024vpc,Pompili:2024yec}.
One way to mitigate such truncation errors is to marginalize over the higher PN order terms that are neglected in the inspiral phase~\cite{Read:2023hkv,Owen:2023mid}. 
Another way to mitigate such errors is to enhance the inspiral model through phenomenological parameters (introduced also in the modeling of the intermediate and merger-ringdown phase) that are fitted or ``calibrated'' to NR simulations~\cite{Khan:2015jqa}.
The calibration of these phenomenological parameters, however, can also introduce systematic errors in parameter estimation, because the fit is over the high-dimensional space of highly correlated phenomenological parameters. 
A recent study investigated systematic errors due to inferring parameters using \texttt{IMRPhenomD} for injections generated with \texttt{IMRPhenomXAS}~\cite{Kapil:2024zdn}. 
The \texttt{IMRPhenomXAS} model is more accurate because it uses a larger catalog of NR simulations for the calibration. 
Reference~\cite{Kapil:2024zdn} found that, indeed, the calibration of the phenomenological parameters can introduce significant systematic errors at high SNR. 
A similar study investigated biases in a population of events due to using the \texttt{IMRPhenomXPHM} family of waveforms against \texttt{SEOBNRv5PHM} injections, reaching similar conclusions~\cite{Dhani:2024jja}.

Corrections to the simplest inspiral models are also sourced from effects such as eccentricity and spins. 
When the signal is sufficiently eccentric and long enough, using a quasicircular waveform model can result in significant biases~\cite{Favata:2013rwa,Moore:2019vjj,Favata:2021vhw,Cho:2022cdy}.
These biases can also propagate to the non-GR sector and contaminate inspiral tests of GR~\cite{Bhat:2022amc,Saini:2022igm,Narayan:2023vhm}.
Including eccentricity also improves constraints on deviations from GR when the eccentricity is large enough~\cite{Moore:2020rva}.
Spins affect the GW signal through the components that are aligned (or anti-aligned) with the orbital angular momentum, which lead to longer (or shorter) signals by correcting the binding energy and luminosity~\cite{Blanchet:2013haa}. 
The spin components that are orthogonal to the orbital angular momentum modulate the GW signal through spin precession.
With just the effect of aligned (or anti-aligned) spins, spinning binaries can be clearly distinguished from nonspinning binaries when the magnitudes of the dimensionless spins are larger than 0.05 at SNR of 10~\cite{Chatziioannou:2014bma}.
At the same SNR, including precession allows for spinning binaries to be clearly distinguishable for even lower dimensionless spins~\cite{Chatziioannou:2014bma}.
Conservatively, when a spinning binary can be clearly distinguished from a nonspinning binary, systematic errors can be incurred.
Inaccuracies in modeling spin precession can thus also result in biases in the measurement of source parameters, particularly the spins and their tilts~\cite{Khan:2019kot}.
Waveform modulations due to spin precession can be confused with modulations due to eccentricity when the signal is short, as shown in~\cite{Romero-Shaw:2022fbf}.
Due to coupling between spins and non-GR parameters, the effect of spin precession can enhance (in terms of PN order) non-GR effects~\cite{Loutrel:2022tbk,Loutrel:2022xok}, leading to potentially tighter constraints on the non-GR parameters.
These studies further motivate the need for modeling inspirals containing spin precession and eccentricity to perform accurate parameter estimation, which is an active area of research~\cite{Arredondo:2024nsl,Fumagalli:2023hde} (see also~\cite{Liu:2023ldr} for an EOB approach to modeling eccentricity and spin-precession).

Incorporating the effects of higher-order, nonquadrupolar modes (caused mainly by mass asymmetry) is also important for accurate parameter estimation, as it can help improve the evidence for asymmetric masses in a binary~\cite{Chatziioannou:2019dsz,LIGOScientific:2020stg}.
When higher modes are neglected, there can be significant biases for edge-on sources, that are highly asymmetric even at moderate SNR~\cite{Kalaghatgi:2019log}.
Even when the inclination is not edge-on, an asymmetric, heavy source with large spins can introduce significant biases through the higher mode content of the waveform~\cite{Shaik:2019dym}.
The interplay of spin precession effects with higher modes can result in improved measurements when both effects are included~\cite{Krishnendu:2021cyi}.
The effects induced by higher modes can bias parametrized inspiral tests of GR, as shown in Ref.~\cite{Mehta:2022pcn}.
The authors of Ref.~\cite{Pang:2018hjb,Puecher:2022sfm,Maggio:2022hre} also showed that there can be significant biases in the parametrized deviations that enter the merger and ringdown; they attribute this to the fact that higher modes affect the late inspiral and merger-ringdown more strongly than the early inspiral.

Going past the inspiral, the merger of the binary is modeled using NR.
When NR simulations are directly used for parameter estimation, a sparse placement in the template bank can induce systematic errors~\cite{Ferguson:2022qkz}.
A way around this problem is through NR surrogate waveforms~\cite{Blackman:2015pia,Varma:2018mmi}.
An important goal of current research is improving the quality of NR and surrogate waveforms in order to overcome systematic errors.
In Ref.~\cite{Mitman:2022kwt}, the authors showed how to fix the reference frame of NR waveforms through the Bondi-van der Burg-Metzner-Sachs transformations to match the frame of PN and EOB waveforms.
This crucial ingredient was used in Ref.~\cite{Yoo:2023spi} to show that memory effects can be better extracted from NR simulations, allowing for a more accurate NR-surrogate family of waveforms.
After merger, modeling the ringdown brings its own set of issues that can produce systematic error.
In recent years, incorporating nonlinear aspects of the ringdown in GR has been shown to be important for accurate inference~\cite{Cheung:2022rbm,Mitman:2022qdl}.
Identifying the quasinormal mode content of a waveform in the presence of nonlinearities  is a challenging problem and it can lead to significant systematic errors, even within GR~\cite{Baibhav:2017jhs,Giesler:2019uxc,Ma:2023vvr,Ma:2023cwe,Baibhav:2023clw,Nee:2023osy,Cheung:2023vki,Takahashi:2023tkb,Redondo-Yuste:2023seq,Redondo-Yuste:2023ipg,Ma:2024qcv,Qiu:2023lwo,Zhu:2023mzv,Zhu:2023fnf,Zhu:2024rej,May:2024rrg,Pitte:2023ltw,Toubiana:2023cwr,Bhagwat:2019dtm,Silva:2022srr}.
There are several proposals to parametrize the dependence of modifications to the quasinormal mode frequencies induced by modified gravity~\cite{Tattersall:2017erk,Tattersall:2018map,Cardoso:2019mqo,McManus:2019ulj,Volkel:2022aca,Franchini:2022axs,Hirano:2024fgp,Maselli:2019mjd,Carullo:2021dui}, as well as general frameworks to compute quasinormal mode frequencies for rotating black holes in some of the most interesting modifications of general relativity~\cite{Cano:2020cao, Cano:2021myl, Li:2022pcy, Hussain:2022ins, Cano:2023tmv, Cano:2023jbk, Chung:2024ira}.
However, studies of optimal ways of extracting valuable bounds from hypothetical deviations in the data are in their infancy~\cite{Volkel:2022khh,Maselli:2023khq,Yi:2024elj,Gupta:2024gun}.

Finite-size effects, which are significant for tidally deformed neutron star-neutron star binaries, also introduce corrections to the GW signal.
During the inspiral, conservative tidal effects enter at 5PN order relative to the leading-order point-particle phase (see e.g.~\cite{Damour:1986ny,Mora:2003wt,Blanchet:2013haa}). 
Incorporating higher-order PN contributions is crucial for accurate measurement of the tidal effects~\cite{Yagi:2013baa,Favata:2013rwa,Wade:2014vqa}. 
The subleading PN terms of the tidal contribution must be accurately modeled to mitigate systematic errors~\cite{Hinderer:2009ca}, and allow for the measurement of the individual tidal deformabilities~\cite{Vines:2011ud}.
Further, inaccuracies in the binary Love relations used for inference of the individual tidal deformabilities can also be a source of systematic error for certain equations of state, and it is essential to marginalize over the uncertainties for robust parameter estimation of the individual tidal deformabilities~\cite{Carson:2019rjx}.
Incorporating dynamical tidal effects can also be important for future detectors, and systematic biases are possible when these effects are not taken into account~\cite{Pratten:2019sed}.
Another source of systematic bias could come from the tidal dissipation of neutron stars~\cite{Ripley:2023qxo}, recently constrained in Ref.~\cite{Ripley:2023lsq}.
For next-generation detectors, since tidal dissipation can be more accurately measured, neglecting it from the waveform can introduce systematic biases. 
These biases are induced by correlations between the conservative and dissipative tidal deformabilities (see e.g.~Fig. 4 of Ref.~\cite{Ripley:2023lsq}).

Systematic biases can also result from corrections to the generation and propagation of GWs due to an external astrophysical environment.
Typically, one would expect modifications to the generation of GWs due to the external environment to be more relevant for space-based detectors than for ground-based detectors.
This is because of the typically longer timescales associated with the effects induced by the external environment, and the fact that the effects modulate the GWs in the early inspiral regime.
A concrete example is a hierarchical triple, where a third body perturbs an inner binary and modulates the GWs emitted by the inner binary.
In these systems, a myriad of effects can be sourced, with Doppler effects~\cite{Yunes:2010sm,Chamberlain:2018snj,Robson:2018svj}, de Sitter precession~\cite{Yu:2020dlm}, and Kozai-Lidov oscillations~\cite{Deme:2020ewx,Rohit_Nico,Gupta:2019unn} as examples.
Any or potentially all of these can introduce systematic biases in parameter estimation.
Another effect from the external environment that modulates the GWs in the early inspiral is due to accretion~\cite{Speri:2022upm}.
In Ref.~\cite{Zwick:2022dih}, the authors surveyed a wide range of astrophysical environmental corrections and showed them to be important for accurate parameter estimation.
Neglecting such environmental effects can bias tests of GR, as shown in Ref.~\cite{Kejriwal:2023djc} through a linear signal analysis.
Intervening compact objects and galaxies affect the propagation of GWs through lensing.
Tests of GR can become biased due to neglecting lensing effects~\cite{Mishra:2023vzo,Wright:2024mco}.
Thus, systematically accounting for such astrophysical environmental effects is important for accurate GW inference.

Even with an accurate gravitational waveform, systematic biases can be incurred from the priors used for parameter estimation.
In addition to accurate waveforms that model the spins, astrophysically motivated priors are also needed to infer the spin distribution of the observed binary black holes~\cite{Callister:2022qwb,Galaudage:2021rkt}.
Similarly, when performing such hierarchical inferences~\cite{thrane-2019} to obtain the mass distribution for the observed population of binary black holes, biases arise from the modeling and priors for the astrophysical population~\cite{Cheng:2023ddt}.
As shown in~\cite{Zevin:2020gxf}, such biases can affect the inference of the astrophysical formation channel for a given event.
Such problems persist when stacking multiple GW observations to better constrain deviations from GR~\cite{Isi:2019asy,Isi:2022cii,Magee:2023muf}.

Finally, systematic biases can arise simply because our underlying theory of gravity is incorrect.
When a particular symmetry of GR is broken, the solutions to the resulting field equations will typically be modified.
GW solutions can thus carry an imprint of the modifications to GR, which typically affect the inspiral, merger and ringdown of the GW signal (see~\cite{Yunes_2009,Berti:2015itd,Yunes_2016} for a comprehensive ``bestiary'' of modified gravity effects).
If these modifications to GR occur in nature, then neglecting them when estimating parameters from a GW observation can result in \emph{fundamental theoretical bias}~\cite{Yunes_2009} that can, in fact, be in the \emph{stealth} regime~\cite{Cornish:2011ys,Vallisneri:2013rc}.
In this regime, the systematic errors are larger than any statistical errors, but the data is not informative enough to prefer the non-GR model over the GR one. This is because strong correlations and degeneracies between the waveform parameters may allow the GR parameters to ``absorb'' the modified gravity effect in the signal, thus mimicking the non-GR model.

The detailed summary of the state of the field provided above may paint a dark picture of our ability to infer astrophysics, nuclear physics and theoretical physics from GW observations, but care should be taken to not exaggerate this point. 
Obviously, whenever one analyzes real data contaminated by noise with approximate models one risks incurring systematic errors. 
For most sources in the first four observing catalogs (O1--O4), however, the SNRs were low enough that statistical errors dominated over systematic ones, and thus, most inferences made to date are safe. 
However, as current-generation detectors are improved (and definitely by the time of next-generation ground- and space-based detectors), the average SNRs of events will increase, reducing the statistical errors, and thus enhancing the importance of systematic errors when making inferences. 
For this reason, it is crucial to study systematic errors and develop methods to ameliorate them.
Towards this goal, this paper is dedicated to the study of systematic biases in tests of GR due to neglecting effects of spin precession or higher modes in the waveform model. 

\section{Review of waveform models} \label{sec:waveform_review}

In the following, we provide a concise summary of the phenomenological IMR waveforms in GR used in our analysis.
A more comprehensive review can be found, e.g., in Ref.~\cite{Isoyama:2020lls}.

\subsection{\texttt{IMRPhenomD}}
The \texttt{IMRPhenomD} model is a phenomenological description of the quasicircular, nonprecessing, dominant mode of the radiation from binary black hole mergers. 
The amplitude and phase are piecewise functions, with the inspiral regime defined by $f < f_1$, the intermediate regime defined by $f_1 \leq f \leq f_2 $, and the merger-ringdown regime defined by $f>f_2$.
The transition frequencies are given by $f_1 = 0.018/m$ and $f_2 = 0.5f_{\mathrm{rd}}$, where $f_{\mathrm{rd}} = 0.071/m_{\rm tot}$, and $m_{\rm tot}$ is the total mass of the source.
In total, the \texttt{IMRPhenomD} model has 11 waveform parameters, namely $\{\mathrm{RA}, \mathrm{DEC}, \psi, \iota, \phi_{\mathrm{ref}}, t_c, D_L, m_1, m_2, \chi_{1z}, \chi_{2z} \}$.
The component masses enter the phase through the chirp mass $\mathcal{M}$ and symmetric mass ratio $\eta$, given by
\begin{align}
    \eta = \dfrac{m_1 m_2}{m^2}, \quad \mathcal{M} = \eta^{3/5} m_{\rm tot},
\end{align}
where $m_{\rm tot}$ is the total mass.
The chirp mass enters the phase at leading PN order, while the symmetric mass ratio enters the phase at 1PN order.
The spin components along the orbital angular momentum enter the phase through the aligned spin effective parameter $\chi_{\mathrm{eff}}$ (see for example~\cite{Ajith:2009bn}):
\begin{align}
    \chi_{\mathrm{eff}} = \dfrac{\chi_{1z} m_1 + \chi_{2z} m_2}{m_1+m_2}. \label{eqn:chi_eff}
\end{align}
The effects of the aligned spins enter the phase at 1.5PN order through the spin-orbit coupling~\cite{Blanchet:2013haa}.
When the spins are aligned (anti-aligned) with the orbital angular momentum, the spin-orbit coupling increases (decreases) the merger time, compared to when the spins are zero (this is easily seen by inspecting Eq. (1.3) of~\cite{Poisson:1995ef}). 
\subsection{\texttt{IMRPhenomP}}
The phenomenological precessing waveforms \texttt{IMRPhenomPv2/3} are obtained by the ``twisting up'' approach~\cite{Hannam:2013oca,Schmidt:2012rh,Khan:2019kot}, that relates the waveforms in the coprecessing frame and inertial frame. 
In the spin-weighted spherical harmonic basis, the GW strain $h(t) = h_{+}(t ) - i h_{\times} (t)$ is given by~\cite{Blanchet:2013haa}
\begin{align}
    h(t) = \sum_{\ell \geq 2} \sum_{|m| \leq \ell} h_{\ell m}(t)  \ {}_{-2}Y_{\ell m} (\theta, \phi),
\end{align}
where $h_{\ell m}(t)$ are the modes of the GW strain $h(t)$, and $\ {}_{-2}Y_{\ell m}$ are the spin-weighted spherical harmonics.
In the phenomenological prescription for \texttt{IMRPhenomPv2}, only the dominant mode $\ell = 2$ is included by twisting-up the \texttt{IMRPhenomD} waveform:
\begin{align} 
\begin{split}
    \tilde{h}^{\mathrm{PhenomPv2}}_{2,m}(f) &= e^{-im\phi_z(f)} \sum_{|m'|=2} \Big [ e^{i m' \zeta(f)} \\
    & \times d^2_{m',m}(-\theta_L(f)) \tilde{h}_{2,m'}^{\mathrm{PhenomD}}(f) \Big ],\\
    &= e^{-im\phi_z(f)} \sum_{|m'|=2} \Big [ d^2_{m',m}(-\theta_L(f))  \\
    & \times A_{2m'} (f) e^{i \Phi_{2m'}(f) + i m'\zeta(f)}  \Big ]\,,
\end{split}
\end{align}
where we have adopted the Euler angles notation $\{ \phi_z, \zeta, \theta_L \}$ from Ref.~\cite{Chatziioannou:2017tdw}. 
To leading PN order, the angles $\phi_z(f)$ and $\zeta(f)$ scale as $f^{-1}$, implying that the secular contributions to the nonprecessing phase enter at 1PN order (relative to the leading PN phase).
We emphasize here that while the Euler angles introduce a correction to the nonprecessing phase at 1PN order, the binding energy receives a correction due to the corotation only at 2PN order (see Ref.~\cite{Blanchet:2013haa} for a review).

\subsection{\texttt{IMRPhenomHM}}
The \texttt{IMRPhenomHM} waveform is constructed using the \emph{quadrupole mapping}~\citep{London:2017bcn}, wherein the amplitude and phase of the subdominant modes are built using the dominant $(2,2)$ mode.
In the frequency domain, the mode components of \texttt{IMRPhenomHM} are given by~\cite{London:2017bcn}
\begin{align}
\begin{split}
    \tilde{h}^{\mathrm{PhenomHM}}_{\ell m}(f) &= A_{\ell m}(f) e^{i \Phi_{\ell m}(f)} \\
               &= |\beta_{\ell m} (f)| A_{22}(f_{22}^A) e^{i [\kappa \phi_{22}(f_{22}^{\Phi}) + \Delta_{\ell m}]},    
\end{split}
\end{align}
where $\beta_{\ell m}(f)$ is a rescaling of the dominant $(2,2)$ mode amplitude, $\kappa$ is a piecewise constant, $\Delta_{\ell m}$ is a phase shift, and $f_{22}$ is a linear mapping to $f$ (see~\citep{London:2017bcn} for details).
For optimization, the amplitude and phase are fitted to NR by varying the transitional frequency, resulting in a different value $f_{22}^A$ and $f_{22}^{\Phi}$.

\section{Review of ppE tests of GR with GWs} \label{sec:review_ppE}

The ppE framework is a theory-agnostic way to test GR by introducing deviations from a GR waveform in the form of an asymptotic PN series, with dimensionless coefficients that can later be mapped to specific theories.
The structure of the ppE waveform depends on the physics that is present in the GR waveform. 
The ppE framework for precessing waveforms and for waveforms with higher modes is described in Refs.~\cite{Loutrel:2022xok} and~\cite{Mezzasoma:2022pjb}, respectively.
The simplest structure of the ppE waveform is when the deviations are introduced only for the dominant mode, without coupling the deviation coefficients to the spins.
The ppE waveform can be parametrized in many different ways~\cite{Yunes_2009,Perkins:2022fh}. 
The ppE framework applied to nonspinning quasicircular inspirals admits the following waveform structure:
\begin{align}
\tilde{h}_{\mathrm{ppE}}(f) = \tilde{h}_{\text{GR}} (1+\alpha u^a) e^{i \beta u^b}, \label{eqn:waveform_ppE_full}
\end{align}
where $u=(\pi \mathcal{M}f)^{1/3}$ and where, for simplicity, we have shown only the leading PN contributions in the amplitude and phase.
The ppE parameters $\alpha$ and $\beta$ control the amplitude and phase modification, respectively.
The integer powers $a$ and $b$ (or ``ppE indices'') control the PN order of the amplitude and phase modification, respectively.
Typically, the corrections to the phase are more significant for data analysis~\cite{Perkins:2022fh}, so we ignore the amplitude modifications. 
While several modifications can be considered simultaneously, it is still fruitful to look at the deviations one at a time~\cite{Perkins:2022fh} by fixing $b$ and placing bounds on $\beta$. 
Setting $b=\{ -7,-3, -1\}$ corresponds to $-1$PN, 1PN, and 2PN ppE deviations, respectively.
A catalog of modified theories with their corresponding ppE mapping can be found in Refs.~\cite{Yunes_2009,Yunes_2016}.

In \texttt{GWAT}, the ppE waveforms are implemented by introducing a ppE correction to the base GR waveform in the inspiral regime. 
The ppE correction is then propagated through the intermediate and merger-ringdown regimes by requiring $C^1$ continuity at the transition frequencies (compare Ref.~\cite{Agathos:2013upa}).
The GW phase has the structure
\begin{align}
\Phi(f) = \begin{cases}
    \Phi_{\mathrm{ins}}(f) = \Phi_{\mathrm{TF2}}(f)+\Phi_{\mathrm{Phenom}}(f) , & f \leq f_1, \\
    \Phi_{\mathrm{int}} = c_0 + c_1 f + \dots, & f_1\leq f \leq f_2,\\
    \Phi_{\mathrm{MR}} = d_0 + d_1 f + \dots, & f \geq f_2
    \end{cases}
\end{align}
where $\Phi_{\mathrm{ins}}(f)$, $\Phi_{\mathrm{int}}$, and $\Phi_{\mathrm{MR}} $ are the inspiral, intermediate, and merger-ringdown phases.
The parameters $c_0$, $c_1$, $d_0$, and $d_1$ are fitted by enforcing $C^1$ continuity at the transition frequencies $f_1$ and $f_2$.
When the inspiral phase acquires a correction of the form $\Phi_{\mathrm{ins}}(f) \rightarrow \bar{\Phi}_{\mathrm{ins}}(f)= \Phi_{\mathrm{ins}}(f) + \Delta \Phi(f)$, where $\Delta \Phi(f) = \beta u^b$, there will be corrections to the intermediate and merger-ringdown phases $\Phi_{\mathrm{int/MR}}(f) \rightarrow \bar{\Phi}_{\mathrm{int/MR}}(f)$.
This can be summarized as
\begin{subequations}
\begin{align}
    \bar{\Phi}_{\mathrm{int}}(f) &= \Phi_{\mathrm{int}}(f) + \Delta c_0 + \Delta c_1 f, \\
    \bar{\Phi}_{\mathrm{MR}}(f) &= \Phi_{\mathrm{MR}}(f) + \Delta d_0 + \Delta d_1 f,
\end{align}
\end{subequations}
where $\Delta c_0$, $\Delta c_1$, $\Delta d_0$, and $\Delta d_1$ are corrections that will be fitted by $C^1$ continuity, and will depend on the added phase $\Delta \Phi(f)$.
Enforcing $C^1$ continuity at $f=f_1$ and at $f=f_2$, we obtain
\begin{subequations}
\begin{align}
 \Delta c_0 &= \beta \left(1-\dfrac{b}{3} \right) (\pi \mathcal{M}f_1)^{b/3},\\
       \Delta c_1  &= \beta \dfrac{b}{3} (\pi \mathcal{M}f_1)^{b/3}/f_1, \\
    \Delta d_0 &= \Delta c_0, \\
    \Delta d_1 &= \Delta c_1.
\end{align}    
\end{subequations}

\section{Review of linear signal approximation and linear waveform difference approximation for computing systematic error} \label{sec:LSA}

For the sake of completeness, here we review the calculationn of systematic errors in the linear signal approximation following Refs.~\cite{Cutler:2007mi,Flanagan:1997kp}.
We also discuss in detail the errors in~\cref{eqn:systematic_error_explicit_main_LO_unexpanded,eqn:systematic_error_explicit_main_LO_expanded}.
The goal is to perturbatively solve~\cref{eqn:max_likelihood_condition_main} by linearizing the waveform for small systematic errors. 
The maximum likelihood point in terms of the injected point is $\ml{\lambda}^i=\tr{\lambda}^i+\Delta_L \lambda^i$.
Linearizing $h_M(\bfml{\lambda})$ gives us
\begin{align}
h_M(\bfml{\lambda}) & \sim h_M (\bftr{\lambda})+\Delta_L \lambda^j \partial_{\lambda^j} h_M(\bftr{\lambda}) + \mathcal{O}((\Delta_L \lambda^i)^2),
\end{align}
where corrections quadratic in the systematic error are neglected. 
In the linear approximation to the waveform about the maximum likelihood point, it is also true that 
\begin{align}
h_M(\bftr{\lambda}) & \sim h_M(\bfml{\lambda})-\Delta_L \lambda^j \partial_{\lambda^j} h_M(\bfml{\lambda}) + \mathcal{O}((\Delta_L \lambda^i)^2),     \label{eqn:linearized_template}
\end{align}
resulting in 
$
    \partial_{\lambda^j} h_M(\bftr{\lambda}) = \partial_{\lambda^j} h_M (\bfml{\lambda}) 
$.
Plugging~\cref{eqn:linearized_template} into~\cref{eqn:max_likelihood_condition_main} yields
\begin{multline}
    \left( \partial_{\lambda^i} h_M(\bfml{\lambda}) | d - h_M(\bftr{\lambda}) \right) \\
    - \Delta \lambda^j \left( \partial_{\lambda^i} h_M(\bfml{\lambda}) | \partial_{\lambda^j} h_M(\bfml{\lambda}) \right) = 0. \label{eqn:max_likelihood_condition_2}       
\end{multline}
Recognizing that the Fisher matrix is given by
\begin{align}
    \Gamma_{ij}(\bfml{\lambda}) = \left( \partial_{\lambda^i} h_M(\bfml{\lambda}) | \partial_{\lambda^j} h_M(\bfml{\lambda}) \right),
\end{align}
we obtain for the systematic error:
\begin{align}
\begin{split}
    \Delta_L \lambda^i &= \left( \Gamma^{-1}(\bfml{\lambda}) \right)^{ij} \left(\partial_{\lambda^j} h_M(\bfml{\lambda}) |  d - h_M(\bfml{\lambda}) \right) \\
    &= C^{ij}(\bfml{\lambda}) \left(\partial_{\lambda^j} h_M(\bfml{\lambda}) |  h_S(\bftr{\theta}) - h_M(\bftr{\lambda}) \right), \label{eqn:systematic_error_implicit}    
\end{split}
\end{align}
where $C^{ij} =  (\Gamma^{-1})^{ij}$ is the Fisher-covariance matrix. This recovers Eq. (12) of~\cite{Cutler:2007mi}. 
As mentioned in the main text, the original application of~\cite{Cutler:2007mi} was to \emph{postdict} the systematic error given an estimate of $\bfml{\lambda}$.
We are however interested in \emph{predicting} the systematic error given $\bftr{\lambda}$.
Since the partial derivatives of the waveform are identical at $\bftr{\lambda}$ and at $\bfml{\lambda}$, it follows that $\Gamma_{ij}(\bftr{\lambda}) = \Gamma_{ij}(\bfml{\lambda})$. 
Consequently, we obtain for the systematic error,
\begin{align}
   \Delta_L \lambda^i &= C^{ij} (\bftr{\lambda}) \left(\partial_{\lambda^j} h_M(\bftr{\lambda}) |  h_S(\bftr{\theta}) - h_M(\bftr{\lambda}) \right). \label{eqn:systematic_error_explicit}
\end{align}
Recall that the true signal and waveform are related through
\begin{subequations}
\begin{align}
    A_S(\bftr{\theta};f) &= A_M(\bftr{\lambda};f) + \Delta A(\bftr{\theta};f), \\
    A_S(\bftr{\theta};f) &= \Psi_M(\bftr{\lambda};f) + \Delta \Psi(\bftr{\theta};f).
\end{align}  \label{eqn:amp_phase_diff}  
\end{subequations}
Substituting~\cref{eqn:amp_phase_diff} into~\cref{eqn:systematic_error_explicit} results in
\begin{widetext}
\begin{align}
\begin{split}
    \Delta_L \lambda^i &= C^{ij} (\bftr{\lambda}) \ 4 \Re \int \limits_{f_{\min}}^{f_{\max}} \Bigg [ \dfrac{df}{S_n(f)} \Big( \partial_{\lambda^j} A_M(\bftr{\lambda};f) + i A_M(\bftr{\lambda};f) \partial_{\lambda^j} \Psi_M(\bftr{\lambda};f) \Big)\\
    &\times \left( \left[A_M(\bftr{\lambda};f) + \Delta A(\bftr{\theta};f) \right] e^{-i \Delta \Psi(\bftr{\theta};f)} - A_M(\bftr{\lambda};f)\right) \Bigg ].
\end{split}
\end{align}      
\end{widetext}
We can now expand in $\Delta \Psi$ and $\Delta A$ and keep terms to NLO, with the result
\begin{widetext}
    \begin{multline}
    \Delta_L \lambda^i = C^{ij} (\bftr{\lambda}) \ 4  \int \limits_{f_{\min}}^{f_{\max}}  \dfrac{df}{S_n(f)}   \Bigg( A_M^2(\bftr{\lambda};f) \partial_{\lambda^j} \Psi_M (\bftr{\lambda};f) \Delta \Psi (\bftr{\theta};f) + \partial_{\lambda^j} A_M (\bftr{\lambda};f) \Delta A (\bftr{\theta};f) \\
    + \left[-\dfrac{1}{2} A_M (\bftr{\lambda};f) \partial_{\lambda^j} A_M (\bftr{\lambda};f) (\Delta \Psi (\bftr{\theta};f) )^2 +  A_M (\bftr{\lambda};f) \partial_{\lambda^j} \Psi_M (\bftr{\lambda};f) \Delta \Psi (\bftr{\theta};f) \Delta A (\bftr{\theta};f)  \right]\Bigg ). \label{eqn:sys-error-nlo}
    \end{multline}
\end{widetext}
At LO, for $\Delta A = 0$, we have that 
\begin{multline}
\Delta_L \lambda^i = C^{ij} (\bftr{\lambda}) \ 4 \int \limits_{f_{\min}}^{f_{\max}} \Big [ \dfrac{df}{S_n(f)} A_M^2(\bftr{\lambda} ; f) \\
\times \partial_{\lambda^j} \Psi_M(\bftr{\lambda} ; f)  \Delta \Psi (\bftr{\theta} ; f)  \Big ], \label{eqn:systematic_error_explicit_main_LO_expanded_phase}    
\end{multline}
at LO in $\Delta \Psi$. 
Likewise at LO, for $\Delta \Psi =0$, we have that
\begin{align}
\begin{split}
\Delta_L \lambda^i &= C^{ij} (\bftr{\lambda}) \ 4 \int \limits_{f_{\min}}^{f_{\max}} \Big [ \dfrac{df}{S_n(f)} \Delta A (\bftr{\theta} ; f)  \partial_{\lambda^j} A_M(\bftr{\lambda} ; f) \Big ], \label{eqn:systematic_error_explicit_main_LO_expanded_amp}
\end{split}
\end{align}
When the amplitude and phase parameters are uncorrelated, as in the toy example of Sec.~\ref{subsec:toy}, the Fisher and Fisher-covariance matrices will be block diagonal~\cite{Cutler:1994ys}.
With $\Delta A=0$ (as we had in the toy example), the LO terms for the amplitude parameters in~\cref{eqn:sys-error-nlo} will vanish, as they depend on $\partial_{\lambda^j} \Psi_M$ contracted with $C^{ij}$.
The nonvanishing contribution to the systematic error in the amplitude parameters will be at NLO, as they depend on  $\partial_{\lambda^j} A_M$ contracted with $C^{ij}$.
To LO in $\Delta \Psi$, we will not have any systematic error in the amplitude parameters, which justifies the results of Sec.~\ref{subsec:toy}.
We emphasize that we have not computed all terms at NLO: there are additional NLO corrections to~\cref{eqn:systematic_error_explicit}, 
so keeping terms beyond LO in~\cref{eqn:sys-error-nlo} will introduce potentially uncontrolled remainders.

We now show how to minimize the effective cycles to obtain the linear scaling between $\mathcal{N}_e$ and $\Delta \Psi$. 
Defining the functional $G_{{\mathcal{N}}_e} (\Delta t, \Delta \phi)$ given by
\begin{align}
    G_{{\mathcal{N}}_e}(\Delta t, \Delta \phi) = \int \limits_{f_{\min}}^{f_{\max}} \dfrac{df}{S_n(f)} A_M^2(\bftr{\lambda} ; f) \left( \Delta \Phi (\bftr{\theta} ; f) \right)^2, 
\end{align}
we analytically minimize over $\Delta t, \Delta \phi$ to obtain
\begin{subequations}
\begin{align}
\Delta t &= -\dfrac{\langle f \rangle \langle \Delta \Psi (\bftr{\theta}) \rangle - \langle f \Delta \Psi (\bftr{\theta}) \rangle \rho^2} {2 \pi (\langle f \rangle^2 - \langle f^2 \rangle \rho^2)}, \\
\Delta \phi &= - \dfrac{\langle f^2 \rangle \langle \Delta \Psi (\bftr{\theta}) \rangle - \langle f \rangle \langle f \Delta \Psi (\bftr{\theta}) \rangle}{\langle f \rangle^2 - \langle f^2 \rangle \rho^2},    
\label{eqn:delta_t_dela_phi}
\end{align}    
\end{subequations}
where $\langle \cdot \rangle$ is defined as
\begin{align}
    \langle x \rangle \equiv \int \limits_{f_{\min}}^{f_{\max}} \dfrac{df}{S_n(f)} A_M^2(\bftr{\theta};f) x(f). \label{eqn:snr_measure}
\end{align}
Substituting~\cref{eqn:delta_t_dela_phi} into~\cref{eqn:eff_cycles_def} thus gives us the effective cycles with the minimization. 
Since $\Delta t$ and $\Delta \phi$ scale linearly with $\Delta \Psi$, we see that $\mathcal{N}_e$ scales linearly with $\Delta \Psi$.

\section{Characterization of significant systematic biases using the toy model} \label{app:bias_char_toy}

The bounds given by~\cref{eqn:sig-stat-thresh,eqn:sig-stat-dist} are sensitive to the PN order of the ppE parameter, the strength of the $\kappa$-effect, and the relative PN order with respect to the $\kappa$-effect, as we explain below.

The bound $(\Delta_L^{\mathrm{ppE}} \beta/ \sigma_{\beta})_{\mathrm{dist}}$ does not depend on the SNR, while the bound $(\Delta_L^{\mathrm{ppE}} \beta/ \sigma_{\beta})_{\mathrm{thresh}}$ depends on the SNR through $\sqrt{\log (1/\sigma_{\beta}^2)}$.
Since $\sigma_{\beta}$ decreases inversely with the SNR $\rho$, the bound $(\Delta_L^{\mathrm{ppE}} \beta/ \sigma_{\beta})_{\mathrm{thresh}}$, in turn, grows as $\sqrt{\log \rho^2}$.
This growth arises from the Occam penalty: as the ppE parameter becomes better constrained, a larger Occam penalty is incurred.
Therefore, given a threshold $BF_{\rm thresh}$, the ratio of systematic to statistical error (on $\beta$) must compensate for the ppE model to be favored over the GR model.
From~\cref{eqn:BF_toy_example_specific}, we know that the ratio of systematic to statistical error grows linearly with SNR.
This growth is faster than the growth of the bound $(\Delta_L^{\mathrm{ppE}} \beta/ \sigma_{\beta})_{\mathrm{thresh}}$. 
Thus, for a sufficiently large SNR, the inequality $(\Delta_L^{\rm ppE} \beta)< (\Delta_L^{\rm ppE} \beta)_{\rm dist}$ can be violated, and so can the inequality $(\Delta_L^{\rm ppE} \beta)< (\Delta_L^{\rm ppE} \beta)_{\rm thresh}$.
We also see this from $FF_{\rm ppE}$ being SNR-invariant, but $BF_{\rm ppE,GR}$ increasing exponentially with the SNR squared [Eq.~\eqref{eqn:BF_sys_error_accuracy}], and $1-FF_{\rm ppE}^{\rm dist}$ decreasing with SNR squared [Eq.~\eqref{eqn:FF_criteria}].
Because of these arguments, the characterization of the bias will always flow upward and rightward in~\cref{fig:stealth-overt} with increasing SNR.

Let us now discuss how the bias behaves with the type of GR effect neglected in the ppE model.  
Neither~\cref{eqn:sig-stat-dist} nor~\cref{eqn:sig-stat-thresh} depend on the strength of the $\kappa$-effect, but $(\Delta_L^{\mathrm{ppE}} \beta/ \sigma_{\beta})$ grows linearly with $\kappa$ [\cref{eqn:BF_toy_example_specific}].
Thus, increasing the magnitude of the $\kappa$-effect has a similar effect on the inequalities as increasing the SNR: one flows upward and rightward in~\cref{fig:stealth-overt} with increasing $\kappa$. 

We can similarly study how the bias behaves with the PN order that we attempt to constrain with the ppE model. The threshold $(\Delta_L^{\rm ppE} \beta)_{\rm thresh}$ depends on the ppE deviation through how well the latter is constrained as compared to the prior range.
Now, ppE deviations of higher PN order are more poorly 
constrained than ppE deviations of lower PN order, for a given SNR and $BF_{\rm thresh}$. However, the prior range allowed for $\beta$ grows as the PN order of the ppE deviation increases [Eq.~\eqref{eqn:relative_Newtonian_ppe_bound}]. The ratio, and therefore, the bound $(\Delta_L^{\rm ppE} \beta)_{\rm thresh}$, remains approximately constant, but with a slight decrease as $b$ increases. 
On the other hand, the quantity $(\Delta_L^{\rm ppE} \beta)_{\rm dist}$ increases with $b$, as one can see by evaluating Eq.~\eqref{eqn:sig-stat-dist}. 
Thus, the characterization of the bias will flow right and down in~\cref{fig:stealth-overt} with increasing PN order of the ppE deviation. 

A slightly curious case occurs when the PN order of the ppE deviation matches exactly the PN order of the neglected GR effect. 
When $b=k$ (with $\kappa_{2,\mathrm{tr}} = 0$), we have that $I_{\mathcal{M}\beta}(k,k) = \det \hat{\Gamma}(k) I(7-2k)$, which implies that $(\Delta^{\mathrm{ppE}}_L \beta/ \sigma_{\beta})_{\mathrm{dist}}$ diverges. Physically, this occurs because the ppE effect one is attempting to constrain can completely absorb the GR effect one has incorrectly neglected to include. Due to this complete degeneracy between $\kappa$ and $\beta$, one finds that the ratio  $(\Delta_L^{\rm ppE} \beta)_{\rm dist}$ diverges, implying that the horizontal line in ~\cref{fig:stealth-overt} moves up, and one is either in the Weak Inference of No GR Deviation~I or the Incorrect Inference of GR deviation subregimes. 

Given the above flows, we can then understand the different subregimes of bias and their flow more clearly, as follows.
\begin{itemize}[leftmargin=*]
			\setlength{\itemsep}{10pt}
			\setlength{\parskip}{0pt}
			\setlength{\parsep}{0pt}
\item[] 
\textbf{Strong Inference of No GR Deviation.} This subregime is the best possible outcome when it comes to systematic biases when testing for GR deviations.
A non-GR inference will not be incorrectly claimed, because the residual SNR test and the model selection test both fail.
However, as the SNR increases, $(\Delta^{\mathrm{ppE}}_L \beta/ \sigma_{\beta})<(\Delta^{\mathrm{ppE}}_L \beta/ \sigma_{\beta})_{\rm thesh}$ will be violated, while $(\Delta^{\mathrm{ppE}}_L \beta/ \sigma_{\beta})>(\Delta^{\mathrm{ppE}}_L \beta/ \sigma_{\beta})_{\rm dist}$ will be reinforced. 
Consequently, with increasing SNR (or also with increasing strength of the $\kappa$-effect), the characterization of the bias will flow from the Strong Inference of No GR Deviation regime to the Weak Inference of No GR Deviation~II regime (upward and rightward in~\cref{fig:stealth-overt}).
\item[] 
\textbf{Weak Inference of No GR Deviation I.} This subregime is less desirable because one may be tempted to conclude a GR deviation has been detected, when in reality there is none. However, although the SNR residual test is passed, the model selection test is not, so we would not conclude that we have detected a GR deviation.  
However, as the SNR increases (or the $\kappa$ effect is increased), one flows upward and rightward in~\cref{fig:stealth-overt}, possibly landing in the Weak Inference of No GR Deviation II or the Incorrect Inference of GR Deviation region. 
\item[] 
\textbf{Weak Inference of No GR Deviation II.} 
This subregime is similar to the previous one, except that one would avoid claiming a false GR deviation because the SNR residual test would not be passed, although the model selection test would be. 
With increasing SNR or for a larger $\kappa$-effect, the inequalities $(\Delta^{\mathrm{ppE}}_L \beta/ \sigma_{\beta})> (\Delta^{\mathrm{ppE}}_L \beta/ \sigma_{\beta})_{\rm dist}$ and $(\Delta^{\mathrm{ppE}}_L \beta/ \sigma_{\beta})>(\Delta^{\mathrm{ppE}}_L \beta/ \sigma_{\beta})_{\rm thesh}$ are both reinforced, meaning that we would stay in this subregime.
When changing the PN order of the ppE parameter, it is possible that the ppE term becomes degenerate with the $\kappa$-term. 
In this scenario, the ppE parameter can absorb the $\kappa$-effect, and the bias can be characterized as an Incorrect Inference of GR Deviation (i.e.~one would flow downward in~\cref{fig:stealth-overt}).
\item[] 
\textbf{Incorrect Inference of GR Deviation.} The worst possible outcome is when the bias is in this subregime, because both the residual SNR and the model selection tests would be passed, potentially leading one to think that the inferred ppE deviation is real, when in reality the signal is consistent with GR.
In this case, increasing the SNR or the magnitude of the $\kappa$ effect only makes matters worse, and the only way out would be to improve the GR sector of the ppE model. 
\end{itemize}

\section{Convergence and robustness checks} \label{sec:convergence}

In the PTMCMC routine of \texttt{GWAT}, the proposal is generated from a carefully tuned mixture of jumps.
The majority of the proposals are based on jumps along the eigenvectors of the Fisher matrix ($60 \%$ for the coldest temperature) and differential evolution ($25\%$ for the coldest temperature).
Gaussian jumps are also used in the proposal mixture, and they are mainly used when the Fisher jumps fail~\cite{Perkins:2022fh}.
We set the number of chains per temperature with \texttt{ensembleN}$=10$, the number of temperature rungs (that range from $1/T = 0$ to $1/T=1$) with \texttt{ensembleSize}$=30$, and the number of independent samples per chain with \texttt{independentSamples}$=2000$.
We arrived at these sampler settings after performing several test runs to ensure convergence and robustness, which we outline in the following.

\begin{figure}[t]
\centering
\includegraphics[width=0.45\textwidth]{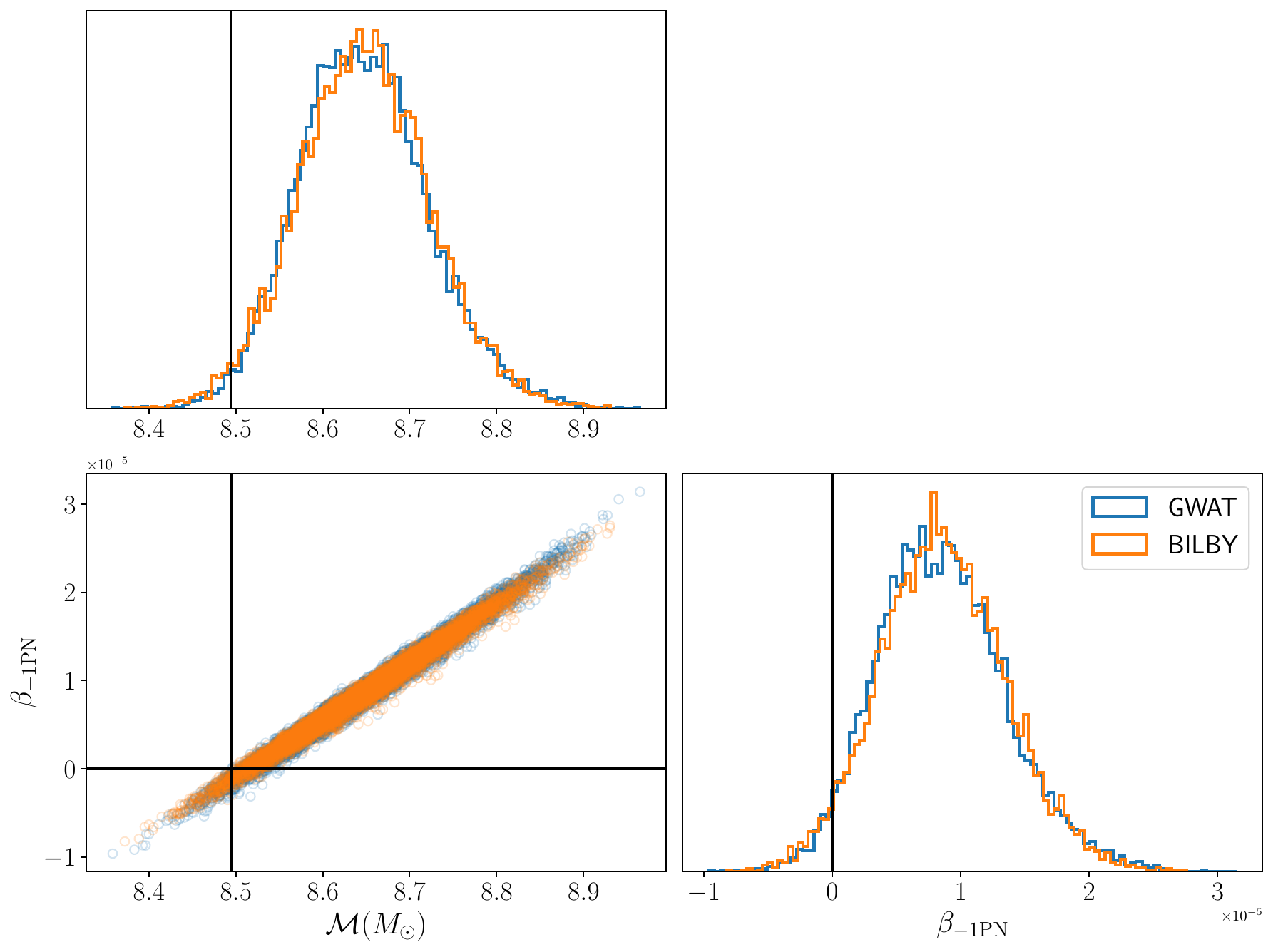}
\caption{Comparison of \texttt{GWAT} and \texttt{BILBY}, showing that the two samplers yield statistically consistent results for the systematic bias.
The injected signal is an edge-on $(12,8)\msun$ binary with $\chi_p^{\mathrm{inj}}=0.9$.}
\label{fig:corner_m1PN_edgeOn_sampler}
\end{figure}

We initially performed several test injection--recovery runs using the PTMCMC sampler with \texttt{IMRPhenomD} and varied both \texttt{ensembleN} $\in \{6,8,10\}$, \texttt{ensembleSize} $\in \{20,25,30\}$ and \texttt{independentSamples} $\in \{1000,2000\}$. 
For each run, we computed the average autocorrelation length and ensured that it is much smaller than the number of independent samples.
The total number of independent samples roughly corresponds to \texttt{ensembleN}$\times$\texttt{ensembleSize}, which we consistently got for all the runs.
An indication of a run that has not converged is when insufficient independent samples are harvested, with autocorrelation length being comparable to the number of independent samples.
We did not find significant differences in the evidences for each run and chose the highest resolution setting of \texttt{ensembleN}$=10$, \texttt{ensembleSize}$=30$, and \texttt{independentSamples}$=2000$ for performing the runs shown in Sec.~\ref{sec:bias_prec}.
We also checked for convergence in the biased case of recovering with \texttt{IMRPhenomD}, given the $(12,8) \msun$ edge-on injection with $\chi^{\mathrm{inj}}=0.9$.
We fixed \texttt{independentSamples} to $2000$ and compared the run with \{\texttt{ensembleN}$=8$, \texttt{ensembleSize}$=25$\} against the \{\texttt{ensembleN}$=10$, \texttt{ensembleSize}$=30$\} run. We found that there was statistical consistency with the harvested chains.
Using 30 temperature rungs is standard with PTMCMC methods, as discussed also in Ref.~\cite{Veitch:2014wba}.
Following Refs.~\cite{Veitch:2014wba} and~\cite{Toubiana:2023cwr}, we also check the $\chi^2$ nature of the log-likelihood distribution.
We find that the variance of the log-likelihood distribution from the runs is consistent with the $\chi^2$ prediction.
Further, we also find that computing the maximum log-likelihood based on the $\chi^2$ estimate (as suggested by~\cite{Toubiana:2023cwr}) is consistent with a direct $\mathrm{argmax}[\log \mathcal{L}]$ estimate.

When performing parameter estimation using the \texttt{BILBY} pipeline, we use the nested sampling~\cite{Skilling:2006gxv} algorithm of the \texttt{dynesty}~\cite{dynesty} sampler with the following sampler configuration options: \texttt{sample}$=$`\texttt{rwalk}', \texttt{nlive}$=2000$, \texttt{nact}$=10$, \texttt{dlogz}$=0.01$, \texttt{bound}$=$`\texttt{live}', and \texttt{maxmcmc}$=5000$.
We again arrived at these sampler settings after performing several convergence checks, which we detail below. 

\begin{figure}[t]
\centering
\includegraphics[width=0.45\textwidth]{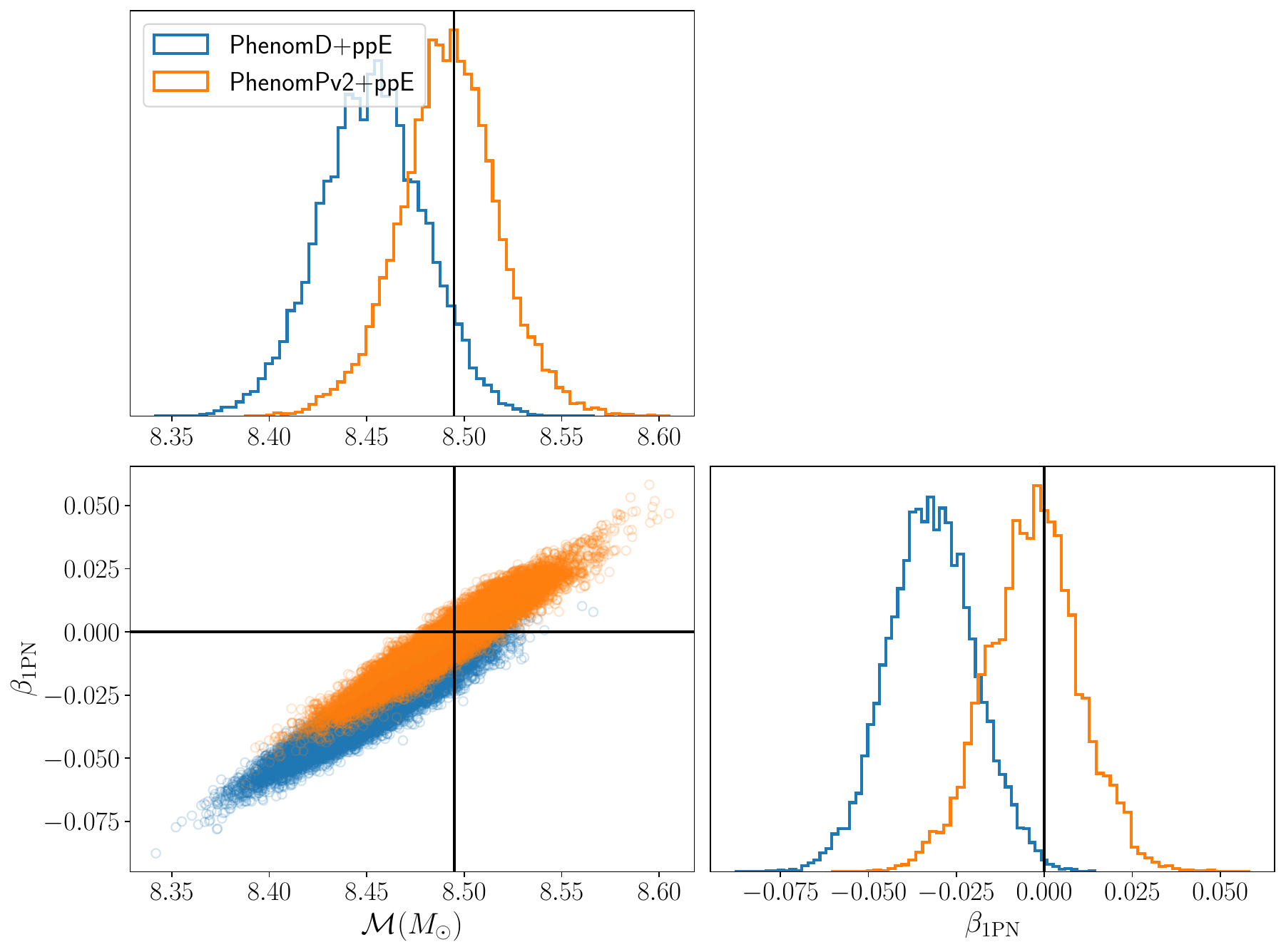}
\caption{Comparison of PhenomD and PhenomPv2 recovery in testing 1PN deviation.
The precessing ppE model recovers the injection accurately, while the nonprecessing ppE model does not.}
\label{fig:corner_chirpM_beta1PN_Pv2_recovery}
\end{figure}

We performed parameter estimation runs (independent from those performed using \texttt{GWAT}) using the \texttt{BILBY} pipeline, where we used the \texttt{dynesty} sampler.
In~\cref{fig:corner_m1PN_edgeOn_sampler}, we show that both the \texttt{BILBY} and \texttt{GWAT} pipelines yield consistent results for the recovery of $\mathcal{M}$ and $\betappe{-1}$, given an edge-on $(12,8)\msun$ injection with $\chi_p^{\mathrm{inj}}=0.9$.
In addition, we checked the results for the 1PN and 2PN recoveries for this injection.

\begin{figure*}[t]
    \centering
    \includegraphics[width=0.8\textwidth]{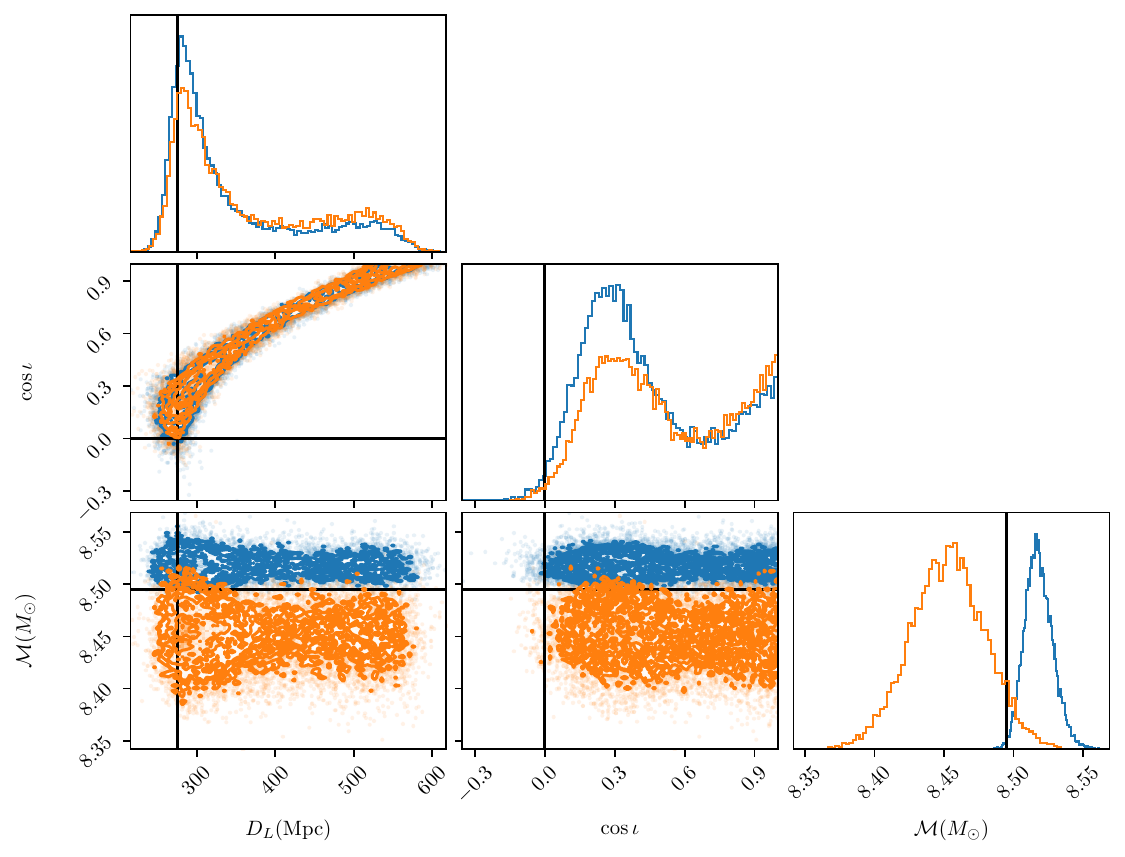}
    \caption{Marginalized 2d and 1d posteriors of $\{ D_L , \cos \iota, \mathcal{M} \}$ recovered using the ppE and GR models for the edge-on $(12,8) \msun$ injection with $\chi_p^{\mathrm{inj}} = 0.9$.
    Observe that the marginalized posteriors in the ppE (with 1PN deviation) recovery are broader as the model contains an additional parameter.
    This figure is a companion to~\cref{fig:corner_extrinsic_intrinsic_edgeOn_precession}.
    }
    \label{fig:corner_GR_ppE_compare}
\end{figure*}

Finally, we also performed a sanity check by recovering with \texttt{IMRPhenomPv2}+\texttt{ppE} for the precessing injections.
We used \texttt{GWAT} for this analysis.
We sampled on $\{a_1, \cos \theta_1, \phi_1\}$ for the primary spin, and likewise for the secondary, using agnostic uniform priors for these spin parameters.
For the other parameters of the model, we followed the analysis outlined in Sec.~\ref{sec:bayesian_framework}.
In~\cref{fig:corner_chirpM_beta1PN_Pv2_recovery} we show the recovered $\mathcal{M}$ and $\betappe{1}$ posteriors for both the precessing ppE and nonprecessing ppE models.
The precessing ppE model \texttt{IMRPhenomPv2}+\texttt{ppE} recovers the injected parameters accurately, as it should.

\section{Additional details on systematic biases in PTMCMC/nested sampling parameter estimation} \label{sec:additional-figures}

In this appendix, we provide additional figures and tables to support the main text.

In~\cref{fig:corner_GR_ppE_compare} we show an example of how the ppE and GR recoveries compare with one another, complementing the results shown in Sec.~\ref{sec:bias_prec}.
We consider an edge-on $(12,8) \msun$ injection with $\chi_p^{\mathrm{inj}} = 0.9$.
The marginalized posteriors in the parameters corresponding to the ppE recovery are larger than those of the GR recovery, owing the the ppE model having an extra parameter.
Consequently, the bias in $\mathcal{M}$ is less significant in the ppE recovery than in the GR recovery.
In addition, the sign of the bias for $\mathcal{M}$ switches because the injected phase is partially absorbed by $\betappe{1}$, as we showed in the toy example (see Sec.~\ref{subsec:toy}).
Note that there is bimodality in the inclination even when using the GR recovery.

\begin{figure*}[ht!]
    \centering
    \includegraphics[width=\textwidth]{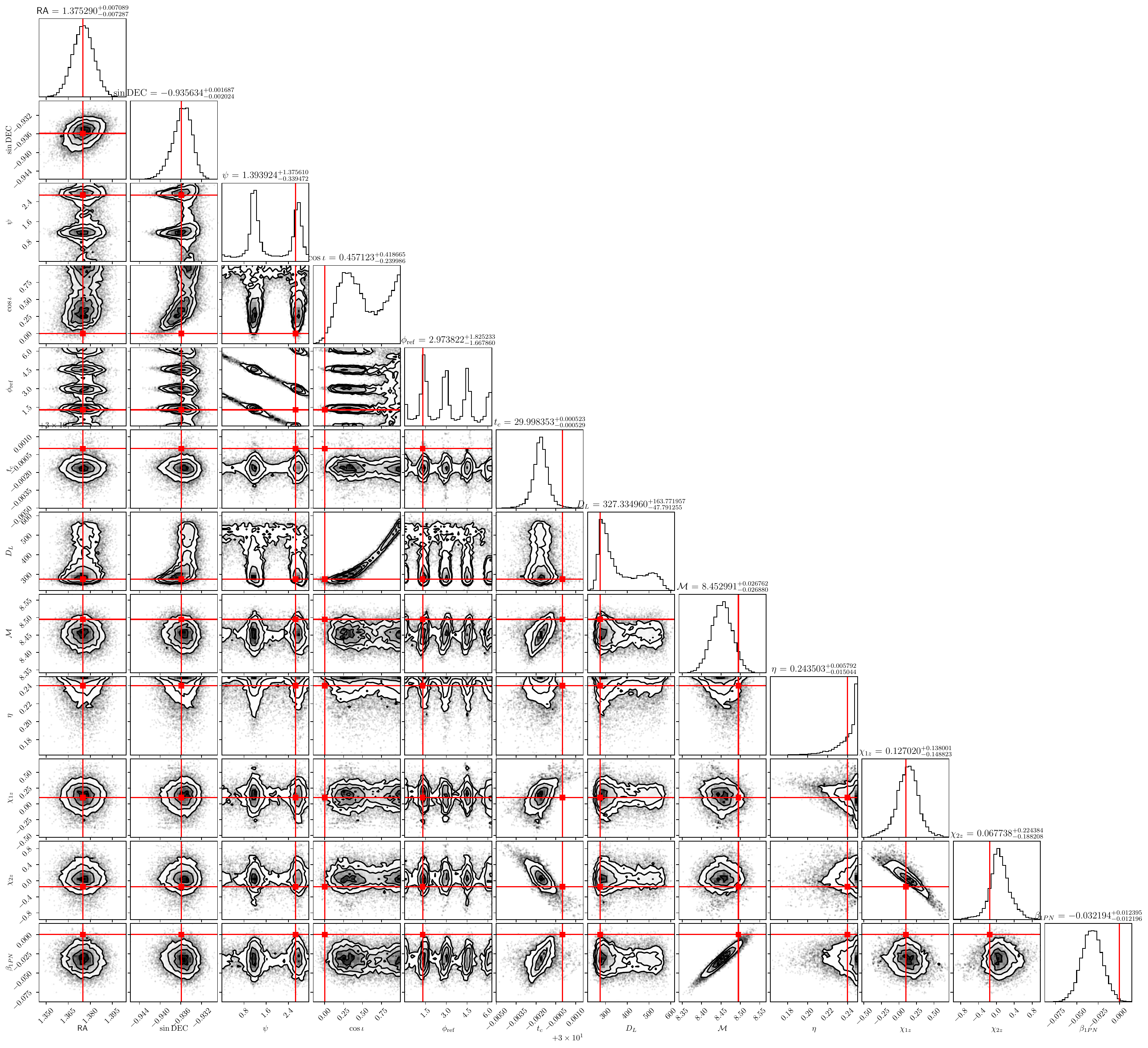}
    \caption{Marginalized 2d and 1d posteriors of all parameters recovered in the ppE model (with 1PN deviation) from the edge-on $(12,8) \msun$ injection with $\chi_p^{\mathrm{inj}} = 0.9$.
    The injected values are shown with red solid lines, and the marginalized posteriors are shown with black histograms.
    This figure is a companion to~\cref{fig:corner_extrinsic_intrinsic_edgeOn_precession}.
    }
    \label{fig:full_corner}
\end{figure*}

In~\cref{fig:full_corner}, for completeness, we show the full corner plot with 1d and 2d marginalized posteriors of all model parameters when testing for a 1PN ppE deviation, given an edge-on injected signal with $(12,8)\msun$ masses and $\chi_p^{\rm inj}=0.9$.
The 2d marginalized posteriors show the correlations between different parameters.
How well the injected parameters (in red solid lines) are recovered is shown by the 1d marginalized posteriors.
Note that apart from the systematic biases in the intrinsic parameters, which was the focus of the main text, there are also systematic biases in the extrinsic parameters.

\begin{figure}[ht!]
\centering
\includegraphics[width=0.45\textwidth]{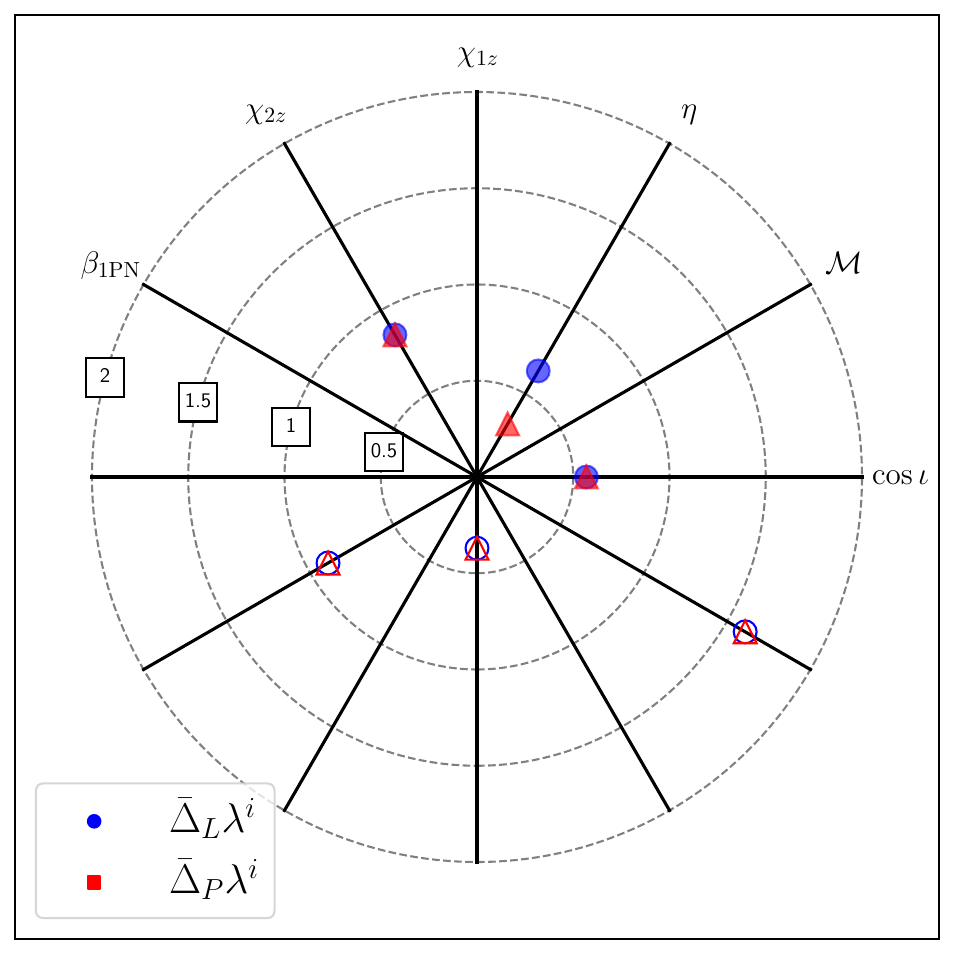}
\caption{Comparison of maximum likelihood (blue circles) and maximum posterior based systematic errors (orange squares). 
The spoke-wheel plot shows the normalized systematic errors in recovered parameters $\BF{\lambda} = \{ \cos \iota, \mathcal{M}, \eta, \chi_{1z}, \chi_{2z}, \betappe{1} \}$ for an edge-on $(12,8) \msun$ injection with $\chi_p^{\mathrm{inj}} = 0.9$. 
Observe that the two definitions of systematic error are consistent with all parameters except $\eta$ because the posterior on $\eta$ peaks at the edge of the prior (the prior on $\eta$ becomes singular at $\eta =1/4$).
This figure is a companion to~\cref{fig:wheel_plot_chip}.
}
\label{fig:wheel_ML_MP}
\end{figure}

As mentioned in the main text, the 1d marginalized posterior on $\eta$ in~\cref{fig:full_corner} is peaked at the edge of the prior, mainly due to the well-known divergence of the Jacobian in transforming from $\{m_1,m_2\}$ to $\{ \mathcal{M},\eta \}$. 
Recall that the Jacobian factor in the log prior is proportional to $\log \left[\mathcal{M}^2 /(\eta^{6/5}\sqrt{1-4\eta})  \right]$, which diverges as $\eta \rightarrow 1/4$.
Due to this divergence of the prior, even when the injected signal has no spin precession, the 1d marginalized posterior on symmetric mass ratio will not typically peak at the injection.
Therefore, there is a sampling bias in the recovery of the symmetric mass ratio, and this becomes clear when we compute the posterior-based systematic error $\Delta_P \lambda^i$, as opposed to the likelihood-based systematic error $\Delta_L \lambda^i$.

In~\cref{fig:wheel_ML_MP} we show the comparison of the absolute ratio of systematic to statistical error (normalized systematic error) across parameters $\{ \cos \iota, \mathcal{M}, \eta, \chi_{1z}, \chi_{2z}, \beta_{\rm 1PN} \}$, when using the two definitions of systematic error.
The normalized posterior-based systematic errors $\bar{\Delta}_P \lambda^i$ are shown by the red triangles, while the normalized likelihood-based systematic errors $\bar{\Delta}_L \lambda^i$ are shown by the blue circles.
The injected signal is an edge-on $(12,8)\msun$ system with $\chi_{p}^{\rm inj}=0.9$.
The normalized likelihood-based systematic errors for this injection-recovery analysis were shown in~\cref{fig:wheel_plot_chip}.
From~\cref{fig:wheel_ML_MP}, we see that both $\bar{\Delta}_L \lambda^i$ and $\bar{\Delta}_P \lambda^i$ are consistent with one another for all parameters, except for the symmetric mass ratio.
As argued above, the discrepancy in the case of symmetric mass ratio is due to the divergent nature of the prior.

\begin{figure}[ht!]
\centering
\includegraphics[width=0.45\textwidth]{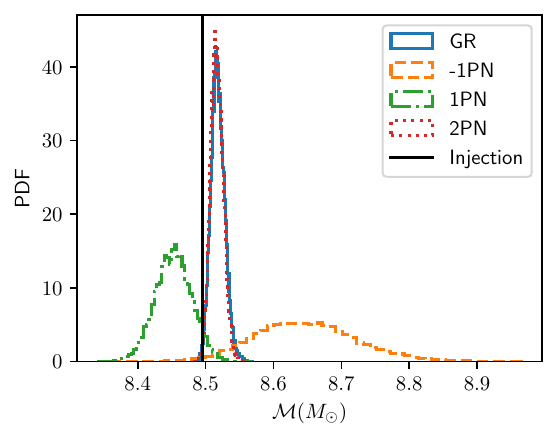}
\caption{Bias in chirp mass when recovering with different ppE deviations.
Observe that when recovering with a 2PN deviation, the chirp mass posterior is entirely consistent with the GR recovery.
The systematic and statistical errors of the chirp mass increase with the decreasing PN order of the ppE deviation.
This figure is a companion to~\cref{fig:chirpM_beta_bias_diff_ppE} and to~\cref{tab:stat-sig-prec-ppE}.
}
\label{fig:chirpM_bias_diff_ppE}
\end{figure}

When studying the dependence of the biases on the PN order of the ppE deviation, we showed with~\cref{fig:chirpM_beta_bias_diff_ppE} and with the calculations in Sec.~\ref{subsec:toy} the importance of correlations between $\mathcal{M}$ and $\betappe{(5+b)}$.
In~\cref{fig:chirpM_bias_diff_ppE} we supplement that discussion by showing how the 1d marginalized posterior of chirp mass varies with the PN order of the ppE parameter.
As predicted by the toy model, the statistical error in measuring $\mathcal{M}$ decreases with the PN order of the ppE deviation.
Including a -1PN ppE deviation results in the widest posterior on $\mathcal{M}$ (orange dashed histogram), while including a 2PN ppE deviation results in the tightest posterior on $\mathcal{M}$ (red dotted histogram).
In fact, when including a 2PN ppE deviation, the posterior on $\mathcal{M}$ is consistent with that of the GR recovery (absence of any ppE deviation) due to the weak correlation between $\mathcal{M}$ and $\betappe{2}$.
The sign of the systematic error for the different recoveries is again due to the correlations between the chirp mass and ppE deviation.

\clearpage

\bibliography{astrophysical-waveform-systematics}

\end{document}